%% file: main.tex
\documentclass{imag-ms-template}

\usepackage{chngcntr}
\usepackage[mathcal]{eucal}
\usepackage{faktor}
\usepackage[symbol]{footmisc}
\usepackage[T1]{fontenc}    
\usepackage{hyperref}
\usepackage[utf8]{inputenc} 
\usepackage{lineno}
\usepackage{microtype}      
\usepackage{nicefrac}       
\usepackage{pifont}
\usepackage{suffix}
\usepackage{tensor}
\usepackage{tikz-cd}
\usepackage{tikz-qtree}     
\usepackage{url}            
\usepackage{verbatim}
\usepackage{wrapfig}
\usepackage{xr}             
\usepackage[savepos,totpages,titleref,dotfill,counter,user]{zref}             
\usepackage{zref-xr,zref-user}             

\hypersetup{
    colorlinks,
    linkcolor={red!50!black},
    citecolor={blue!50!black},
    urlcolor={blue!80!black}
}

\usepackage{subfiles}                           

\usepackage[export]{adjustbox}
\usepackage{multirow}
\usepackage[table,xcdraw]{xcolor}

\newcommand{\cmark}{\textcolor{green}{\ding{51}}}
\newcommand{\xmark}{\textcolor{red}{\ding{55}}}

\newcommand*{\figuretitle}[1]{%
    {\centering
    \large
    \textbf{#1}
    \par\medskip}

}

\newcommand{\cD}{\mathcal{D}}

\newcommand{\cF}{\mathcal{F}}

\newcommand{\cH}{\mathcal{H}}

\newcommand{\cN}{\mathcal{N}}

\newcommand{\cP}{\mathcal{P}}

\newcommand{\cS}{\mathcal{S}}

\newcommand{\bbN}{\mathbb{N}}

\newcommand{\bbR}{\mathbb{R}}

\newcommand{\bbZ}{\mathbb{Z}}

\newcommand{\sD}{\mathscr{D}}

\newcommand{\sP}{\mathscr{P}}

\renewcommand{\hat}{\widehat}
\renewcommand{\tilde}{\widetilde}
\renewcommand{\phi}{\varphi}
\renewcommand{\epsilon}{\varepsilon}
\renewcommand{\bar}{\overline}

\providecommand{\norm}[1]{\left\lVert #1 \right\rVert}
\providecommand{\abs}[1]{\left\lvert #1 \right\rvert}

\DeclareMathOperator{\corr}{corr}

\DeclareMathOperator{\im}{im}

\DeclareMathOperator{\VR}{VR}
\DeclareMathOperator{\PH}{PH}
\DeclareMathOperator{\PD}{PD}
\DeclareMathOperator{\dgm}{dgm}
\DeclareMathOperator{\Jac}{Jac}
\DeclareMathOperator{\Pers}{Pers}

\DeclareMathOperator{\Sym}{Sym}

\definecolor{ForestGreen}{RGB}{34,139,34}
\numberwithin{theorem}{section}
\numberwithin{equation}{subsection}

\begin{document}

\title{Using topological data analysis to compare inter-subject variability across resting state functional MRI brain representations}

\author{
 Ty Easley,$^{1,4\ast}$ Kevin Freese, Elizabeth Munch,$^{2,3}$ Janine Bijsterbosch$^{1\ast}$\\
 {\small $^{1}$Mallinkrodt Institute of Radiology, Washington University in Saint Louis,}\\
 {\small $^{2}$Dept of Computational Mathematics,
 Science and Engineering, Michigan State University}\\
 {\small $^{3}$Dept of Mathematics, Michigan State University}\\
 {\small $^{4}$Dept of Mathematics, Washington University in Saint Louis}\\
 {\small $^\ast$Correspondence:  tyoeasley@gmail.com, janine.bijsterbosch@wustl.edu}\\
}
\date{}

\maketitle

\begin{abstract}
In neuroimaging, extensive post-processing of resting-state functional MRI (rfMRI) data is necessary for its application and investigation in relation to brain-behavior associations. Such post-processing is used to derive brain representations, lower dimensional feature sets used for brain-behavior association studies. A brain representation involves a choice of dimension reduction (a parcellation into regions or networks) and a choice of feature type, such as spatial topography, connectivity matrix, amplitude. However, widespread variability in rfMRI brain representations has hindered both reproducibility and knowledge accumulation across the field. Brain representation choice effects measurements of inter-subject variability, which muddies the comparison and integration of findings. We leveraged persistent homology on the subject-space topologies induced by 34 different brain representations to enable direct comparison of brain representations in the context of individual differences. Our findings reveal the importance of considering feature type when comparing results derived from different brain representations, suggesting best practices for assessing the replicability and generalizability of brain-behavior research in rfMRI data.
\end{abstract}

\newpage

\section{Introduction}

Resting-state functional MRI (rfMRI) captures rich imaging of human whole-brain neural activity in the absence of external demands, enabling the investigation of individual differences in the principles of functional organization.
After acquisition, rfMRI data typically undergoes extensive preprocessing (\cite{glasserMinimalPreprocessingPipelines2013}) before deriving a reduced-dimension feature set. 
A choice of reduced-dimension neural feature set -- or "brain representation" -- consists of a spatial decomposition within each subject's neural data (e.g., parcellation into areas or networks) coupled with a choice of feature set extracted from that decomposition (such as spatial maps of topography or matrices of connectivity; \cite{Bijsterbosch2020-nd}). 
Although the resulting brain representations are derived from the same dataset, direct comparison is often impossible due to extensive differences in the type and dimensionality of the resulting feature spaces. 
The wide range of brain representations carry varied strengths, focus on different questions, and emphasize complementary aspects of functional organization. 
However, this proliferation of practices also produces adverse effects: it creates challenges to clinical translation, silos basic research efforts, and slows the accumulation of knowledge across the field (\cite{Bijsterbosch2020-nd}). 
This inability to compare brain representations results in widespread analytical variability analogous to that observed in task-based fMRI (\cite{botvinik-nezerVariabilityAnalysisSingle2020}). 
In this paper, we articulate a comparison framework for brain representations and employ it to highlight relationships between a cross-section of widely used methods.

Prior efforts to compare brain representations anchor their comparisons of brain representations in brain-behavior relationships.
In \cite{dadiBenchmarkingFunctionalConnectomebased2019}, more than 240 different analysis pipelines were benchmarked by behavior classification accuracy between the extreme upper and lower thirds of a behavioral measure. The results showed little variation in overall classifier performance between function-derived brain representations.
A similar picture emerges from \cite{kongComparisonGradientsParcellations2023}, which found similar brain-behavior associations across brain representations, with the exception of a diffusion gradient method (\cite{Margulies2016-na}). 
These findings suggest a large degree of behavior-predictive variance is preserved across different brain representations. 

Importantly, relegating brain representation comparisons to behavioral data provides only a limited window.
Although informative, brain-behavior associations suffer from low effect sizes (\cite{marekReproducibleBrainwideAssociation2022a}) potentially driven by limited reliability of behavioral phenotypes (\cite{nikolaidisSuboptimalPhenotypicReliability2022}), placing a ceiling on the strength of brain-behavior associations as a result of limited joint reliability (\cite{gellBurdenReliabilityHow2023}).
Furthermore, \cite{liMovingProcessingAnalysisrelated2024}'s recent research into the analytical variability introduced by preprocessing pipeline choices suggests the need for evaluations of consensus that move beyond the horizon of behavioral test-retest reliability. Our approach quantitatively compares brain representations without anchoring such comparisons in behavior. 

Another avenue for the comparison of brain representations seeks unified parameterizations underpinning multiple brain representations.
For example, Bolt et al \cite{boltParsimoniousDescriptionGlobal2022a} showed that the variance captured by several different brain representations is retained by the first three components ("spatiotemporal patterns") of a complex-valued principal component analysis (CPCA).
Furthermore, \cite{pangGeometricConstraintsHuman2023} used a basis of structural eigenmodes (i.e., eigenmodes the Laplace-Beltrami operator applied to structural brain data) to describe functional activity, finding that linear combinations of $\approx 200$ structural eigenmodes were able to explain a high degree of variance in subject-averaged functional data.
While a new technique for parsimoniously expressing functional data can yield import insights, it may not provide an articulation of convergence or divergence across methodological choices.
Rather than filter a comparison of brain representations through downstream outcomes (e.g., behavior) or a tertiary analysis of neuroimaging data (creating another brain representation), we aim to compare brain representations directly.

Beyond behavioral or parsimonious comparisons, the absence of a ground-truth knowledge of functional organization also renders the utility of comparison of brain representations in terms of "quality" uninformative. To facilitate quantitative comparisons without relying on behavior, we aim to characterize relationships between the individual difference information preserved by different brain representations.
Assessing the degree of preservation of inter-subject variability between different brain representations is nontrivial, even when relatively low-loss translations between brain representations are possible (\cite{jamisonKrakencoderUnifiedBrain2025}).
For example, our prior work revealed that over 60\% of inter-subject variance in functional connectivity network matrices could be reproduced by inter-subject spatial network organization with constituent timeseries fixed to a group average (\cite{bijsterboschRelationshipSpatialConfiguration2018, Bijsterbosch2019-lo}), showing that very different brain representation can be fruitfully compared by their subject embeddings. 
Building on this work, we develop a quantitative framework leveraging persistent homology to enable comparison across brain representations in terms of the constraints that each brain representation imposes on the observable structure of individual differences.

Our application of persistent homology to neuroimaging data differs from prior work that has applied topological data analysis (TDA) in neuroimaging. 
Previous studies used TDA for direct analyses of structural neuroimaging data (\cite{kanariTopologicalRepresentationBranching2018,sizemoreCliquesCavitiesHuman2018,kanariObjectiveMorphologicalClassification2019}), multimodal neuroimaging data (\cite{giustiTwoCompanyThree2016}), or structure-function correspondence (\cite{reimannCliquesNeuronsBound2017a}).
For example, prior work has used persistent homology to define "scaffolds" of functional networks (\cite{Petri2014-iq,guerraHomologicalScaffoldMinimal2021,nguyenVolumeoptimalPersistenceHomological2024}), and to uncover both "transition hub" brain networks that tracked with task changes and key topological features of brain dynamics' state space (\cite{saggarNewApproachReveal2018,saggarPrecisionDynamicalMapping2022,zhangTemporalMapperTransition2023}).
Instead of applying TDA directly to neuroimaging brain representations, we compare their embeddings of individual variability by computing the persistent homology of the varying subject-by-subject distances matrices they produce.

Prior work has explored persistent homology as a means of comparing varying embeddings of a fixed dataset, but typically under assumptions or within a data regime that do not hold in our problem setting.
For example, prior work in \cite{Paul2017-px,Rieck2015-gw} compares dimension reductions of known manifold embeddings in a low-dimensional, high signal-to-noise ratio regime, which is typically unavailable for brain representations.
Additionally, our approach is similar in motivation and concept to the Representation Topology Divergence (RTD) (\cite{barannikovRepresentationTopologyDivergence2022a,trofimovLearningTopologyPreservingData2023}), which measures differences in the persistence of topological features between different embeddings of the same data. 
However, the RTD presumes a higher degree of topological similarity between embeddings than is likely to hold in our case, and does not afford a comparable avenue for making statistical claims.
In some ways similar to the Representation Topology Divergence, \cite{yoonPersistentExtensionsAnalogous2023}'s method of "analogous bars" also computes the topology of the \textit{difference} between two different finite metric spaces, as opposed to computing the topology of each and measuring a corresponding difference.
Unfortunately, because large numbers of low-persistence bars characterize our data regime, their approach was not computationally tractable in our setting.
Each of these works share important motivational and methodological overlap with our own, but could not be applied directly in our context.
Our approach specializes to the high-dimensional, low-SNR regime of our data by analyzing distributions of persistence bars rather than individual homology generators.

This paper introduces a novel quantitative framework to enable direct comparison between rfMRI brain representations based on subject embeddings. Based on our prior work in \cite{bijsterboschRelationshipSpatialConfiguration2018, Bijsterbosch2019-lo}, we hypothesized that brain representations would capture similar subject embeddings. Our results do indeed show some consistencies (e.g., between amplitude features from different parcellations), but also uncover key differences between brain representations (e.g., spatial maps and network matrices). 
In light of our findings, we emphasize the necessity of a principled treatment of neuroimaging features as intrinsically tied to the brain representations from which they are derived, so that conclusions drawn from neuroimaging data under a given brain representation are always understood as just one piece of a broader inferential context.

\section{Materials and Methods}

\begin{sidewaysfigure}[h]
    \centering
    \fbox{\includegraphics[width=\linewidth,trim={10, 0, 10, 0}]{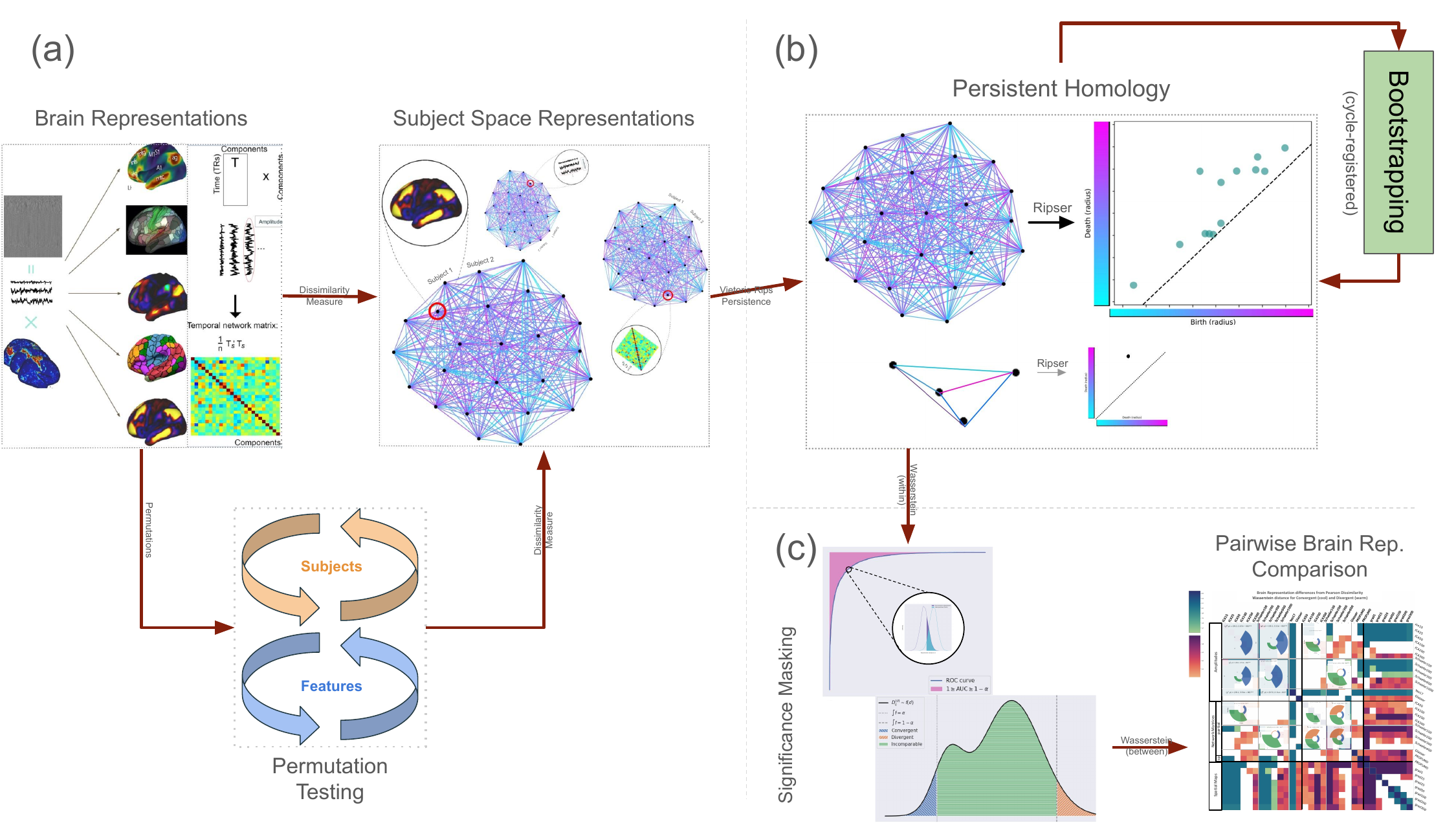}}
    \caption[Schematic of methods.]{A schematic summary of the analysis pipeline. (a) Human Connectome Project (HCP) data is condensed into multiple brain representations, (see the list in Table \ref{tbl:brep_params}). Null versions of these brain representation datasets are also computed. (b) Each brain representation gives rise to an "embedding of subject space" whose persistent homology is computed over bootstraps and the full data. (c) Wasserstein distances between the diagrams of empty, null, and real data are compared in order to mask brain representations that do not demonstrate statistically significant topological signal. Pairwise Wasserstein distances between brain representation diagrams are computed and compared to the distribution of distances between the corresponding null diagrams.}
    \label{fig:methods_pipeline}
\end{sidewaysfigure}

Our goal in this work was to compare brain representations on measurements of inter-subject variability.
An overview of the key analysis steps is provided in figure \ref{fig:methods_pipeline}. Our code is publicly available on GitHub at \cite{Easley_Brain_Representations_2024}.

\subsection{Dataset}\label{sec:dataset}

The data for this study consists of preprocessed resting-state functional MRI data from $N=1003$ Human Connectome Project young adult subjects (HCP-YA; \cite{Glasser2016-ak}).
HCP-YA resting state fMRI data were acquired on a 3T Siemens scanner using the HCP protocol ($TR=0.72$ sec, $TE=33$ ms, isotropic voxel size$=2$mm, multiband factor $8$; \cite{Glasser2016-ak}), and preprocessed according to the Human Connectome Project pipeline (\cite{glasserMinimalPreprocessingPipelines2013}), which includes MSMAll alignment (\cite{robinsonMSMNewFlexible2014}) and ICA-FIX cleanup (\cite{griffantiICAbasedArtefactRemoval2014,salimi-khorshidiAutomaticDenoisingFunctional2014}).
After this preprocessing pipeline, each subject’s data consists of a matrix of 91,282 “grayordinates” (a fixed atlas of graph vertices on the surface of the brain), with 4 runs of 1200 time samples at each grayordinate.
We treat this as our pre-representation source data, giving a source embedding dimension of $D = 91282 \times 1200 \times 4 \approx 10^9$.

\subsection{Data Ethics}

Informed consent was obtained as part of the HCP-YA data collection process (see \cite{bannierOpenBrainConsent2021} for details) for scientific research and online data sharing, and Human Connectome Project protocols were followed for data usage in this project. 
The authors have no known competing interests to declare.

\subsection{Representations of resting-state fMRI data}\label{sec:Breps}

In this section, we detail the choices of  brain representation explored in this work.
The sets of options explored are summarized in in Table \ref{tbl:brep_params}, with corresponding dissimilarity functions specified in figure \ref{fig_sppl:dist_tree}. For the following, denote the HCP-YA dataset $\hat{\cS} = \left\{ X_i \in \bbR^D \right\}_{i=1}^N$ as a set of $N$ spatiotemporal scans $X_i \in \bbR^{V \times t}$, where $V$ is the number of spatial vertices and $t$ the number of timepoints.

\subsubsection{Feature types in resting-state fMRI}

From each brain representation, one or more feature types were computed to reflect the typical use of brain representations in the neuroimaging literature.
The feature types considered in this work are as follows: (1) “amplitude,” the average power of the time signal in a given spatial component; (2) “partial correlation network matrix” (or "partial netmat"), the variance-normalized precision matrix; (3) “map,” the spatial membership weights of a given spatial component in grayordinate space; and (4) “spatial network matrix”, the matrix of pairwise correlations between maps for each spatial component.
The decomposition rank, feature types, and number of features for each brain representation is summarized in Table \ref{tbl:brep_params}.

For the following feature descriptions, we assume a parcellation into $r$ components separates a signal (i.e., some rfMRI scan) $X \in \bbR^{V \times t}$ into low-rank components
\[
X = S \times T,
\]
where $S \in \bbR^{V \times r}$ is the "set of spatial maps" and $T \in \bbR^{r \times t}$ is the "timeseries." We refer to $r$ as the "rank" of the decomposition.
Note that since subject data are encoded in terms of features, it is the feature number and \textbf{not} the parcellation’s decomposition rank that denotes the dimension $d$ of the target embedding space in the mapping $\hat{\varphi}: \hat{S} \to \bbR^d$.
Concretely, $s$ is the number of voxels or grayordinates ("spatial points") in the signal and $t$ is the number of sequential measurements collected at each spatial point ("time point").
When multiple runs exist for a given subject, they are concatenated to a single timeseries $T$.
Note that this model encompasses the "hard parcellation" methods as well, with $r$ the number of parcels, the row vector $[S]_i$ of $S$ given by the (spatial) indicator function of the $i^\text{th}$ parcel, and the $i^\text{th}$ timeseries row $[T]_i$ as the average timeseries of all voxels within that parcel.
We give explicit definitions of the feature types $(1-5)$ given above in terms of $S$ and $T$.

\paragraph{Amplitudes of timeseries components.} The amplitudes $a(T)$ of the timeseries component $T$ is the square root of its component-wise variance: $a_i = \sigma([T]_i)$.
As before, $[T]_i$ is the $i^\text{th}$ row of the timeseries matrix.
The number of amplitude features $a_i$ is equal to the rank of the parcellation.

\paragraph{Functional connectivity (partial correlation networks).}\label{sec:pnets_def} 
The partial correlation $\Pi(T)$ is defined row-wise on $T$.
The $ij$ entry $\Pi_{ij}$ corresponds to the correlation between the residuals of $[T]_i$ and $[T]_j$ after regressing against all $[T]_k$ with $i \neq k \neq j$.
Writing the time correlation matrix $\Sigma \coloneqq \rho(T)$ and the diagonal matrix $R_{ii} \coloneqq \Sigma^{-1}_{ii}$, we may define the partial correlation matrix $\Pi$ as 
\[
\Pi = R^{-1/2} \Sigma^{-1} R^{-1/2}.
\]
The numerical stability of $\Pi$ decreases with increasing $r$, because $\Sigma$ is almost surely non-invertible as $r \to \min(s,t)$.
For this reason, we exclude the partial correlation matrices of the 1000-rank Schaefer parcellation from our analysis.
We implicitly regularize all other partial correlation matrices by inverting $\Sigma$ with the Moore-Penrose pseudo-inverse, which coincides with the $L^2$-regularized inverse in the limit $\lambda \to 0$.
To redistribute the set of partial correlation values to a Gaussian, we apply \cite{Fisher1915-yq}'s z-transformation, otherwise known as the inverse hyperbolic tangent function $\tanh^{-1}$.

\paragraph{Component-wise weight maps of the cortex ("spatial maps").} These are simply the column vectors $[S]_i$ of $S$, the components of the spatial decomposition.

\paragraph{Correlation structure of spatial components ("spatial network matrices").} The "spatial network matrices" are defined as the Pearson correlation between vectorized and $z$-transformed spatial maps: $\Sigma^{(s)} \coloneqq \corr(S^T)$.

\subsubsection{Parcellations and low-rank decompositions of resting-state fMRI}

The brain representations we consider can roughly be grouped by their underlying models of brain function.
We characterize the first group of methods as "hard parcellations" seeking to cluster neural activity into spatially contiguous cortical “parcels.” 
In the parcellation family, we have \cite{Yeo2011-zo}’s parcellated networks, \cite{glasserMultimodalParcellationHuman2016a}’s multimodal parcellation (MMP-1.0), and \cite{schaeferLocalGlobalParcellationHuman2018}’s local-global parcellation.
We also sample from a family of low-rank matrix factorization methods that parse non-contiguous networks of functional activity.
In the factorization family, we include Independent component analysis (ICA; \cite{beckmannProbabilisticIndependentComponent2004,Comon1994-fd,varoquauxGroupModelStable2010a}), PROFUMO (\cite{harrisonLargescaleProbabilisticFunctional2015a,Harrison2020-zb}), which parses “functional modes” of brain activity from hierarchical Bayesian signal models, and the “principal gradient(s)” (\cite{Margulies2016-na}), a diffusion embedding method that organizes brain function through cortical geometry.
Table \ref{tbl:brep_params} summarizes the brain representations chosen for this study, and each parcellation is described in greater detail below.

\begin{table}
\begin{adjustbox}{width=\textwidth}
\begin{tabular}{|c|c|c|c|}
\hline 
\textbf{Parcellation $\phi$}    & \textbf{Rank(s) $r$} & \textbf{Considered Feature Type(s)}   & \textbf{Feature Number(s) $d_i$}   \\ \hline
PROFUMO   & 31   & maps, spatial network matrices & $91282 \times 33$, ${31 \choose 2}$  \\ \hline
\begin{tabular}[c]{@{}c@{}}Dual-regression \\ spatial ICA \end{tabular} & \begin{tabular}[c]{@{}c@{}}15, 25, 50, \\ 100, 200, 300\end{tabular}   & \begin{tabular}[c]{@{}c@{}}amplitudes, partial correlation networks\end{tabular} &  $r$, ${r \choose 2}$, ${r \choose 2}$  \\ \hline
Glasser & 360  & \begin{tabular}[c]{@{}c@{}}amplitudes, partial correlation networks\end{tabular} & $360$, ${360 \choose 2}$, ${360 \choose 2}$  \\ \hline
Schaefer & \begin{tabular}[c]{@{}c@{}}100, 200, 300,\\ 600, 1000 \end{tabular} & \begin{tabular}[c]{@{}c@{}}amplitudes, partial correlation networks\end{tabular} &  $r$, ${r \choose 2}$, ${r \choose 2}$ \\ \hline
Yeo & 17   & \begin{tabular}[c]{@{}c@{}}amplitudes, partial correlation networks\end{tabular} & $17$, ${17 \choose 2}$, ${17 \choose 2}$ \\ \hline
\begin{tabular}[c]{@{}c@{}}Gradient \\ (diffusion embedding)\end{tabular} & \begin{tabular}[c]{@{}c@{}}1, 15, 25, 50, \\ 100, 200, 300\end{tabular}    & maps & $91282 \times r$    \\ \hline
\end{tabular}
\end{adjustbox}
\caption[Brain representations studied]{The combinations of brain representation, decomposition rank parameters, and feature types investigated in the present work. The feature dimension(s) -- i.e., ambient embedding dimension -- is shown as either a number or as a function of the rank parameter $r$ for the feature types associated with in a given parcellation or reduction algorithm. For the parcellations listed, the rank $r$ is the number of spatial components or parcels.} 
\label{tbl:brep_params}
\end{table}

\include{appendix/brainrep_algs}

\subsubsection{Measuring inter-subject variability}\label{sec:dist_fns}

For each brain representation method, decomposition rank within a given representation, and considered feature type, we compute pairwise distances between all subjects.
Each feature type under consideration is structured either as a vector (maps, amplitudes) or a symmetric positive semidefinite (SPSD) matrix (network matrices).
The primary comparisons were performed using correlation because it enabled comparisons across feature types, which was calculated for vectorized and network matrices as described below. For each combination of brain representation, rank parameter, and feature type shown in Table \ref{tbl:brep_params}, we compute pairwise dissimilarity according to whichever of the two correlation-based measures is relevant.
The subject-pairwise matrix of dissimilarities then forms the dissimilaritymatrix used to compute the persistent homology.
We consider the impact of other choices of dissimilarity function in supplemental investigation \ref{sec:metric_supp_res}.

\paragraph{Correlation dissimilarity for vectorized datatypes.}

Suppose $s_i$ and $s_j$ are data vectors in a reduced feature space $\bbR^d$, and let $\rho(s_i, s_j)$ denote their Pearson correlation.
We then define the "correlation divergence"
\begin{equation}
\delta_{v}^2(s_i, s_j) = 1 - \rho^2(s_i, s_j)   \label{eqn:corr_dissim},
\end{equation}
and corresponding "correlation dissimilarity" $\delta_v$. Since $s_i, s_j \in \bbR^d$ for large values of $d$, the angle $\theta_{ij} \in [0,\pi]$ such that $\cos\theta_{ij} = \frac{\langle s_i, s_j \rangle}{\norm{s_i}\norm{s_j}}$ is generically small. In the case of centered data (that is, $s \mapsto s - \frac{1}{d}1^Ts$), we have $\rho^2(s_i, s_j) = \cos^2 \theta_{ij}$ and $\delta_{v}(s_i, s_j) = \sin^2 \theta_{ij} \approx \theta^2_{ij}$. Thus, when $s_i$ and $s_j$ are centered, we can interpret $\delta_v$ as the approximate angular distance between $s_i$ and $s_j$.

\paragraph{Correlation dissimilarity for network matrix datatypes.}

In the SPSD matrix case, we slightly modify the Pearson divergence $\delta_{v}$ for the correlation matrix case by precomposing it with Fisher's z-transformation \cite{Fisher1915-yq}, otherwise known as the inverse hyperbolic tangent function (as described in \S \ref{sec:pnets_def}).
We write
\begin{equation}
\delta_{pd_2}(A_i, A_j) = \tanh^{-1*}\delta_{v}(\hat{a}_i, \hat{a}_j), \label{eqn:dM2}
\end{equation}
where $\hat{a}_i$ is the half-vectorization of the symmetric matrix $A_i$ (diagonal excluded).
This precomposition is necessary for correlation matrices, as it redistributes the correlation values along a Gaussian before re-correlating them.

The choice of dissimilarity measure ("distance metric") between subject pairs heavily informs subsequent analysis. Other "metric" definitions and an analysis of their employment can be found in supplement \ref{sec:supp_inv}.

\subsection{Persistent homology of subject space}\label{sec:TDA}

Our investigation employs persistent homology to compare the topological structure of different representations of subject space. 
We give a brief overview of the relevant tools in this section; see Appendix \ref{appendix:alg_top} for a more comprehensive introduction.
In summary, we compute the Vietoris-Rips persistence using Ripser (\cite{Bauer2021-an, Bauer2022-sq}) and compare diagrams using the Wasserstein distance $W_{p,q}$ with $p=q=2$.

\subsubsection{Vietoris-Rips persistent homology}

We wish to associate a notion of "shape" to a given realization of a dataset. The \textbf{\textit{Vietoris-Rips persistent homology}} allows this by associating summary features $H_k$ of the topological data $VR_r$ to a choice of brain representation $\hat{\cS}_i$ and dissimilarity measure $\delta_j$. We summarize this process as follows:
\begin{align*}
X_{ij} &\coloneqq (\hat{\cS}_i, \delta_j), \\
\VR_r(X_{ij}) &\coloneqq \left\{ S \subset \hat{\cS}_i \mid \delta_j(x,y) < r \text{ for all } x,y \in \hat{\cS}_i \right\}, \\
H_k(\VR_r(X_{ij})) &\coloneqq k^\text{th} \text{ simplicial homology group of } \VR_r(X_{ij}), \\
\PH(X_{ij}) &\coloneqq \left\{ H_k(\VR_r(X_{ij})) \right\}_{r \in \bbR} \text{ and incl. maps}
\end{align*}
See definition \ref{def:VR_cplx} for more details. 
Since the main body of our manuscript restricts to a single choice of $\delta_j$, we will typically suppress the second subscript and refer to the realization of the dataset $X$ under brain representation $\varphi_k$ as $X_k$.

\subsubsection{Wasserstein distance}

Several of our analyses rely on the comparison of persistence results derived from different subject embeddings from brain representation data. 
For example, such comparisons are needed to assess the distribution of bootstrapped results, permuted results, and to pairwise compare brain representations (see below).
For all of these comparisons, we adopted the Wasserstein distance. 
Given a realization $X_i$ of subject space, we computed pairwise Wasserstein distances $W_p(\dgm(X_i), \dgm(X_j))$ to give an initial difference measurement between topological features of brain representation pairs. 
See \S \ref{sec:Wp_dgm} for background on the Wasserstein distance.

\subsection{Existence and stability of topological structure in subject space}

Resting-state functional MRI often faces low SNR and/or small effect sizes in relation to behavior \cite{marekReproducibleBrainwideAssociation2022a}.
As such, it is important to perform rigorous null comparisons to assess the structure induced by our brain representations. 
To verify (a) the existence of topological signal in our data and (b) its relationship to inter-subject variability, we compared the persistence properties of cycle-registered bootstraps and structure-conserving null models.

\subsubsection{Topological bootstrapping}

We implemented topological bootstrapping to assess the stability of the topological features of our data.
Because it is possible (and, in fact, generic) for multiple data elements to define the same homology generator, bootstrapping (\cite{abu-mostafaLearningData2012}) is less straightforward in persistent homology than in many other modes of analysis.
However, \cite{Reani2021-mu} recently demonstrated a cycle registration method that uses image persistence (\cite{Cohen-Steiner2009-yj}) to register homology generators in co-embeddable spaces. 

\paragraph{Implementation} 
The specific challenges of our problem context required a mild extension of existing topological bootstrapping implementations.
Our implementation of the topological bootstrap builds on \cite{Garcia-Redondo2022-bv}'s work, which efficiently integrates cycle registration with Ripser and Ripser-image (\cite{Bauer2021-an,Bauer2022-sq}) and refines the cycle affinity measures proposed by \cite{Reani2021-mu}.
We extended their implementation by broadening the conditions under which topological bootstrapping may be applied to include the case of arbitrary, precomputed metrics.
We conducted cycle registration using $n=1000$ bootstraps per dataset at a sub-sampling ratio of $r=0.90$ (no replacement); we considered $k=1$-dimensional cycle registration. 
Our implementation is branched from \cite{Garcia-Redondo2022-bv} and available on Github (\cite{Easley_Co_homology_cycle-match_for_2024}).

See Appendix \S\ref{sec:topo_bts} for further details on the topological bootstrap. 
Additionally, in Appendix \S \ref{sec:def_prevwt_Wp}, we construct a modified Wasserstein metric between diagrams that have each undergone cycle-matched bootstrapping.

\subsubsection{Structured null models of brain representations}\label{sec:null_model}

Because our brain representation "subject space" datasets are outputs of varying dimension-reduction algorithms on the HCP-YA dataset, they are likely to share substantial structure.
We constructed null models that break specific information structures while retaining as many of the dataset's empirical properties as possible.
Conservative null models give the most stringent standard for the observation of statistically significant differences between real and null data properties (\cite{vasaNullModelsNetwork2022}).

We generated our null via permutation testing (\cite{nicholsControllingFamilywiseError2003}).
Let $\Sym(N)$ denote the symmetric (i.e., permutation) group on $N$ elements.
For the remainder of this section, fix some choice of reduced-dimension $\hat{\cS}_k$ (with reduced ambient embedding dimension $d_k$) and associate to it the $N \times d_k$ matrix $X_k$ of the coordinates of all $N$ points sampled from $\cS_k$.
We considered two types of null distributions: one that removes inter-subject variability (or "community") structure by subject-shuffling, and the other that removes the global structure induced by the dimensionality reduction ("projection" structure) by feature-shuffling. We drew a total of $R=1000$ samples from each null distribution.

\paragraph{Subject-shuffling permutations.}
By breaking community structure in the data without changing its projection structure, we tested the sensitivity of a given brain representation to inter-subject variability.
To operationalize this, we held the \textit{order} of features fixed while shuffling a given feature between subjects independently for each feature in a brain representation. 
After shuffling, the resulting permutation contains the same number and type of features, but with random mixtures of subjects' contribution to each permutation. 
We can give a more formal and precise description of this procuedure:
Choose a length-$d_k$ sequence $(\sigma) \coloneqq \sigma_1, \ldots, \sigma_{d_k}$ of permutations with no fixed points (called "derangements") in $\Sym(N)$. 
A null dataset $X^{(\sigma)}$ corresponding to the sequence $(\sigma)$ is given by the map
\begin{equation}\label{eqn:subj_perms}
    (\sigma) : [X]_{ij} \mapsto [X]_{\sigma_j(i) j}
\end{equation}
Choosing $R$ such sequences $(\sigma)$ from $\Sym(N)$ defines an approximate empirical "subject-null" distribution of data in $\hat{\cS}_k$.

However, due to familial relationships between many members of the Human Connectome Project, not all permutations $\sigma_j \in \Sym(N)$ of subject contributions satisfy the exchangeability criterion necessary for data permutation to generate a valid null distribution (\cite{winklerMultilevelBlockPermutation2015}).
The HCP-YA dataset contains a complex set of family relationships, including $149$ genetically confirmed monozygotic twin pairs and $94$ genetically confirmed dizygotic twin pairs, non-twin siblings and half-siblings.33
Family relationships in the HCP-YA dataset invalidate simple swapping as a viable permutation approach, since not all data are exchangeable.
To properly account for family relationships in permutations, we implemented \cite{winklerMultilevelBlockPermutation2015}'s \textit{exchangeability blocks} to restrict to choices of permutation sequences $(\sigma)$ in $\Sym(n)$ that assume exchangeability holds only within (not across) families.

\paragraph{Feature-shuffling permutations.}
To complement the subject-shuffled null, we also constructed a null retaining community structure while removing projective structure: the "feature-shuffled" null. 
To generate "feature-type" null distributions, we held each subject's feature set fixed while independently shuffling feature order for each subject. 
Data significance with respect to the feature-shuffled null is explored in the supplemental investigation \ref{sec:supp_inv}; see \S \ref{sec:feat_shuff} for more information on the construction of the feature null.

\subsubsection{Topologically non-null data in a brain representation} \label{sec:solosig_criteria}

\begin{figure}[ht]
    \centering
    \includegraphics[width=\linewidth]{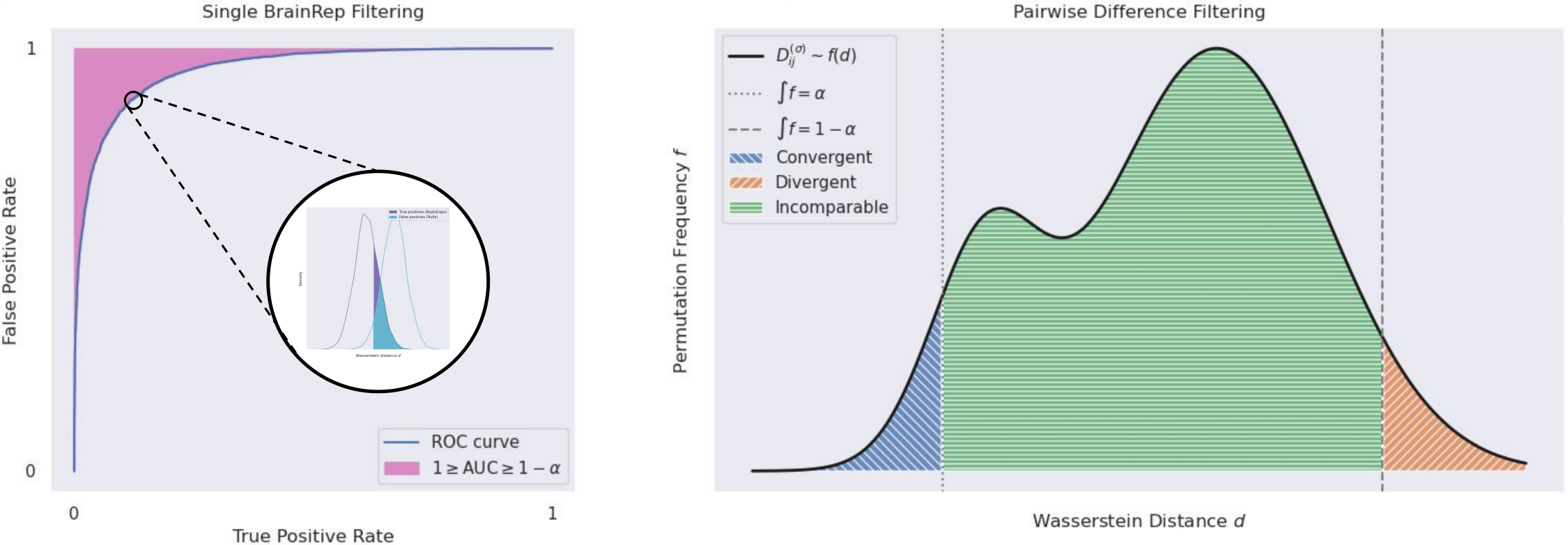}
    \caption[Wasserstein distance ROC curve.]{\textbf{Left}: A schematic showing the computation of the true and false positive rates (TPR and FPR resp.) from distributions of Wasserstein distance to the bootstrap distribution and (subject-type) null distribution from original data at some threshold $\tau$, showing a $0.05$-significantly non-null example. \textbf{Right}: The two-sided significance test of difference between a pair of brain representations.}
    \label{fig:Wp_ROC}
\end{figure}

To assess whether a given embedding of subject space contained non-null topological signal, we compared bootstrapped and null model-derived persistence modules through the respective distributions of their Wasserstein distance (see section \ref{sec:Wp_dgm}) to the persistence diagram of the data. 
Setting $X_k = (\hat{\cS}_k, d_k)$ for some brain representation, we define random variables $\hat{X_k}$ (a bootstrap of $X_k$) and $X_k^{(\sigma)}$ (the null embedding generated by the permutation sequence $(\sigma)$). 
We define real random variables $\hat{D}_{X_k}, \hat{D}_{0,k}, D^{(\sigma)}_{X_k}, D^{(\sigma)}_{0,k}$ from the $p$-Wasserstein distances between the persistence diagrams of the initial (deterministic) $X_k$, $\hat{X_k}$, $X_k^{(\sigma)}$, and the empty diagram $\Delta$:
\begin{align*}
\hat{D}_{X_k} &\sim W_p\left(\dgm\left(X_k\right), \dgm\left(\hat{X}_k\right)\right), &\hat{D}_{0,k} &\sim W_p\left(\Delta, \dgm\left(\hat{X}_k\right)\right), \\
D^{\left(\sigma\right)}_{X_k} &\sim W_p\left(\dgm\left(X_k\right), \dgm\left(X_k^{\left(\sigma\right)}\right)\right),     &D^{\left(\sigma\right)}_{0,k} &\sim W_p\left(\Delta, \dgm\left(X_k^{\left(\sigma\right)}\right)\right).
\end{align*}
The probability $P\left(\hat{D}_{X_k} < D_{X_k}^{(\sigma)}\right)$ is the probability that $\hat{X}_k$ is closer than $X_k^{(\sigma)}$ is to $X_k$. Similarly, the probability $P\left(\hat{D}_{0,k} > D_{0,k}^{(\sigma)}\right)$ is the probability that $\hat{X}_k$ has greater total $p$-persistence than $X^{(\sigma)}_k$ (intuitively, that the null contains "less" topological data). We define the individual difference topology of $X_k$ as being \textbf{\textit{$\alpha$}-significantly non-null} if
\begin{align}
P\left(\hat{D}_{X_k} < D_{X_k}^{(\sigma)}\right) &\geq 1-\alpha \nonumber \\
P\left(\hat{D}_{0,k} > D_{0,k}^{(\sigma)}\right) &\geq 1-\alpha \label{eqn:solosig_criteria}
\end{align}
both hold. Intuitively, we are imposing the requirement that $\dgm({X_k})$ both contain topological information not present in the null data $\dgm(X_k^{(\sigma)})$ and represent it stably. 
These probabilities were calculated by the area under (AUC) the receiver-operator characteristic curve (ROC; \cite{fawcettIntroductionROCAnalysis2006}) where $\hat{D}_\bullet$ is the distribution of "true positives" and $D_\bullet^{(\sigma)}$ the distribution of "true negatives".
See figure \ref{fig:Wp_ROC} (left) for a schematic of the calculation of points along this ROC curve.

\subsection{Stable differences in subject-space topology between brain representations}\label{sec:sigpair_criteria}

The primary goal of this paper was to enable direct comparison between rfMRI brain representations based on subject embeddings. 
We calculated Wasserstein distances between the diagrams of all pairs of brain representations with topologically non-null signal (i.e., those verifying the significance criteria $\ref{eqn:solosig_criteria}$).
To assess the statistical significance of differences, these pairwise diagram distances were compared against distributions of distances between paired null diagrams.
Enumerating the brain representation-reduced datasets $X_i$, we first define the "true" distance $d_{ij}$ between the non-null diagrams and the real random variable $D^{(\sigma)}_{ij}$ of distances between persistence diagrams of nulls $X_i^{(\sigma)}$ and $X_j^{(\sigma)}$:
\begin{align*}
d_{ij} &\coloneqq W_p\left( \dgm(X_i), \dgm(X_j) \right), \\
D^{(\sigma)}_{ij} &\coloneqq W_p\left( \dgm\left(X_i^{(\sigma)}\right), \dgm\left(X_j^{(\sigma)} \right) \right).
\end{align*}
In particular, note that the same null permutation sequence $\sigma$ was applied to both subject space embeddings $X_i$ and $X_j$ when computing distance between diagrams of nulls.
When $X_i$ and $X_j$ have different embedding dimensions (i.e., number of features) and thus permutation sequences of different length, sequences match up to truncation.
A two-tailed significance test between the value $d_{ij}$ and the random variable $D^{(\sigma)}_{ij}$ can suggest that $X_i$ and $X_j$ are either:
\begin{enumerate}[label=(\roman*)]
    \item "Convergent" if 
    \begin{equation}
    d_{ij} \leq D^{(\sigma)}_{ij} \label{eqn:conv_criterion}        
    \end{equation}
    for almost every null permutation $\sigma$ (left-significant);
    \item "Divergent" if 
    \begin{equation}
    d_{ij} \geq D^{(\sigma)}_{ij}  \label{eqn:div_criterion}
    \end{equation}
    for almost every null permutation $\sigma$ (right-significant);
    \item "Incomparable" if the two-tailed test of $d_{ij}$ against $D^{(\sigma)}_{ij}$ is not significant.
\end{enumerate}
Note that a non-significant result still carries interpretive information.
If both left and right alternative hypotheses are rejected, then we are forced to conclude that the in-data structure removed to create the null model does not significantly affect the distance between $\dgm(X_i)$ and $\dgm(X_j)$.
This case still carries meaningful implications for how the information in $\dgm(X_i)$ and $\dgm(X_j)$ must differ.

\subsubsection{False discovery rate control.}
We control for  multiple comparisons by controlling the false discovery rate (FDR).
Viewing $D^{(\sigma)}_{ij}$ as distributed over permutation sequences $(\sigma)$, we empirically approximate the \textit{uncorrected} $p$-values $\tilde{P}_{ij}\left( D^{(\sigma)}_{ij} \leq d_{ij} \right)$ from null-derived diagram distance distributions over $1000$ choices of $(\sigma)$.
We then modify the distribution $\tilde{P} = \cup_{i<j} \tilde{P}_{ij}$ via \cite{benjaminiControlFalseDiscovery2001}'s step-up procedure to obtain the FDR-controlled distribution $P$.

\subsubsection{Summarizing systematic differences between brain representations}\label{sec:BRgroups_chi2}

As described in section \S \ref{sec:sigpair_criteria} above, pairs of brain representations can topologically structure inter-subject variability in convergent, divergent, or incomparable topological structure of inter-subject ways. To summarize patterns of comparisons between brain representations, we pooled counts of convergent, divergent, or incomparable results to test for systematic relationships as a function of feature type (e.g., all pairs of amplitude representations) and/or parcellation. We conducted $\chi^2$ tests for independence and homogeneity to test against a null hypothesis of a uniform distribution of convergent, divergent, and incomparable comparisons within each subgroup using a significance threshold of $\alpha = 0.05$.

\newpage

\section{Results}

\subsection{Existence and stability of topological signal in subject space}

Our analysis found that a large majority (over 90\%) of brain representations captured non-trivial information about inter-subject variability (Table \ref{tbl:psim_sig_solo-breps}). 
Of the 34 brain representations listed in Table \ref{tbl:brep_params}, 31 satisfied the significance criteria in equations \ref{eqn:solosig_criteria} against the subject-type null at $\alpha = 0.05$. 
The three brain representations that did not exhibit subject-structure significance were the partial correlation matrices for the Yeo parcellation and the 15- and 25-dimensional ICA parcellations (\textbf{bolded} in Table \ref{tbl:psim_sig_solo-breps}).
In particular, only \textbf{\textit{low-rank}} $(<50)$ partial functional connectivity matrices failed our significance criteria.
Notably, as we show in supplemental investigation \ref{sec:supp_inv}, none of these functional connectivity matrices failed significance criteria when their pairwise subject dissimilarity was computed using the geodesic distance on the symmetric positive-definite cone (see table \ref{tbl:sig_solo-breps}).
The significance of difference between original and null data was stable across a range of significance thresholds (see supplementary figure \ref{fig:AUC_significance_threshlines}).

\begin{figure}[h]
    \centering
    \fbox{
    \includegraphics[width=0.48\linewidth]{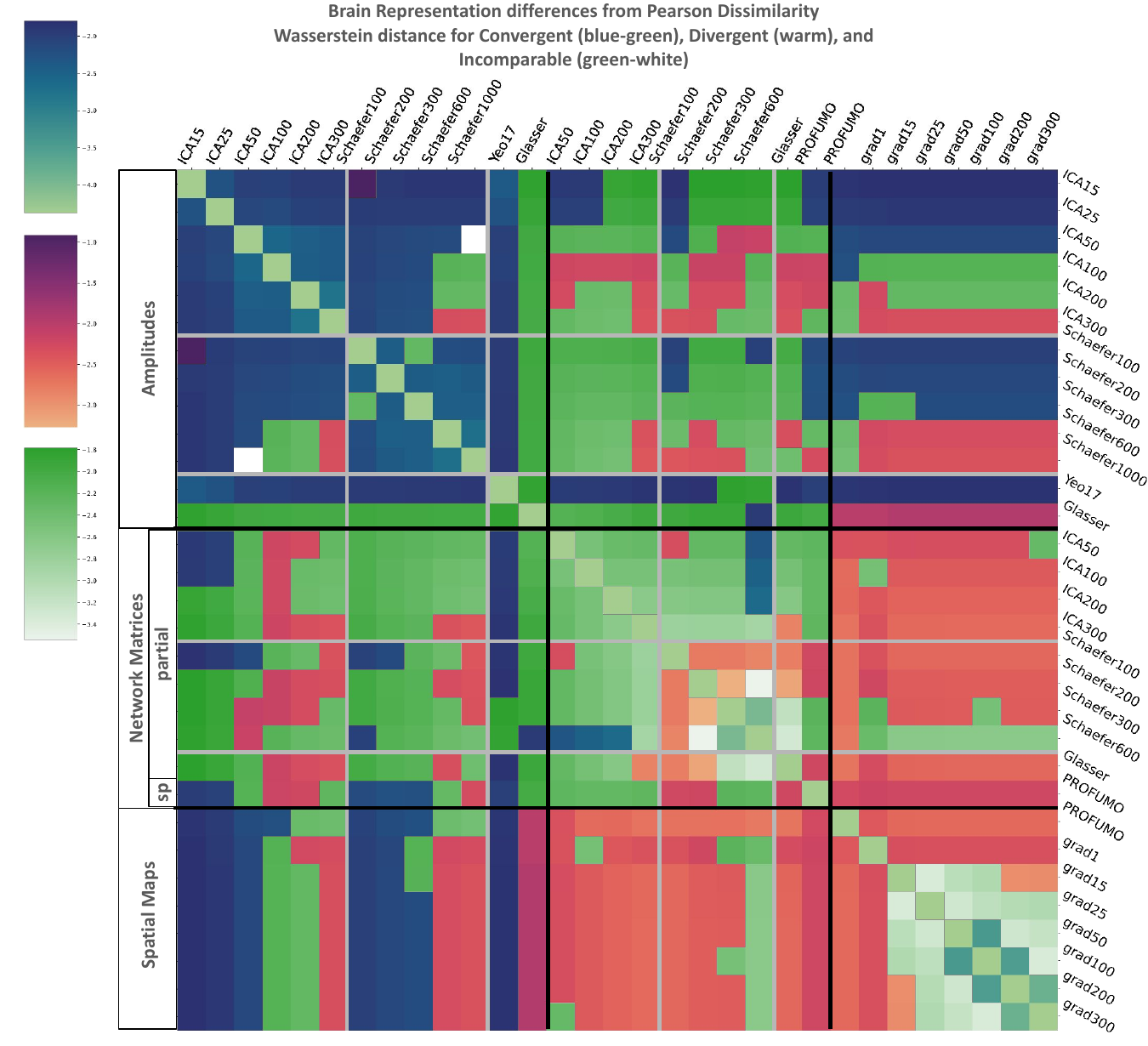}
    \includegraphics[width=0.48\linewidth]{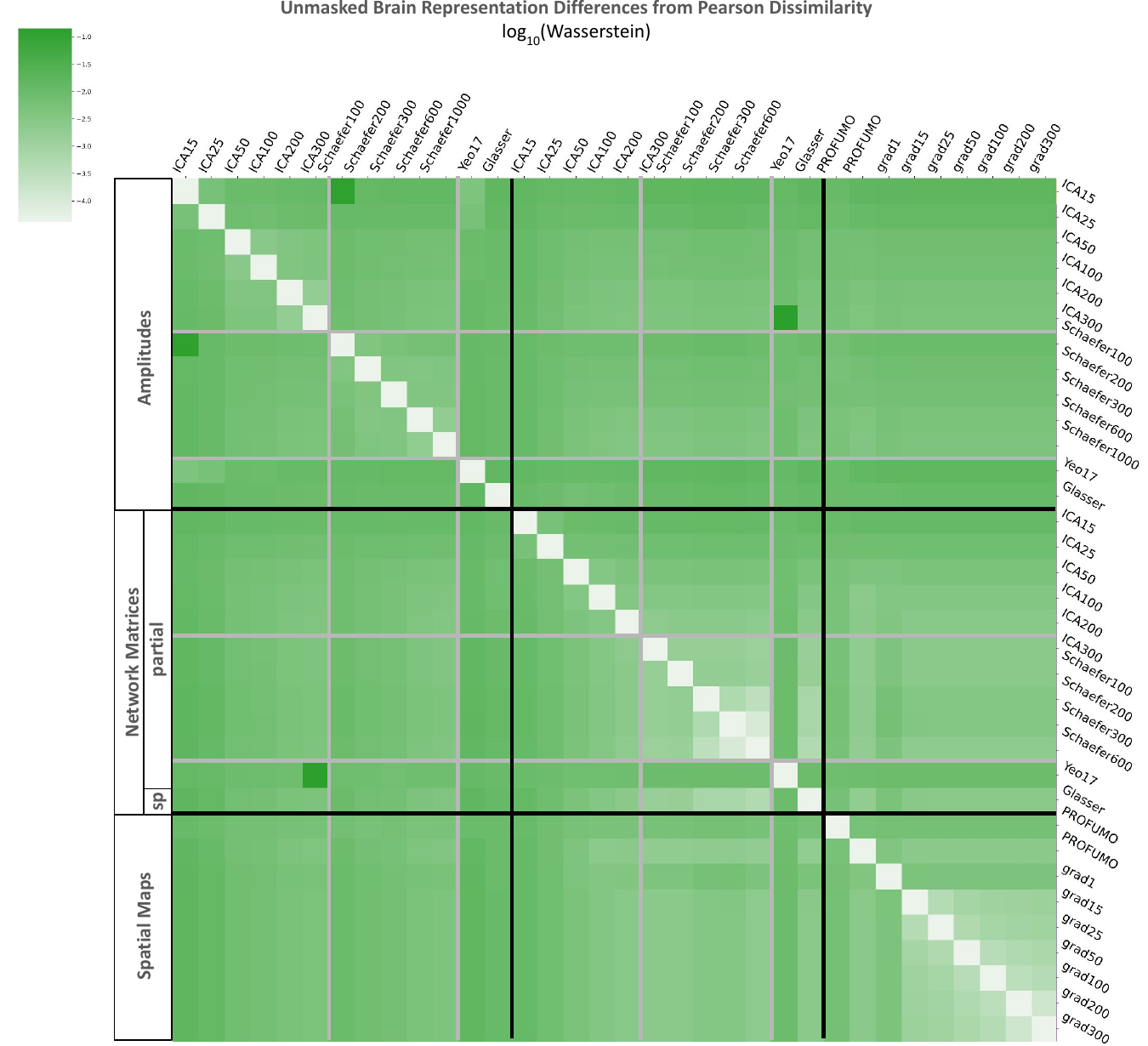}
    }
    \caption[Pairwise Wasserstein distances.]{Pairwise Wasserstein distances between brain representations. The Wasserstein distance between a pair of brain representations is not necessarily a good predictor of the distribution of significance counts. Rather, both the spread and center of the null-distance distribution play larger roles in determining the convergence, divergence, or incomparability of a pair of representations, and their relative importance varies over representation pairs (see figure \ref{fig:masked-unmasked_WpXY_distsummary}). The significance-masked Wasserstein grid \textbf{(left)}, colored by convergence, divergence, or incomparability, is shown next to its unmasked counterparts \textbf{(right)}.}
    \label{fig:masked-unmasked_WpXYconsensus}
\end{figure}

\include{tables/psim_significance_table}

\subsection{Topologically non-null differences between brain representations}\label{sec:res_btwnBRs}

\begin{figure}
    \centering
    \fbox{\includegraphics[width=\linewidth]{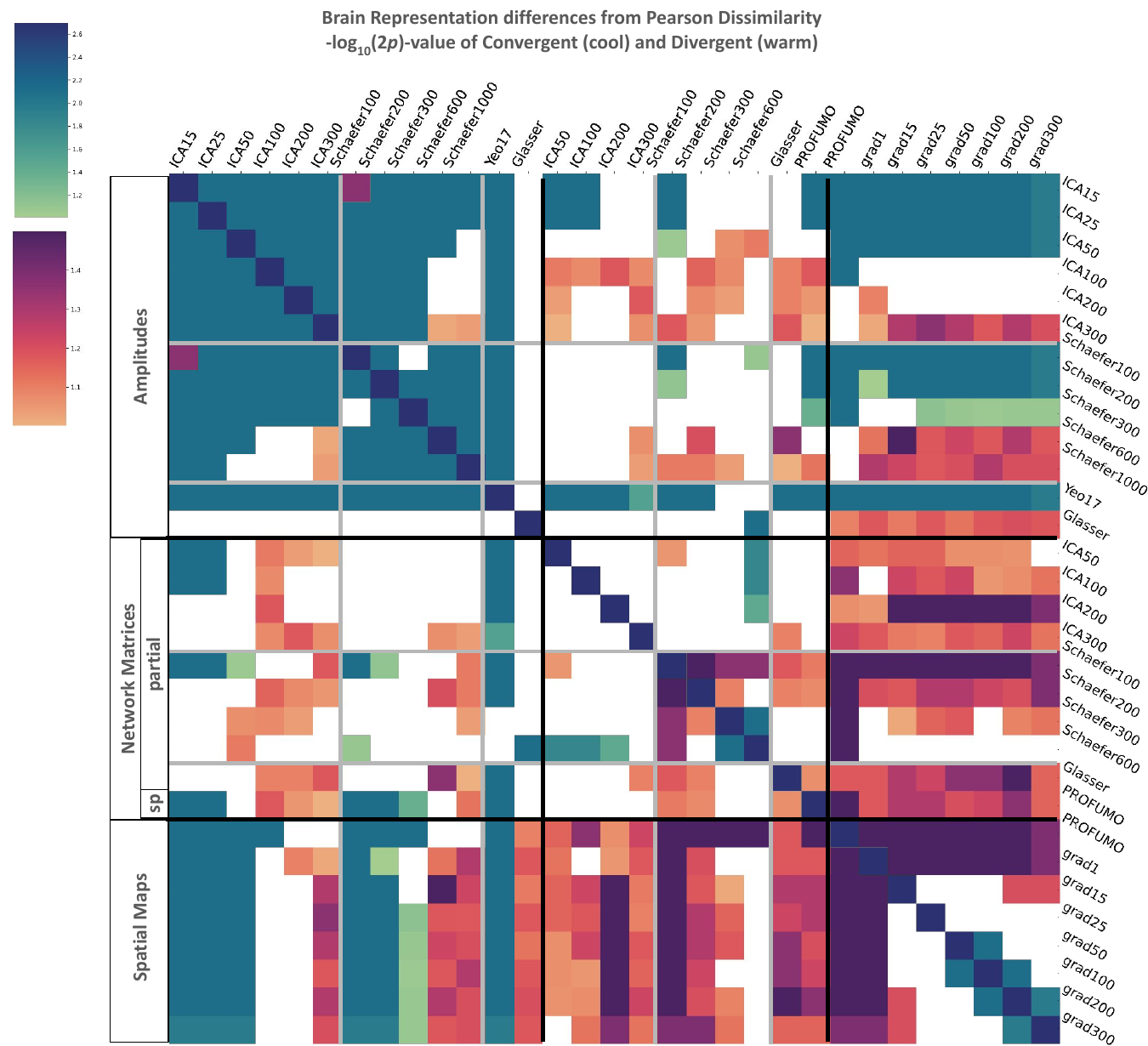}}
    \caption[Significant distances between brain representations.]{Empirical $p$-values associated to Wasserstein distances with respect to subject-null Wasserstein distributions. Distances between brain representations that do not meet one of the $p < \alpha$ similarity or difference significance thresholds are suppressed in this figure. Black lines denote boundaries between feature type, while gray lines denote boundaries between parcellation groups. Grouping brain representations by either feature type or parcellation rejects a null hypothesis assuming independence from consensus-type counts: $(\chi^2,p)=(280,6.7\times10^{-49})$ (feature), $(\chi^2,p)=(213, 5.7\times10^{-19})$ (parcellation).}
    \label{fig:pval_convdiv-heatmap_psim}
\end{figure}

Our results revealed that brain representations vary substantially in their measurements of inter-subject variability. 
When pairwise comparing the 31 brain representations with non-trivial topological structure, 142 brain representation pairs captured significantly convergent inter-subject variability, 129 brain representation pairs captured significantly divergent inter-subject variability, and 165 brain representation pairs were not significantly different from the distribution of null distances (shown in figure \ref{fig:pval_convdiv-heatmap_psim}). 
Notably, the comparison between pair representations is not determined by the Wasserstein distance between the diagrams alone (see figure \ref{fig:masked-unmasked_WpXYconsensus}). Rather, the spread and center of the distribution of distances between nulls plays a larger role;  however, neither the null mean distance nor standard deviation clearly dominate the statistical comparisons between brain representation pairs (see figure \ref{fig:masked-unmasked_WpXY_distsummary}).

\subsubsection{Summarizing systematic differences between brain representations}\label{sec:grp_BRdffs}

\begin{figure}
    \centering
    \fbox{\includegraphics[width=\linewidth]{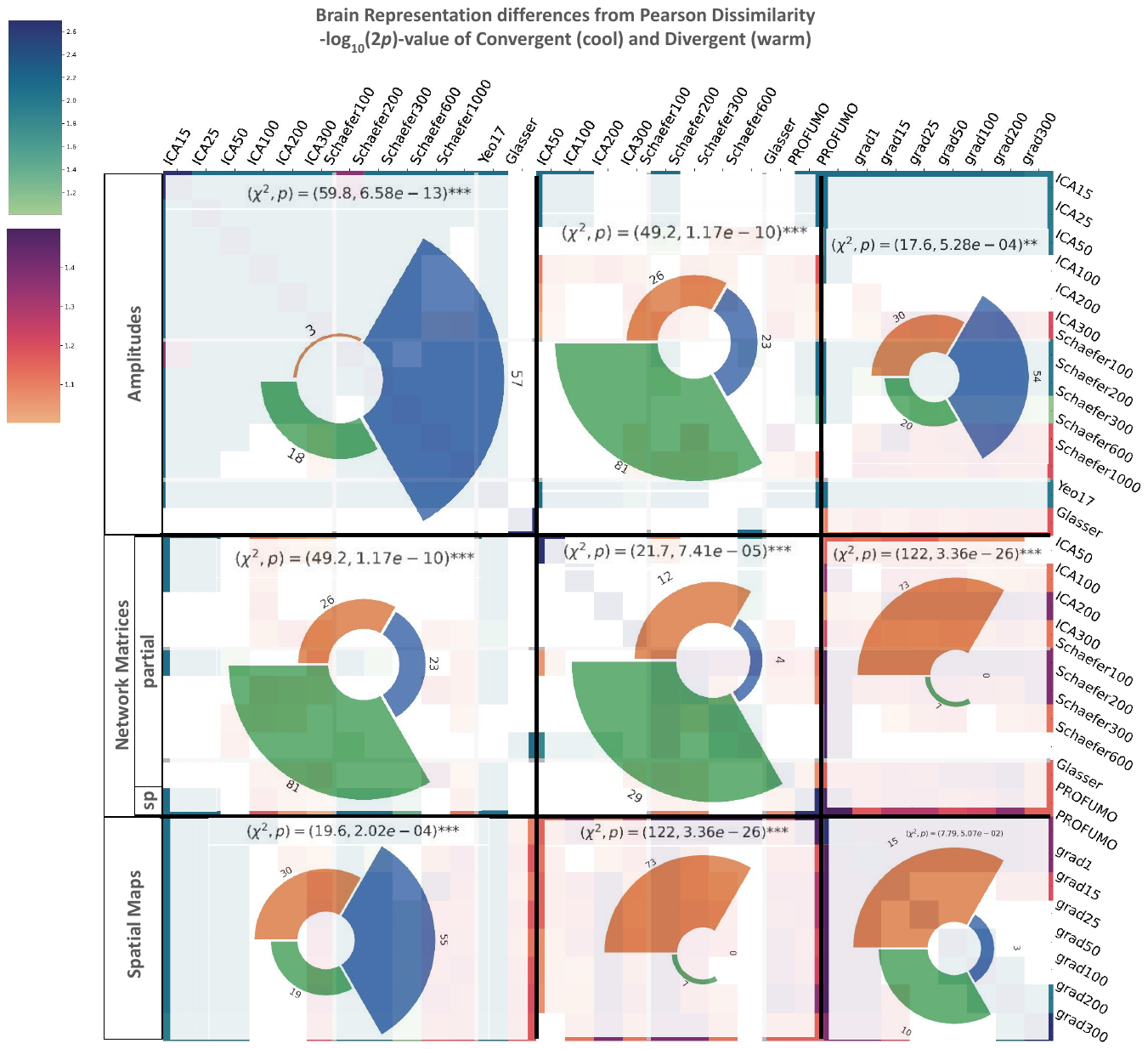}}
    \caption[Significantly similar/different feature types.]{Count distributions of convergent (blue, right), divergent (orange, upper left), and incomparable (green, lower left) brain representation pairs, grouped by parcellation pair. A $\chi^2$ homogeneity test is conducted for (and shown above) each pairwise distribution. Feature type and convergent, divergent, or incomporable count are not independent $(\chi^2,p)=(280,6.7\times10^{-49})$. Network matrices significantly diverge from spatial maps, do not converge among themselves, and tend to be incomparable to amplitudes. Maps tend diverge among themselves and converge with amplitudes. Amplitudes are significantly self-convergent.}
    \label{fig:chi2_feat_psim_overlay}
\end{figure}

\paragraph{Feature choice influences brain representation more than any other decision point.}
The distribution of convergent, divergent, and incomparable comparison counts depended strongly on feature choice $(\chi^2,p)=(280,6.7\times10^{-49})$; see figure \ref{fig:chi2_feat_psim_overlay}. 
Amplitudes most often converged with other amplitudes $(\chi^2, p) \approx (60, 7\times10^{-13})$ and spatial maps $(\chi^2,p) \approx (20, 2\times10^{-4})$. 
However, "incomparable" comparisons dominated between amplitudes and network matrices $(\chi^2,p) \approx (50, 10^{-10})$, indicating that differences between amplitudes and functional connectivity did not statistically differ from between those of shuffled inter-subject variability.
Similarly, differences between partial correlation network matrices across parcellations largely did not differ from those between nulls $(\chi^2,p) \approx (22, 7\times10^{-5})$.
By contrast, spatial maps exhibited a non-significant trend towards divergence $(\chi^2,p) \approx (8, 0.05)$, driven by PROFUMO maps diverging from all gradient diffusion maps and the principal gradient "grad1" (\cite{Margulies2016-na}) diverging from those of all other dimensionalities. Although convergence emerged between pairs of adjacently high-dimensional gradient diffusion maps (i.e., $(50,100)$, $(100,200)$, $(200,300)$), gradient diffusion maps of different dimensions were more often incomparable than not.
Finally, our results indicated that spatial maps tended to produce divergent organizations of inter-subject variability than do partial network matrices $(\chi^2,p) \approx (122, 3.4\times10^{-26})$).
Because these significant differences in subject organization emerged between significantly non-null brain representations, we interpret them as signaling the existence and accessibility of complementary subject variability between spatial maps and network matrices. 

\paragraph{Parcellation and rank also have significant effects on comparison type.}
As shown in figure \ref{fig:chi2_BR_psim}, Convergence across parcellations was rare among our results, with parcellation type having a significant influence on the comparison between brain representations $(\chi^2,p)=(213, 5.7\times10^{-19})$. 
Parcellation rank and type especially influenced comparisons including network matrices.
Schaefer network matrices diverged between high (300-600) and low (100-200) parcellation number, while network matrices produced by the rank-600 Schaefer parcellation converged with the network matrices from all ICA ranks (except rank 300) and \cite{glasserMultimodalParcellationHuman2016a}'s HCP-MMP-1.0 parcellation.
See figure \ref{fig:chi2_feat-parcel_psim_overlay}.
Amplitudes for higher-rank (100-300) ICA parcellations were more likely to be divergent from all network matrices than at lower ranks $(\chi^2,p) \approx (125, 7\times10^{-27})$.
In addition, comparisons between spatial maps and amplitudes were affected by parcellation rank.
Lower-rank parcellations (Schaefer 100-300, ICA 15-50, and Yeo) converged with all gradient diffusion spatial maps, exhibiting significantly different behavior than their higher-rank counterparts $(\chi^2,p) \approx (260, 6\times10^{-56})$.

\section{Discussion}

In this paper, we developed a novel approach to compare different resting state fMRI brain representations that overcomes challenges with different feature sizes and avoids the requirement of using  behavioral information.
Specifically, we assessed the topological structure of between-subject distance matrices induced by different brain representations.
Our results revealed that the majority of brain representations capture non-trivial topological individual difference structure, and we highlighted important differences between brain representations in their organizations of inter-subject variability. 

The large majority of brain representations preserve inter-subject variability information. 
This is consistent with prior work showing only mild differences in predictive accuracy across most brain representations (\cite{dadiBenchmarkingFunctionalConnectomebased2019,kongComparisonGradientsParcellations2023}). 
Only low-rank (15-25) partial network matrices failed to demonstrate significantly non-null topological signal, suggesting that reducing the spatial resolution of a parcellation may limit its ability to capture inter-subject variability in measurements of functional connectivity.
As we show in supplement \ref{sec:supp_inv}, however, none of these low-dimensional functional connectivity matrices failed significance criteria when their pairwise subject dissimilarity was computed using the geodesic distance on the symmetric positive-definite cone. 
Correspondingly, our results suggest that accounting for the geometric structure of the symmetric positive-definite cone is important for recovering subject-specific information from lower-rank measurements of functional connectivity.
See \S \ref{sec:brep_metric_stability} in the supplemental investigation for details.
Overall, we demonstrate in this work that imaging-derived features distill statistically significant information about the organization of inter-subject variability from resting-state fMRI data.

Statistical comparison of the topologies of individual differences between brain representations offers direct insight into the source, nature, and degree of variability in results arising from analytical flexibility. 
Our findings revealed significantly similar (converging) inter-subject topological structure between amplitude features from all brain representations and between spatial maps and amplitudes.
These convergent results suggest, for example, that translating between representations with amplitudes (a la Krakencoder, \cite{jamisonKrakencoderUnifiedBrain2025}) should be possible with relatively high agreement of inter-subject variability.
More surprisingly, we observed significantly different (diverging) structure among PROFUMO and diffusion gradient spatial maps, as well as between nearly all spatial maps and functional connectivity matrices considered.
Crucially, these diverging results suggest that many behavior/diagnosis regression/prediction tasks are too coarse to probe the analytical variability introduced by these brain representation differences.
This finding offers elaboration on prior work analyzing inter-pipeline agreement. 
For example, the divergence between diffusion gradient maps and most other brain representations (see figure \ref{fig:pval_convdiv-heatmap_psim}) is consistent with \cite{kongComparisonGradientsParcellations2023}'s work showing the principal gradient's degraded prediction performance. 
Furthermore, the \cite{liMovingProcessingAnalysisrelated2024}'s observation that pipeline differences propagate more for higher levels of spatial granularity may be consistent with our finding that divergence dominates the behavior of comparisons between spatial maps and functional connectivity matrices, as well as those of vertex-level spatial maps against one another.
Notably, our finding that PROFUMO maps diverge from all functional connectivity matrices holds tension with our previous work showing a 60\% overlap in explained variance between inter-subject spatial network organization and functional connectivity (\cite{bijsterboschRelationshipSpatialConfiguration2018,Bijsterbosch2019-lo}). 
However, these prior results may be partly driven by taking the outer product between spatially overlapping PROFUMO maps and group-averaged timeseries, thereby propagating individual-specific mixing of timeseries. 
Our novel findings suggest that functional spatial organization and functional connectivity may contain more complementary information than previously believed.

Our analyses also revealed that the inter-subject topological structure between some pairs of brain representations that could not be distinguished from null (incomparable). 
In particular, network matrices were largely incomparable to one another and to amplitudes, demonstrating relative insensitivity to differences in inter-subject variability between different parcellations' realizations of functional connectivity. 
This degraded sensitivity suggests that second-order statistics may introduce a compounding of difference between representations of subject space. 
Although care should be taken when interpreting results that do not reject the null hypothesis, these null findings still offer some interpretability.
Specifically, incomparable results occur when the difference between the true and mean null distance is small compared to the variance of null distances. 
A small separation between true and mean null diagram distances suggests insensitivity to inter-subject variability, while a large variance in a null distribution indicates volatility in the comparison of respective embeddings.
However, because all pairwise comparisons were conducted only on significantly non-null brain representations, the existence of an incomparable pair cannot be attributed to a dearth of inter-subject variability information contained in the brain representation.
Instead, an incomparable pair occurs when the variation between subjects within each representation is small compared compared to the difference between subject space embeddings across representations.
This can occur either when decomposition algorithms force subject space into a pair of feature spaces that are much more different than each other than either one is from its subject-shuffled null, or when at least one brain representation's null is volatile with respect to the shuffling (and thus diffusely distributed).

Taken together, these findings have important implications for best practices in brain-behavior association research. 
For example, in applied research, convergent brain representations should be expected to yield more similar biotypes in subtyping research than divergent representations.
Similarly, we would expect convergent representations to result in more similar brain-behavior associations than divergent brain representations.
However, it is important to note that we may not necessarily expect behavioral prediction accuracy to differ more between a pair of divergent representations than between a pair of convergent ones, as prior work would suggest (\cite{dadiBenchmarkingFunctionalConnectomebased2019}).
Instead, a pair of divergent brain representations should be understood as likely to contain a higher proportion of complementary (and thus less mutual) information than a convergent pair; conversely, we understand convergent representations as containing compatible inter-subject variability information. 
Therefore, comparing results derived from convergent representations constitutes a useful check of the stability and robustness of results in the presence of analytical variability.
The degree to which such complementary information is behaviorally relevant is purposefully outside of the scope of this work.

There are several key areas where this study could extend into future work.
First, our analysis can extend to a longer chain of analytically flexible decisions by incorporating preprocessing choices. 
Registration (both for groups and repeated individual scans), normalization, and standardization are especially important decision points in the definition, computation, and interpretation of a brain representation. 
These processing choices have important implications for the inferences available to researchers in a dataset.
Furthermore, a broader scope of brain representations deserve inclusion in this analytical framework.
For example, these could include ICA dual-regression spatial maps, alternate definitions of functional connectivity, and a broader set of multimodal parcellations.
Finally, a broader survey and deeper interrogation of measurements of inter-subject variability may be an important step in characterizing the recoverable information available in resting-state fMRI.
In addition to exploring the brain-behavior mapping directions outlined above, future work may also explore some of the theoretical foundations underpinning our practical mixing of statistical and topological tools.

Ultimately, articulating the relationship between topological features of subject space under different brain representations allows us to extend investigations of shared feature spaces to a much more general setting, observing more complex relations of interpretive consensus between brain representations.
Understanding the consequences of brain representation choice allows us to set expectations and guide best practices for choosing resting state analysis pipelines.

\section*{Funding}

Ty Easley was supported in part by NSF Grant DGE-2139839.
Elizabeth Munch was supported in part by NSF grants CCF 1907591, CCF 2106578, and CCF-2142713.
Janine Bijsterbosch was supported in part by NIH grants NIMH R01 MH128286 and NIMH R01 MH132962.
Computations were performed using the facilities of the Washington University Research Computing and Informatics Facility (RCIF), which has received funding from NIH S10 program grants: 1S10OD025200-01A1 and 1S10OD030477-01.
The authors have no known competing interests to declare.

\newpage

\begin{spacing}{1.0}
\printbibliography[
heading=bibintoc,
title={Rerefences}
]
\end{spacing}

\counterwithin{figure}{section}
\counterwithin{table}{section}

\appendix
\subfile{appendix/supplemental_results}

\include{appendix/homology_and_TDA_primer}
\include{appendix/persistence_weighted_Wp}

\subfile{supplemental_investigation.tex}

\end{document}

%% file: appendix/brainrep_algs.tex
\paragraph{Independent Component Analysis (ICA) with dual regression .}
Independent Component Analysis (ICA; \cite{Comon1994-fd}) was performed on group concatenated resting state data to create a set of group-aggregate maps of independent spatial components and timeseries (\cite{guoUnifiedFrameworkGroup2008}). Subsequently, Dual Regression (DR) was performed to produce subject-specific spatial components by regressing individual data against group-level timeseries (\cite{varoquauxGroupModelStable2010a}). If $A$ is a matrix of group-level independent (spatial) components and $M$ its corresponding mixing matrix, then ICA+DR decomposes the data $X_s$ of subject $s$ data according to
\[
X_s = W_s \left( \Lambda_s M A + R_s \right) + \varepsilon_s.
\]
In the equation above, $\Lambda_s$ describes the subject-specific loadings of group spatial independent components, $R_s$ captures individual deviations from the group model (i.e., residuals), $W_s$ is the product of the two left components of the singular value decomposition of $X_s$, and $\varepsilon_s$ is a random noise term.

\paragraph{Probabilistic functional modes (PROFUMO).}
PROFUMO (\cite{harrisonLargescaleProbabilisticFunctional2015a}) employs a hierarchical Bayesian model of per-subject spatial maps and time-courses to produce low-rank decompositions of rfMRI that emphasize the definition and stability of functional spatial networks of neural activity.
In particular, functional spatial networks are operationalized as realizations of a spatially fixed probability distribution of neural activation.
Furthermore, PROFUMO incorporates resampling reliability into network definitions.
If subjects $s=1,\ldots,n$ each have multiple "runs" (i.e., re-scans) $r=1,\ldots,m$, then scan data $X^{(sr)}$ is decomposed according to
\[
X^{(sr)} = S^{(s)} H^{(sr)} T^{(sr)} + \varepsilon^{(sr)},
\]
where $S^{(s)}$ are the spatial modes associated to subject $s$, and $H^{(sr)}, T^{(sr)}, \varepsilon^{(sr)}$ are (respectively) the amplitudes, time-courses, and noise samples associated to the $r^\text{th}$ sampling instance of subject $s$.
The membership weight $S^{(s)}_{vm}$ of a voxel $v$ in the $m^\text{th}$ mode of  $S^{(s)}$ (for some subject $s$) is distributed according to the densities proscribed by the hierarchical model
\begin{align*}
\left(S^{(s)}_{vm} \; | \; q^{(s)}_{vm} = 1 \right) &= \cN(\mu_{vm}, \sigma^2_{vm} ) \\
\left(S^{(s)}_{vm} \; | \; q^{(s)}_{vm} = 0 \right) &= \cN(0, (\eta^{(s)}_{m})^2 \zeta_v^2 ), \\
q^{(s)}_{vm} &\sim \text{Bernoulli}( \pi_{vm} ), \\
\eta^{(s)}_{m} &\sim \cN(0, \gamma_m^2 ),
\end{align*}
where $\cN(\mu, \sigma^2)$ is the normal distribution with mean $\mu$ and variance $\sigma^2$ (\cite{Harrison2020-zb}).
Note that the parameters $\mu_{vm}, \sigma_{vm}$ and hyperparameters $\zeta_v, \pi_{vm}$, and $\gamma_m$ are set at the group level and do not depend on the subject $s$.
All group-level hyperparameters are drawn from maximally noninformative priors and fit to both subject- and group-level imaging data by Bayesian update via gradient descent.

\paragraph{Connectivity gradient.}
The connectivity gradient initially utilized by \cite{Margulies2016-na} decomposes spatiotemporal scan by calculating a diffusion embedding (\cite{coifmanGeometricDiffusionsTool2005a}) using the Pearson correlation between time-courses of input voxels as a kernel function.
Given voxels $x,y \in \bbR^t$ and their centered timeseries $\bar{x} \coloneqq x - \mu(x)$, their correlation $\rho(x,y) = \frac{\bar{x} \cdot \bar{y}} {\norm{\bar{x}}_2\norm{\bar{y}}_2}$ defines a diffusion kernel
\[
k(x,y) \coloneqq \frac{\rho(x,y)}{ \int_X \rho(x,y) \; dp(y) }
\]
for a given probability density $p(y)$. The eigenmodes of the associated diffusion operator (integrated over the HCP-YA dataset $X$)
\[
K : f \longmapsto \int_{X} k(\cdot,y) f(y) \; dp(y)
\]
then give a reduced-dimension representation of the initial data, which we refer to as the "connectivity gradient."

\paragraph{HCP2016 multimodal parcellation.}
\cite{glasserMultimodalParcellationHuman2016a}'s MMP-1.0 parcellation is a hard parcellation defining structural and functional boundaries derived from high-precision registration of multimodal data. Extending the multimodal surface matching (MSM; \cite{robinsonMSMNewFlexible2014}) technique to emphasize group-level parcellation consensus allows the definition of a parcellation of cortical grayordinates incorporating data from myelin maps, resting-state networks, and anatomical topography.

\paragraph{The Schaefer and Yeo parcellations.}
Both the Schaefer and Yeo parcellations are hard parcellations defined from functional connectivity. After cortical surface registration, \cite{Yeo2011-zo}'s parcellation was defined by clustering cortical grayordinates by patterns of functional connectivity.  \cite{schaeferLocalGlobalParcellationHuman2018}'s approach refined this parcellation by integrating local gradient information (i.e., functional and anatomical boundary regions) with connectivity clustering using a gradient-weighted Markov random field (\cite{zhangSegmentationBrainMR2001}). Under any rank parameter choice, the Schaefer parcellation is a strict refinement of the Yeo parcellation.

%% file: tables/psim_significance_table.tex
\begin{table}[hb]
\centering
\begin{adjustbox}{width=\textwidth}
\begin{tabular}{|l|c|c|c|c|c|c|}
\hline
\textbf{\begin{tabular}[c]{@{}l@{}}Brain Representation\\ (parcel, rank, feat. type)\end{tabular}} & \multicolumn{ 1}{c|}{\textbf{\begin{tabular}[c]{@{}c@{}}nontrivial \\ (subject)\end{tabular}}} & \multicolumn{ 1}{c|}{\textbf{\begin{tabular}[c]{@{}c@{}}nontrivial \\ (feature)\end{tabular}}} & \multicolumn{ 1}{c|}{\textbf{\begin{tabular}[c]{@{}c@{}}$\alpha < 0.05$ \\ (subject)\end{tabular}}} & \multicolumn{ 1}{c|}{\textbf{\begin{tabular}[c]{@{}c@{}}$\alpha < 0.05$ \\ (feature)\end{tabular}}} & \multicolumn{ 1}{c|}{\textbf{\begin{tabular}[c]{@{}c@{}}$\alpha < \varepsilon$ \\ (subject)\end{tabular}}} & \multicolumn{ 1}{c|}{\textbf{\begin{tabular}[c]{@{}c@{}}$\alpha < \varepsilon$ \\ (feature)\end{tabular}}} \\ \hline
ICA 15 amplitudes     & \cmark  & \cmark  & \cmark   & \cmark   & \cmark      & \cmark      \\ \hline
ICA 25 amplitudes     & \cmark  & \cmark  & \cmark   & \cmark   & \cmark      & \cmark      \\ \hline
ICA 50 amplitudes     & \cmark  & \cmark  & \cmark   & \cmark   & \cmark      & \cmark      \\ \hline
ICA 100 amplitudes        & \cmark  & \cmark  & \cmark   & \cmark   & \cmark      & \cmark      \\ \hline
ICA 200 amplitudes        & \cmark  & \cmark  & \cmark   & \cmark   & \cmark      & \xmark     \\ \hline
ICA 300 amplitudes        & \cmark  & \cmark  & \cmark   & \cmark   & \xmark     & \cmark      \\ \hline
Schaefer 100 amplitudes       & \cmark  & \cmark  & \cmark   & \xmark      & \cmark      & \xmark     \\ \hline
Schaefer 200 amplitudes       & \cmark  & \cmark  & \cmark   & \xmark      & \cmark      & \xmark     \\ \hline
Schaefer 300 amplitudes       & \cmark  & \cmark  & \cmark   & \cmark   & \cmark      & \cmark      \\ \hline
Schaefer 600 amplitudes       & \cmark  & \cmark  & \cmark   & \cmark   & \xmark     & \cmark      \\ \hline
Schaefer 1000 amplitudes      & \cmark  & \cmark  & \cmark   & \cmark   & \xmark     & \cmark      \\ \hline
Yeo 17 amplitudes     & \cmark  & \cmark  & \cmark   & \cmark   & \cmark      & \cmark      \\ \hline
Glasser amplitudes       & \cmark  & \cmark  & \cmark   & \cmark   & \xmark     & \cmark      \\ \hline
\textbf{ICA 15 partial corr. }     & \cmark  & \cmark  & \xmark      & \xmark      & \xmark     & \xmark     \\ \hline
\textbf{ICA 25 partial corr. }     & \cmark  & \cmark  & \xmark      & \cmark   & \xmark     & \cmark      \\ \hline
ICA 50 partial corr.      & \cmark  & \cmark  & \cmark   & \cmark   & \xmark     & \cmark      \\ \hline
ICA  100 partial corr.     & \cmark  & \cmark  & \cmark   & \cmark   & \xmark     & \cmark      \\ \hline
ICA 200 partial corr.     & \cmark  & \cmark  & \cmark   & \cmark   & \cmark      & \cmark      \\ \hline
ICA 300 partial corr.     & \cmark  & \cmark  & \cmark   & \cmark   & \cmark      & \cmark      \\ \hline
Schaefer 100 partial corr.    & \cmark  & \cmark  & \cmark   & \cmark   & \xmark     & \cmark      \\ \hline
Schaefer 200 partial corr.    & \cmark  & \cmark  & \cmark   & \cmark   & \xmark     & \cmark      \\ \hline
Schaefer 300 partial corr.    & \cmark  & \cmark  & \cmark   & \cmark   & \cmark      & \cmark      \\ \hline
Schaefer 600 partial corr.    & \cmark  & \cmark  & \cmark   & \cmark   & \cmark      & \cmark      \\ \hline
\textbf{Yeo 17 partial corr. }     & \cmark  & \cmark  & \xmark      & \cmark   & \xmark     & \xmark     \\ \hline
Glasser partial corr.    & \cmark  & \cmark  & \cmark   & \cmark   & \cmark      & \cmark      \\ \hline
PROFUMO spatial corr.       & \cmark  & \cmark  & \cmark   & \cmark   & \xmark     & \cmark      \\ \hline
PROFUMO Maps       & \cmark  & \cmark  & \cmark   & \cmark   & \cmark      & \cmark      \\ \hline
Diff. Grad. 100 Maps       & \cmark  & \cmark  & \cmark   & \cmark   & \xmark     & \cmark      \\ \hline
Diff. Grad. 15 Maps    & \cmark  & \cmark  & \cmark   & \cmark   & \xmark     & \cmark      \\ \hline
Diff. Grad. 1 Maps     & \cmark  & \cmark  & \cmark   & \cmark   & \cmark      & \cmark      \\ \hline
Diff. Grad. 200 Maps       & \cmark  & \cmark  & \cmark   & \cmark   & \xmark     & \xmark     \\ \hline
Diff. Grad. 25 Maps    & \cmark  & \cmark  & \cmark   & \cmark   & \xmark     & \cmark      \\ \hline
Diff. Grad. 300 Maps       & \cmark  & \xmark     & \cmark   & \xmark      & \cmark      & \xmark     \\ \hline
Diff. Grad. 50 Maps    & \cmark  & \cmark  & \cmark   & \cmark   & \xmark     & \cmark      \\ \hline
\end{tabular}
\label{tbl:psim_sig_solo-breps}
\end{adjustbox}
\caption[Table of significant brain representations.]{Non-null topological significance for all brain representations under the Pearson metric (eqn. \ref{eqn:corr_dissim}). A green check-mark \cmark indicates that the brain representation is significant at the given threshold; a red 'X' \xmark indicates that it is not. Brain representations that were not significantly different from the subject-permuted null (at $\alpha = 0.05$) are shown in \textbf{bold text}. Significance with respect to both null distribution types is shown. The "nontrivial" and "$\alpha < \varepsilon$" columns correspond to $p < ( 1-\alpha)$ and $p <  10^{-6}$, respectively, and show when a given brain representation is "never significant" (\xmark  in the "nontrivial" column) or "almost always significant" (\cmark  in the "$\alpha < \epsilon$" column).}
\end{table}

%% file: appendix/supplemental_results.tex
\setcounter{section}{18}

\section{Supplemental Results}

\subsection{Existence and stability of topological signal}

\begin{figure}[h]
    \centering
    \figuretitle{Thresholding sensitivity of topological significance in brain representations}
    \fbox{
    \includegraphics[width=0.4\linewidth]{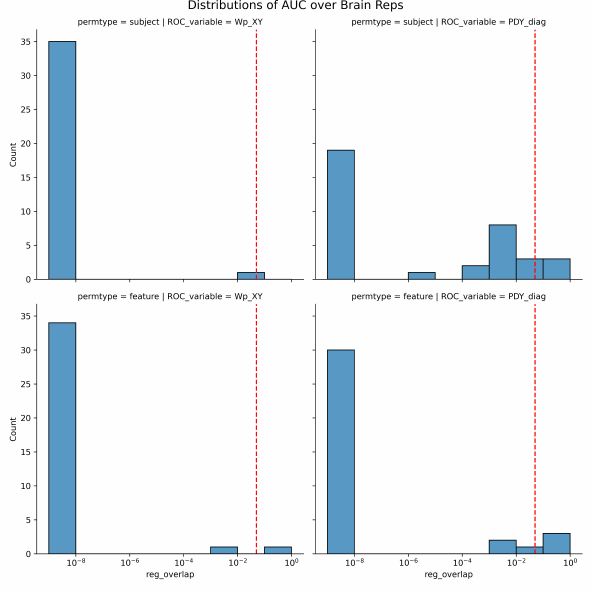}
    \includegraphics[width=0.55\linewidth]{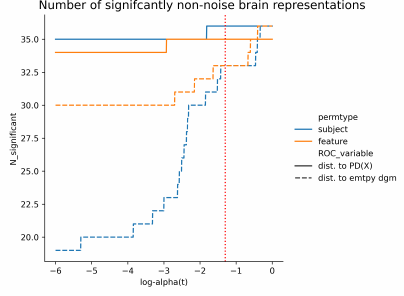}
    }
    \caption[Aggregated per-rep null and bootstrap distances.]{
    \textbf{Left}: Histogram of overlap ($1 - \text{AUC}$) values as defined in \ref{eqn:solosig_criteria}. Significant values are to the left of the dotted red line ($\alpha=0.05$). \textbf{Right}: Number of brain representations $X_i$ meeting significance criteria (eq. \ref{eqn:solosig_criteria}) as a function of threshold value $\alpha$. The left (right) edge of this graph demonstrates that AUC$=1$ (AUC$=0$) is realized by many brain representations. The distribution of $(1 - \text{AUC})$ values was heavily weighted towards extremes (Figure \ref{fig:AUC_significance_threshlines}, bottom left) indicating that, for both types of null distribution, the significance of most brain representations did not depend on the choice of significance threshold.
    }
    \label{fig:AUC_significance_threshlines}
\end{figure}

Figure \ref{fig:AUC_significance_threshlines} shows the significance of non-null signal as a function of threshold type. 
The bootstrap-vs-null original data distance criterion (solid line) is stable with respect to the threshold choice for both types of null distribution, while the bootstrap-vs-null empty diagram distance criterion (i.e., the "total persistence" distributions for both bootstrapped and null data; dotted line) shows a steep drop-off for the subject-permuted null distribution. 
This reflects the amount of structure retained by the subject-permuted null distribution.
While inter-subject variability accounts for a measurable, statistically significant proportion of topological structure in the data, the global structure of a dataset's embedding usually makes more than a negligible contribution.

\subsection{Null distance distributions play a larger role than the true diagram distance in determining convergence, divergence, or incomparability}

\begin{sidewaysfigure}
    \centering
    \figuretitle{Relationship between subject-null distributions and convergence, divergence, or incomparability}
    \fbox{
    \includegraphics[width=0.32\linewidth]{results_figs/WpXY-pairwise_results_subject-null_alpha05.pdf}
    \includegraphics[width=0.32\linewidth]{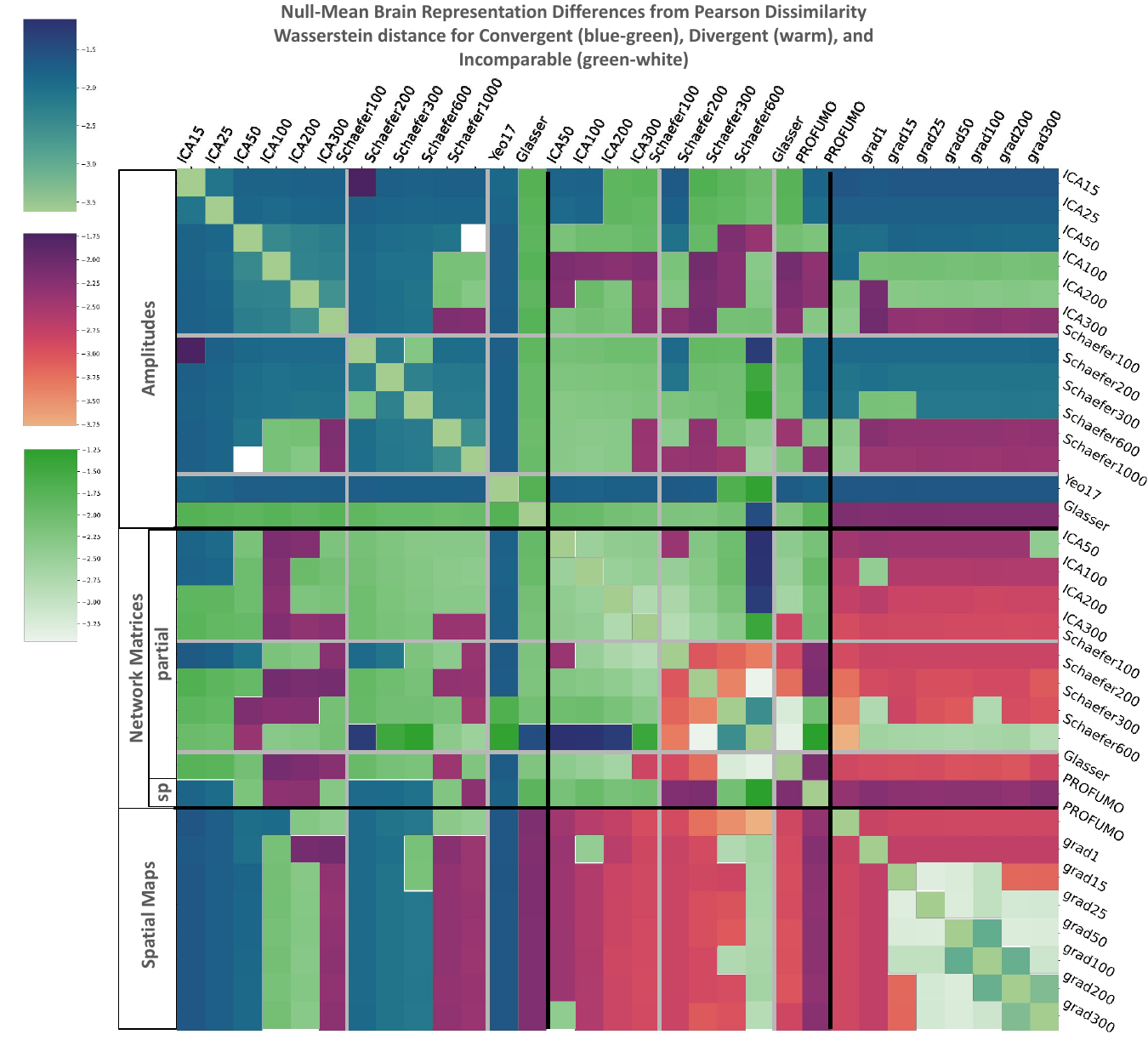}
    \includegraphics[width=0.32\linewidth]{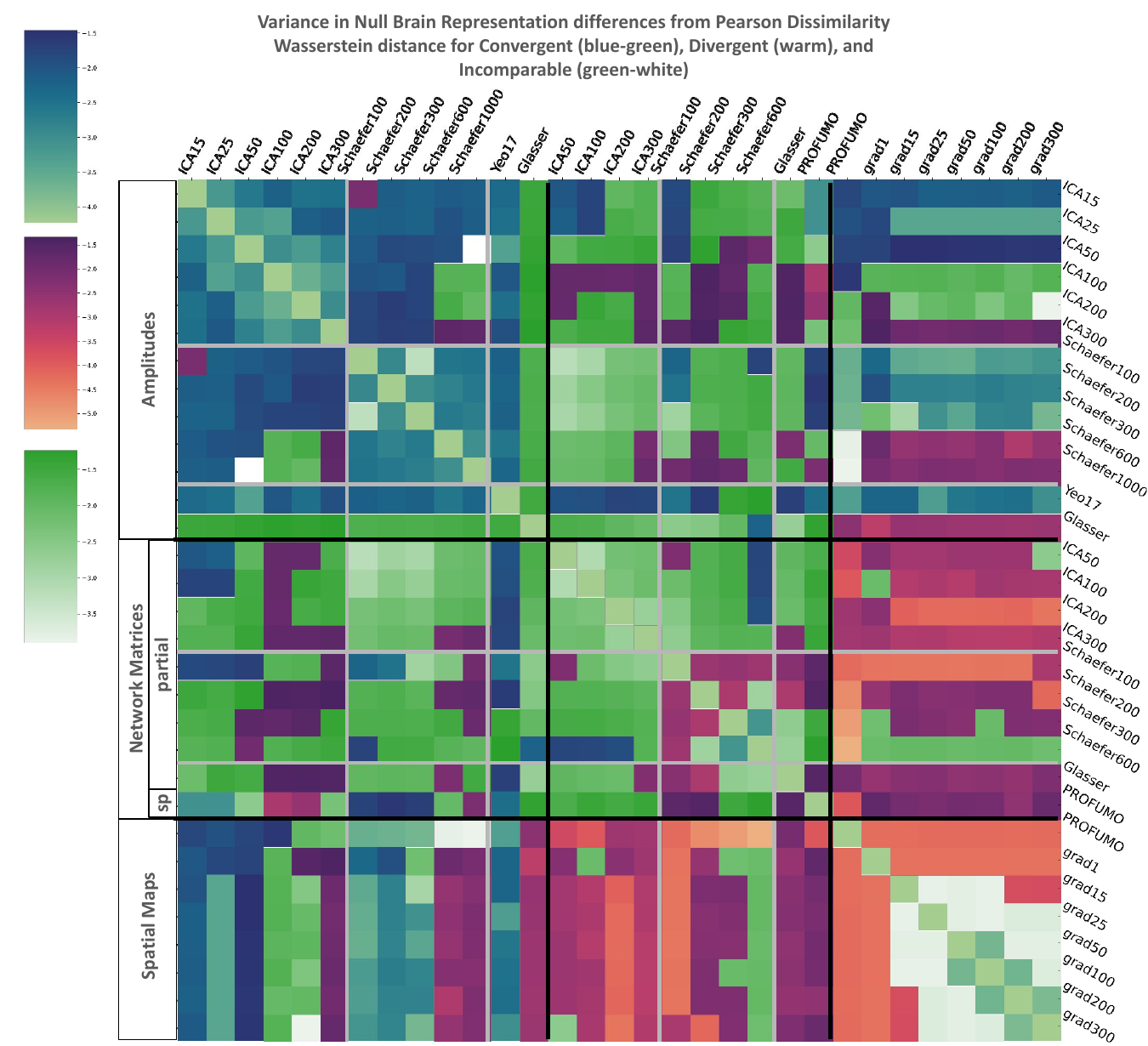}
    }
    \fbox{
    \includegraphics[width=0.32\linewidth]{results_figs/WpXY-pairwise_no-mask.pdf}
    \includegraphics[width=0.32\linewidth]{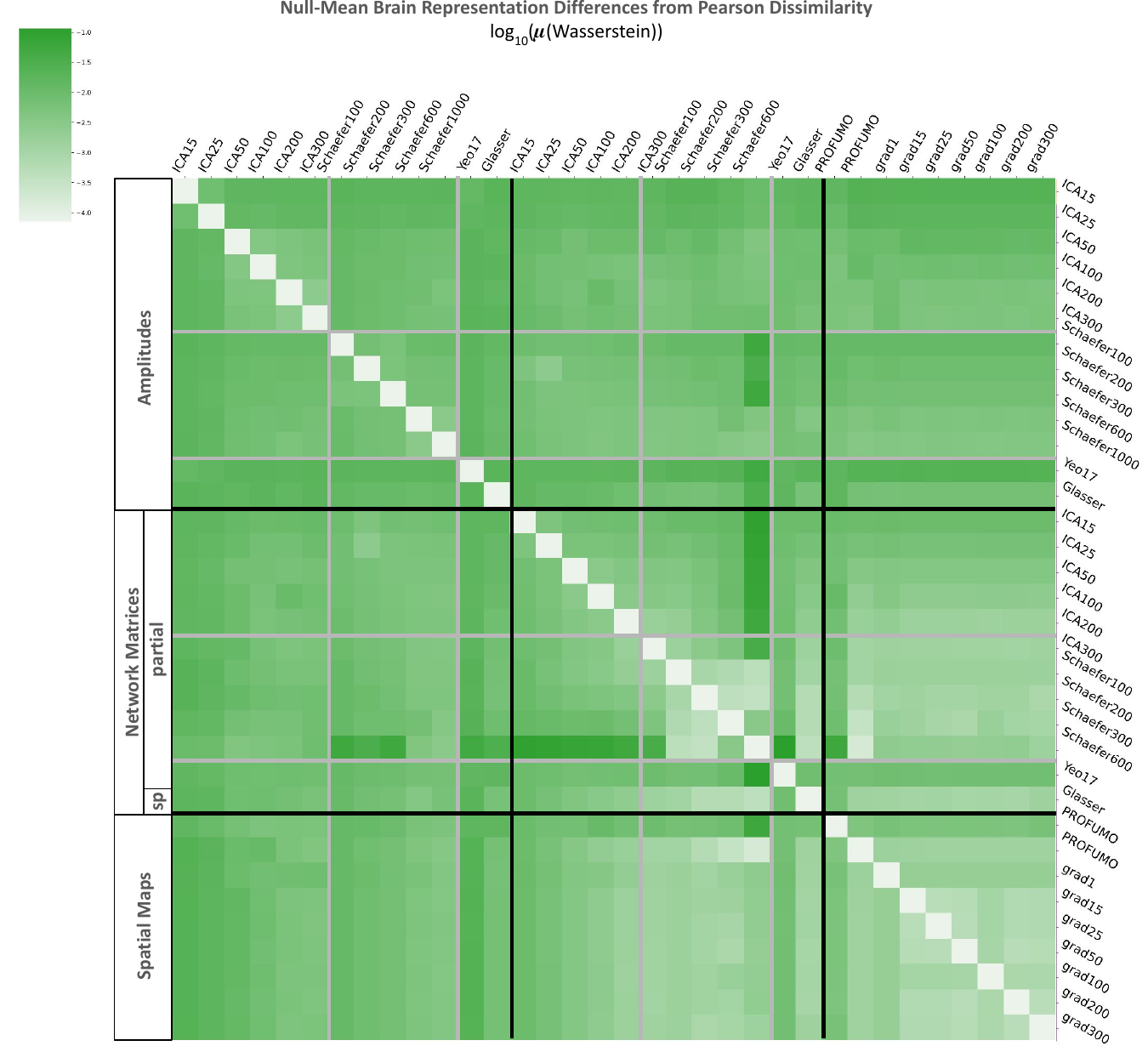}
    \includegraphics[width=0.32\linewidth]{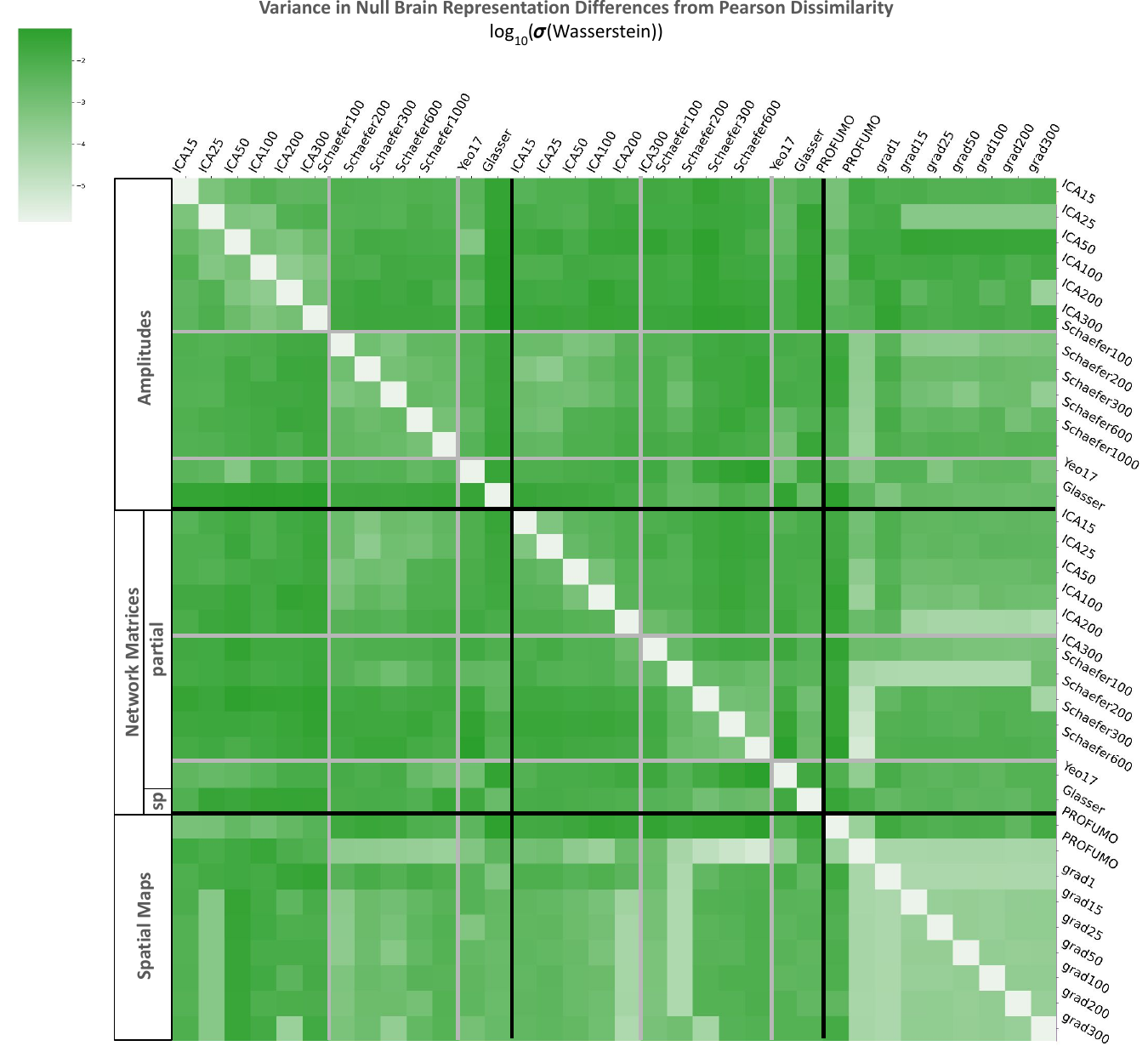}
    }
    \caption[Pairwise Wasserstein value and null mean/variance distribution summaries.]{
    Pairwise Wasserstein values, and the mean and standard distribution of their corresponding null distributions. Distributions are shown over the permuted-subject null, and the convergence, divergence, or incomparability is evaluated with respect to a significance threshold of $\alpha = 0.05$. The Wasserstein distance between a pair of brain representations is not necessarily a good predictor of convergence, divergence, or incomparability. Rather, both the spread and center of the null-distance distribution play larger roles in determining the convergence, divergence, or incomparability of a pair of representations, and their relative importance varies over representation pairs. Significance-masked Wasserstein grids \textbf{(top)}, colored by convergence, divergence, or incomparability, are shown above their unmasked counterparts \textbf{(bottom)}. The grids displayed here show the Wasserstein distance between the brain representations \textbf{(left)}, the mean Wasserstein distance between null-distributed diagrams \textbf{(middle)}, and the square root of the variance of the same \textbf{(right)}.}
    \label{fig:masked-unmasked_WpXY_distsummary}
\end{sidewaysfigure}

On its own, pairwise Wasserstein distance between brain representation diagrams is a poor predictor of convergence, divergence, or incomparability. 
Rather, convergence, divergence, or incomparability is determined by both the spread of the null distance distribution as well as the separation between its center and the true distance,
Furthermore, the relative importance of these contributions varies over representation pairs.
Figure \ref{fig:masked-unmasked_WpXY_distsummary} shows the grids of true diagram distances, mean null diagram distances, and standard deviation of null diagram distances; each grid is shown masked by convergence, divergence, or incomparability (top) and unmasked (bottom).
Because null mean distance, null distance spread, and true distance vary in their relative influence on determining convergence, divergence, or incomparability, we may also conclude the presence or absence of inter-subject variability information is of inconsistent importance to observable differences between rfMRI data under different brain representations.
The assignment of convergence, divergence, or incomparability is a complex interplay between null and true distances between brain representation diagrams, and urges care in its interpretation.
Thoughtfully approached, however, the lens of "convergence, divergence, or incomparability" becomes a powerful unit of analysis of analytical flexibility in resting-state fMRI pipelines.

\subsection{Summarizing systematic differences between brain representations}

Brain representations can be grouped by choices of parcellation, decomposition rank, and feature.
Groupings along any combination of these axes yield statistically significant dependencies of the distrubtion of convergent, divergent, and incomparable counts. 
Figure \ref{fig:chi2_feat-parcel_psim_overlay} shows parcellated subgroups of feature-grouped counts, and figure \ref{fig:chi2_BR_psim} shows the overall effects of grouping by parcellation choice. 
While parcellation choice introduces a significant dependency of convergent, divergent, and incomparable counts, it has a weaker effect than the choice of feature type.

\begin{figure}
    \centering
    \fbox{\includegraphics[width=\linewidth]{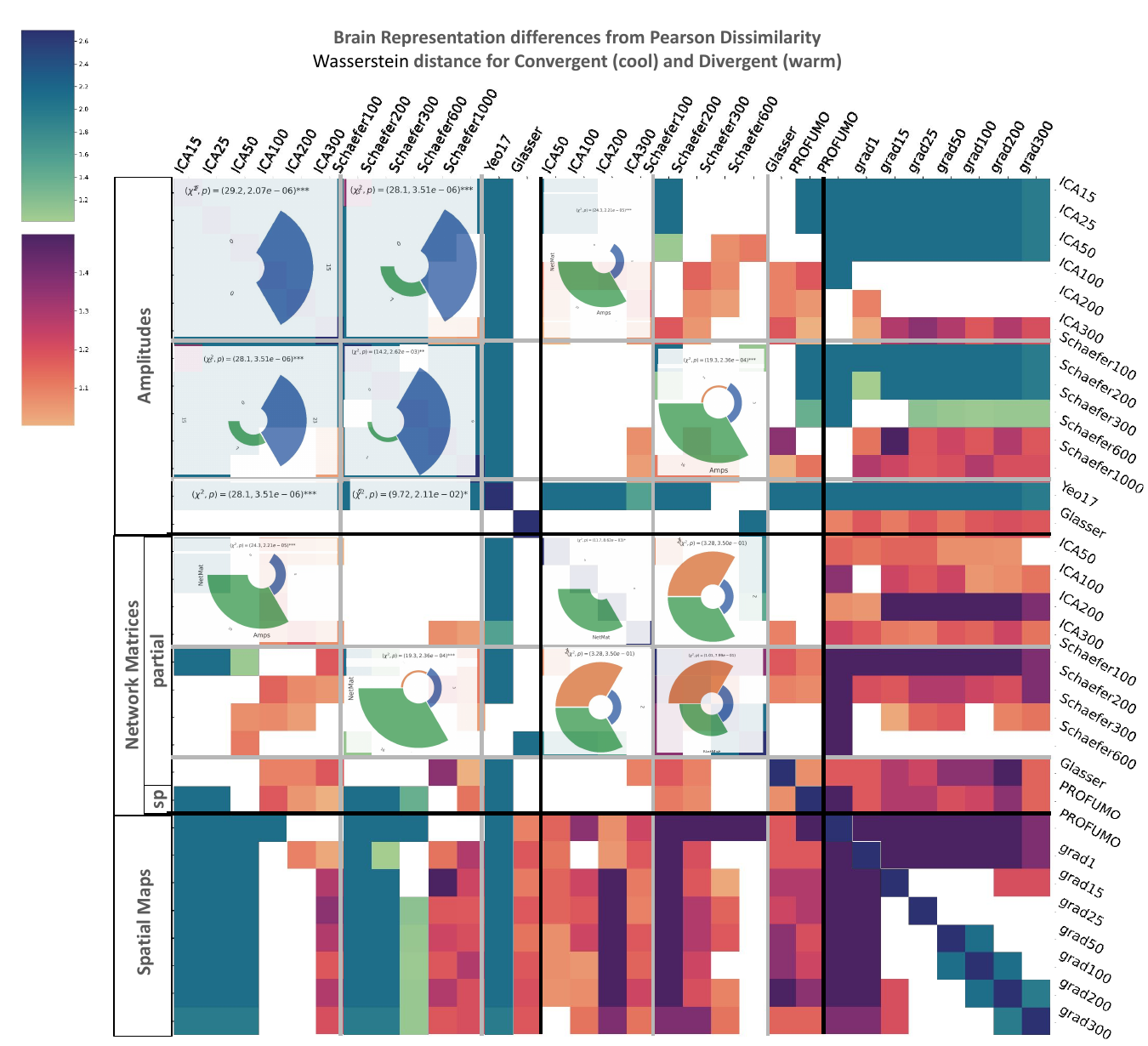}}
    \caption[Significantly similar/different parcellations.]{Count distributions of convergent (blue, right), divergent (orange, upper left), and incomparable (green, lower left) brain representation pairs, grouped by both parcellation and feature type. A $\chi^2$ homogeneity test is conducted for each grouped consensus distribution, its results shown above. The count distributions show differences in cross-representation interaction for functional connectivity matrices originating from different parcellations.}
\end{figure} \label{fig:chi2_feat-parcel_psim_overlay}

\begin{figure}
    \centering
    \figuretitle{Parcellation effects on convergence, divergence, or incomparability distribution}
    \fbox{\includegraphics[width=\linewidth]{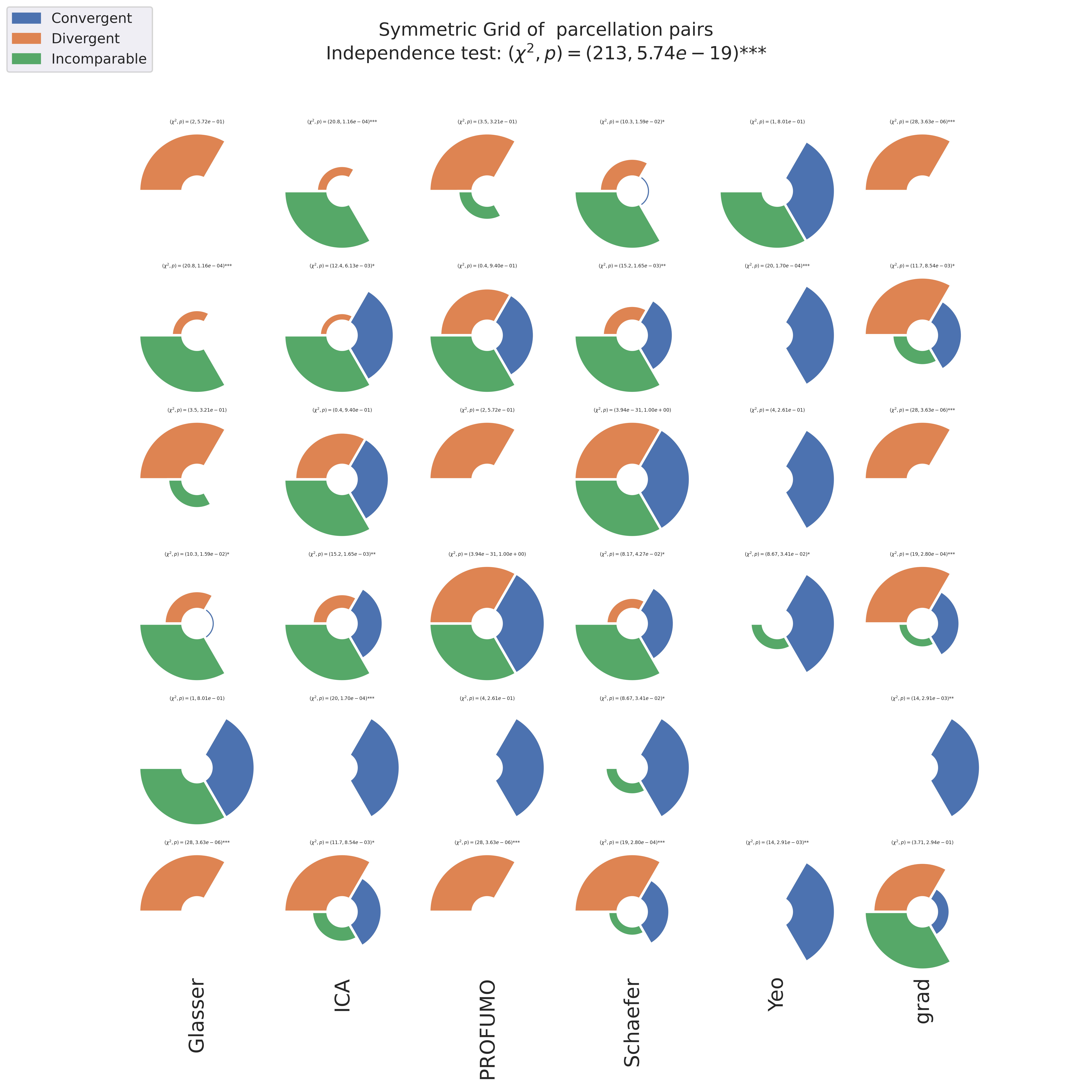}}
    \caption[Significantly similar/different parcellations.]{convergence, divergence, or incomparability distributions as a function of pairs of parcellation choices. Note that the diffusion gradient is significantly divergent with every other parcellation except Yeo, with which it is convergent. Yeo is also significantly convergent with Schaefer and ICA. Schaefer and ICA parcellations are mutually (significantly) non-divergent with one another and within themselves (across different features and ranks). Glasser's parcellation is significantly incomparable with both ICA and PROFUMO parcellations.}
\end{figure} \label{fig:chi2_BR_psim}

\setcounter{section}{0}

%% file: appendix/homology_and_TDA_primer.tex
\section{A Primer on Persistent Homology}\label{appendix:alg_top}

This appendix gives a brief overview of some key concepts, definitions, and tools in topological data analysis (TDA). In particular, we give some exposition on "persistent homology," which is a central tool of this thesis and TDA methods as a whole. We begin with a primer on some algebraic preliminaries.

\subsection{Simplicial Homology}

We begin by introducing the idea of "homology," which is a group structure summarizing a long exact sequence of maps on free abelian groups. We consider the homology of \textit{simplicial complexes} in this subsection.

\subsubsection{Abstract Simplicial Complexes}

A \textit{simplicial complex} is an incredibly flexible type of structure that can be equipped to any type of set. It is a fundamental building block to algebraic topology and the basic object of persistent homology. We begin with an abstract definition of the simplicial complex:
\begin{definition}\label{def:abs_simp_cplx}
Suppose $X$ is a set. An \textit{abstract simplicial complex} $K$ on $X$ is a subset $K \subset 2^X$ of its power set with the property
\[
A \in K \text{ and } A'\subset A \; \implies \; A' \in K.
\]    
\end{definition}
That is, $K \subset 2^X$ is closed under inclusion. 
It's immediately clear that any $A \in K$ can also be regarded as an abstract simplicial complex. It is a special type of subcomplex of $K$ called a \textit{face} and has an associated dimension $\dim(A) = \abs{A} - 1$. 
A face of dimension $k$ is called a $k$-face, and the collection of $k$-faces in $K$ comprise a basis of what is called its $k^\text{th}$ chain group $C_k(K)$. 
In particular, given some abelian group $G$ of coefficients (thought of as a $\bbZ$-module), the chain group $C_k(K)$ is the set of $\bbZ$-linear combinations of $k$-faces of $K$. 

\subsubsection{The Homology Groups $H_n(X)$}

These chain groups $C_k(K)$ give rise to the homology groups $H_k(K)$, whose construction begins by defining the \textbf{\textit{boundary map}}
\begin{align*}
\partial_k : C_k(X) &\to C_{k-1}(X) \\
\left[ v_0 \cdots v_k \right]
& \mapsto \sum_{i=0}^k (-1)^k \left[ v_0 \cdots \hat{v}_i \cdots v_k \right],
\end{align*}
where $\left[ v_0 \cdots \hat{v}_i \cdots v_k \right]$ denotes the subcomplex of $\left[ v_0 \cdots v_k \right]$ formed by removing $v_i$.
By definition, $\partial_0$ is the $0$-map.
Using the boundary map, we can define
\begin{definition}
Given $\partial_k: C_k(X) \to C_{k-1}(X)$, define the $k$-\textbf{cycles} $Z_k(X)$ and $k$-\textbf{boundaries} $B_k(X)$ of $X$ by
\begin{align*}
Z_k &= \ker(\partial_k) \\
B_k &= \im(\partial_k). \\
\end{align*}
Since $\partial_k$ is a group homomorphism and $B_k \subset Z_k$, we can define
\begin{equation}
    H_n(X) \coloneqq Z_n(X) / B_n(X),
\end{equation}
as the \textbf{\textit{$n^\text{th}$} homology group} of $X$.

\end{definition}

\subsubsection{The Vietoris-Rips Complex}

We now introduce a special simplicial structure that can be defined on any metric space $(X,d)$. This simplicial structure, called the \textit{Vietoris-Rips complex}, plays a central role in most practical current uses of persistent homologies.

\begin{definition}\label{def:VR_cplx}
Given a metric space $(X,d)$, we may define the Vietoris-Rips complex (of radius $r$) on $X$ as
\begin{equation}\label{def:VR(r)}
\VR_r(X) \coloneqq \{ S \subset X \mid d(x,y) < r \text{ for all } x,y \in S. \},
\end{equation}
where $N_r(x)$ is the ball of radius $r$ centered at $x \in X$.
\end{definition}

Geometrically, we can interpret this the simplicial complex formed when $k$-faces are defined by a set of $k+1$ points that are all at most a distance $r$ from one another. This is a simplicial complex because sets of pairwise intersubsections are closed under inclusion. 

\subsection{Persistent Homology}

Having given some background on homology, we are now poised to define "persistent homology" and describe some of its basic properties. 

\subsubsection{Filtrations}

We now introduce the concept of a \textit{filtration}, which is just a nested family of simplicial complexes:
\begin{definition}\label{def:filtration}
For a simplicial complex $K$, a \textit{filtration} $\cF$ on $K$ is a collection of subcomplexes $K_\alpha \subset K$ (indexed by $\alpha \in A$ for some totally ordered indexing set $A$) that are totally ordered by inclusion:
\[
\alpha < \alpha' \iff K_\alpha \subset K_{\alpha'} \subset K.
\]
In general, a filtration can be regarded as a function $\cF : A \to 2^K$. 
\end{definition}

\begin{example}
Suppose $K$ is a simplicial complex. Let $K_n$ be the $n$-skeleton of $K$. Then the map $\cF : \bbN \to 2^K$ sending $n \mapsto K_n$ is a filtration on $K$.
\end{example}

\begin{example}[Vietoris-Rips filtration]    
Under definition \ref{def:filtration}, it is immediately clear that, for any nondecreasing function $r: A \to \bbR$ on a totally ordered set $A$ (i.e., $r(\alpha) \leq r(\alpha')$ if and only if $\alpha \leq \alpha'$ for $\alpha, \alpha' \in A$) and metric space $X$, the set $\left\{ \VR_{r(\alpha)}(X) \right\}_{\alpha \in A}$ is a filtration.
In our case, $A \subset \bbR$ and $r$ is simply the identity.
\end{example}

\subsubsection{The Persistence Module}

By definition, a filtration $\cF = \{K_t\}$ of some complex $K$ has a sequence of inclusions $\iota^{st} : K_s \hookrightarrow K_t$ for $s < t$ which in turn induce homology maps $\iota_*^{st} : H_\bullet(K_s) \to H_\bullet(K_t)$. 
\begin{definition}
Suppose $K$ is a simplicial structure on $X$ and $\cF: \bbN \to 2^K$ a filtration on $K$. Set $K_n = \cF(n)$. The set of homology groups $\left\{H_\bullet (K_n) \right\}$ and inclusion-induced maps between them comprise the \textit{persistence module}
\[
\PH_\bullet(X | \cF) \coloneqq \left( \left\{ H_\bullet(K_n) \right\}_n, \left\{\iota_*^{mn}: H_\bullet(K_m) \to H_\bullet(K_n)\right\}_{m < n} \right).
\]
When $\cF$ or $X$ is clear from context, we may (respectively) write $\PH_\bullet(X)$ or $\PH_\bullet(\cF)$.
\end{definition}
If $H_\bullet(X)$ is a vector space over some field $F$, then $\PH_\bullet(X)$ may be regarded as a functor $A \to \text{Vec}(F)$. 

\begin{example}
Suppose $(X,d)$ is a finite metric space and $A \subset \bbR$ is finite. Then $A$ specifies a filtration $\left\{\VR_r(X)\right\}_{r \in A}$ on $X$.
Let $X_r \coloneqq \VR_{r}(X)$ and apply the homology functor to the filtration $\{X_r\}_{r \in A}$. This is called the \textit{Vietoris-Rips persistence module}, written
\[
PH_\bullet(X) = \left\{ H_\bullet(X_r) \right\}_{r \in A}.
\]
The Vietoris-Rips persistence is an important special case and lies beneath all analyses in this work.
\end{example}

\subsubsection{The Wasserstein distance for persistence diagrams}\label{sec:Wp_dgm}

Finally, we briefly describe how the Wasserstein distance is computed between persistence diagrams. The classical $p$-Wasserstein distance is a metric on probability measures (\cite{clementElementaryProofTriangle2007}) with bounded $p^\text{th}$ moments over a separable metric space $(X,d)$.
If $\mu,\nu$ are densities on $X$ with bounded $p^\text{th}$ moments, then their Wasserstein distance $W_p$ is calculated as
\[
W_p(\mu, \nu) \coloneqq \inf_{\gamma \in \Gamma(\mu, \nu)} \left[ \int_{X \times X} d(x, x')^p \; d\gamma(x, x') \right]^{1/p},
\]
where $\gamma \in \Gamma(\mu,\nu)$ is a joint distribution on $X \times X$ with marginals $\mu$ and $\nu$ as its first and second factor.
The Wasserstein distance between persistence diagrams adapts this definition under a few necessary modifications.

Given a finite persistence module $\PH(X)$ (see \ref{sec:TDA}) associated to the birth-death intervals $\{(b_i,d_i)\}$, where $d_i \leq \infty$, we may define its \textit{persistence diagram} as the multiset $\dgm(X)$ in $\bar{\bbR}^2$ given by
\[
\left\{ (a_i,b_i) \in \bar{\bbR}^2 \mid (b_i, d_i) \in \PH(X) \right\} \cup \Delta,
\]
where $\bar{\bbR} = \bbR \cup \{\pm\infty\}$ is the extended real line and $\Delta \coloneqq \{ (x,x) \mid x \in \bbR\} \subset \bbR^2$ is the diagonal graph (of which each point is included with infinite multiplicity).
Since two different persistence diagrams $\dgm(X)$ and $\dgm(Y)$ may in general have different numbers of points (and thus total mass), a generalized Wasserstein distance (\cite{piccoliGeneralizedWassersteinDistance2014}) must be considered.
Following the convention of \cite{Dey2022-id}, we write $\Gamma_{XY}$ to denote the set of "augmented diagram bijections" between the multisets $\dgm(X)$ and $\dgm(Y)$ that map excess points into the diagonal $\Delta$.
If $\norm{\cdot}_p$ denotes the Euclidean $p$-norm on $\bar{\bbR}^2$, then $p,q$-Wasserstein distance between $\dgm(X)$ and $\dgm(Y)$ can be written
\[
W_{p,q}(\dgm(X), \dgm(Y)) \coloneqq \inf_{\gamma \in \Gamma_{XY}} \left( \sum_{x \in \dgm(X)} \norm{x - \gamma(x)}_q^p \right)^{1/p},
\]
where $x = (b,d)$ is a birth-death pair in the diagram.
When $p = q$, we simply write $W_p$. In this work, we choose $p=q=2$.

\subsubsection{Module Structure and Metric Stability of Persistence}

Here, we state two key theorems about persistence diagrams that underwrite the validity of our analysis. The first, from \cite{corbetRepresentationTheoremPersistence2018}, is the essential uniqueness of their decomposition into persistence diagrams:

\begin{theorem}[Persistence Structure Theorem]
Let $R$ be a ring with unity. The category of persistence modules of finitely presented type over $R$ is isomorphic to the category of finitely presented graded modules over $R[t]$.
\end{theorem}

In effect, the structure theorem allows us to treat the persistence \textit{diagram} (a multiset summary of birth-death pairs) as a distinguishing invariant of the persistence \textit{module}. The diagram is much easier to do computations with, and in particular can be shown to be stable with respect to perturbations in the input data:

\begin{theorem}[Cellular Wasserstein Stability, \cite{skrabaWassersteinStabilityPersistence2021}]
Suppose $K$ is a cellular complex and $f,g: K \to \bbR$ are monotone. Let $\dgm(f)$ and $\dgm(g)$ be the persistence diagrams produced from the sublevel set persistence of $f$ and $g$ (respectively). Then
\[
W_p(\dgm(f), \dgm(g)) \leq \norm{f-g}_p.
\]
\end{theorem}

\begin{remark}
We will define the Wasserstein distance in more detail shortly, but it suffices for the moment to understand that it is a metric on the space of persistence diagrams.
Before moving on, we verify that \cite{skrabaWassersteinStabilityPersistence2021}'s cellular Wasserstein stability (Theorem 4.7) applies to our investigation.
Given distance metrics $\hat{d}_f$ and $\hat{d}_g$ on a space $X$, we can define corresponding monotone functions $\hat{f}$ and $\hat{g}$ on the complete graph $G$ on points of $X$: each edge of $G$ is assigned the distance between the points associated to its vertices, and each vertex is assigned the minimum value of an associated edge. 
Let $K_f$ and $K_g$ be the Vietoris-Rips complexes on $X$ corresponding to $\hat{d}_f$ and $\hat{d}_g$, respectively, and set $K$ as the cellular closure of $K_f \cup K_g$. Then $\hat{f}$ and has a monotone extension $f: K \to \bbR$ given by applying
\[
f(\sigma) \coloneqq \max_{\tau \subsetneq \sigma} \hat{f}(\tau)
\]
(and analogously for $g$) along the upward closure $G \to K$. 

This construction allows us to apply the cellular Wasserstein stability theorem to our investigation. 
Explicitly, we may take the dataset $X$ as our underlying finite set, and the pair $(\varphi_i, \delta_j)$ of a brain representation and dissimilarity function (respectively) as together specifying a "metric" function $d_f: X \times X \to \bbR$.
Different choices of brain representation and/or dissimilarity function correspond to different choices of functions $d_f:X \times X \to \bbR$. 
Induction on the construction above allows us to apply the cellular stability theorem to a suitable simplicial closure of $X$ and the constructed monotone functions $f_{ij} \leftrightarrow (\varphi_i, \delta_j)$.
In particular, our persistence comparisons are stable (under the $p$-Wasserstein metric) with respect to perturbations in the values of dissimilarity functions on the dimension-reduced dataset $\hat{X}_i$.
\end{remark}

%% file: appendix/persistence_weighted_Wp.tex
\section{Topological Bootstrapping and the Prevalence-Weighted Wasserstein Metric}\label{sec:prev_Wp}

\subsection{Introduction}

This appendix gives further detail on the topological bootstrap and adds a minor theoretical exploration of ability to inform comparisons between dimensionality reductions. 
We begin by tracing the exposition of \cite{Reani2021-mu}'s topological bootstrapping method as implemented by \cite{Garcia-Redondo2022-bv} and modified by \cite{Easley_Co_homology_cycle-match_for_2024}. 
Next, we define a modified Wasserstein metric $\hat{W}_{p,q}$ that incorporates statistical stability information via the topological bootstrap. 
Finally, we prove a few basic properties of this modified metric.

\subsection{Background on the cycle-matched bootstrap}
We give a brief overview of topological bootstrapping in this subsection.

Because it is possible (and, in fact, generic) for multiple data elements to define the same homology generator, bootstrap sub-sampling (\cite{abu-mostafaLearningData2012}) is less straightforward in persistent homology than in many other modes of analysis. 
However, \cite{Reani2021-mu} recently demonstrated a cycle registration method that uses image persistence (\cite{Cohen-Steiner2009-yj}) to register homology generators found in co-embeddable spaces that allows persistence bootstrapping. 
We describe the cycle registration procedure in the following subsubsection and the bootstrapping procedures it enables in the next.

\subsubsection{Cycle registration in persistent homology}\label{sec:cyc_reg}

Suppose $X$ and $Y$ are topological spaces both equipped with embeddings into some space $Z$. 
Then the inclusion maps $X \xhookrightarrow{\iota_X} Z$ and $Y \xhookrightarrow{\iota_Y} Z$ induce homology maps $\iota_X^*$ and $\iota_Y^*$. 
If $X$ and $Y$ are also equipped with filtrations $X_r$ and $Y_s$, then we may equip $Z$ with a filtration $Z_\tau$ compatible with $X_r$ and $Y_s$ and induce corresponding maps $\iota_{X,r}^*$ and $\iota_{Y,s}^*$ from the graded persistence modules $\PH_*(X)$ and $\PH_*(Y)$ (respectively) to $\PH_*(Z)$ (\cite{Garcia-Redondo2022-bv}). 
From here on, we will use the same filtration index subscript $t$ for $\iota_{t,X}^*$ and $\iota_{t,Y}^*$ to denote that these maps have been refined to compatible filtrations.

\begin{figure}
    \centering
    \fbox{\includegraphics[width=\linewidth, trim={0 115 0 15}, clip]{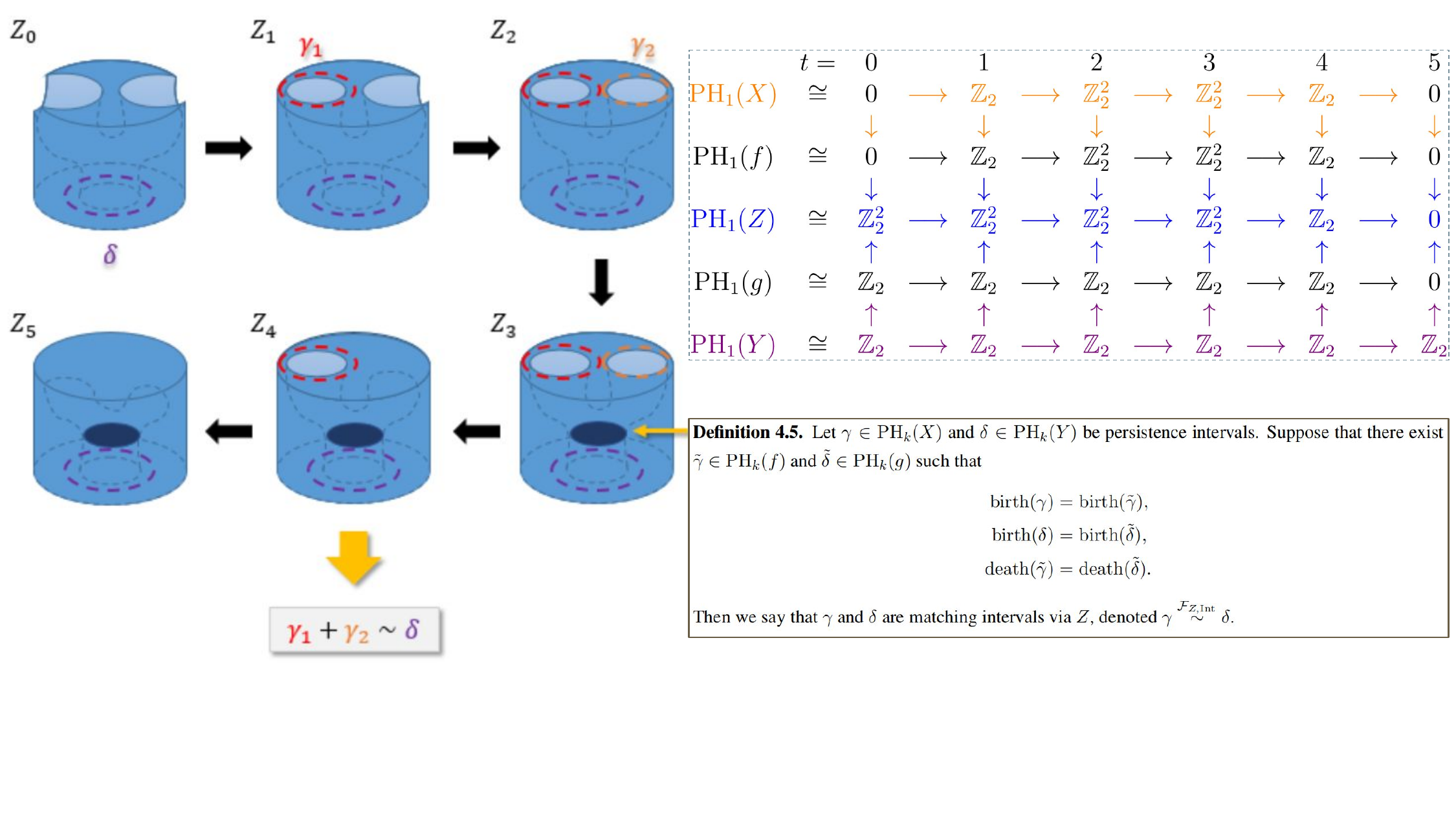}}
    \caption[The topological bootstrap.]{An illustration of the topological bootstrap in a toy-model filtration (figure adapted from \cite{Reani2021-mu}, Figure 10). The explicit chain homomorphisms are shown, tracing the embedding-induced images of homology generators (with coefficients in $\bbZ_2$) for each step in the filtration $\{Z_1, \ldots, Z_5\}$. The top and bottom of the mapping cylinder $Z_t$ correspond to $X_t$ and $Y_t$, respectively, in the chain maps.}
    \label{fig:topo_bts}
\end{figure}

A pair of nontrivial elements in $\PH_k(X)$ and $\PH_k(Y)$ is said to "\textit{match at $t$ via $Z$}" if $\iota_{t,X}^*$ and $\iota_{t,Y}^*$ map them to the same nontrivial element of $\PH_k(Z)$ for some filtration value $t$. 
We say elements in $\PH_k(X)$ and $\PH_k(Y)$ match in the interval $(a,b)$ if they match via $Z$ for all $t \in (a,b)$. This procedure simplifies substantially in the sub-sampling case.
When sub-sampling, we have $Z=X$ and $Y = \widehat{X} \subset X$, and we need only check nontrivial elements of $\PH_k(X)$ for matches in $\PH_k(\widehat{X})$. 

For a matched pair, the affinity score $\alpha$ of the match can be computed from ratios of lengths of filtration intervals for which elements in $\PH_k(X)$ and $\PH_k(Y)$ are matched via $Z$. 
More specifically, let $\tilde{\gamma}, \tilde{\eta} \in \PH_k(Z)$ be image-bars through which the bars $\gamma \in \PH_k(X)$ and $\eta \in \PH_k(Y)$ are matched. 
Writing $\Jac(I,J) = \frac{\abs{I \cap J}}{\abs{I \cup J}}$ to denote the Jaccard index between intervals $I$ and $J$, we define the \textit{\textbf{matching affinity}} between $\gamma, \eta$ as follows:
\begin{align*}
&\alpha : \PH_k(X) \times \PH_k(Y) \to [0,1] \\
&\alpha(\gamma, \eta) = \Jac(\gamma, \tilde{\gamma}) \cdot \Jac(\eta,\tilde{\eta}) \cdot \Jac(\gamma,\eta) \cdot \Jac(\tilde{\gamma}, \tilde{\eta}) 
\end{align*}
We assign $\alpha = 0$ when no match is found over a set of nonzero measure in $\bbR$, and we have $\alpha \in (0,1]$ otherwise. 
The above definition of matching affinity corresponds to "matching affinity C $(\rho_C)$" in \cite{Garcia-Redondo2022-bv}. 
Note that since $X = Z$ and $Y = \hat{X} \subset X$ in the sub-sampling setting, we have $\gamma = \tilde{\gamma} = \tilde{\eta}$ in the definition above. 
In this case, the affinity between matched cycles $\gamma \in \PH_k(X)$ and $\hat{\gamma} \in \PH_k(\hat{X})$ simplifies to
\[
\alpha(\gamma, \hat{\gamma}) = \Jac(\gamma, \hat{\gamma})^2.
\]
We also note that in the sub-sampling case, Redondo et al's matching affinities "$A$ ($\rho_A$)" and "C ($\rho_C$)" coincide, while their matching affinities "B ($\rho_B$)" and "D ($\rho_D$)" satisfy $\rho_B = \rho_D = \rho_C^{1/2}$ (\cite{Garcia-Redondo2022-bv}). 
Consequently, the choice between matching affinity definitions should not yield qualitatively different results in the sub-sampling setting. 

\subsubsection{Topological bootstrapping and the prevalence score}\label{sec:topo_bts}
We now confine our attention to the bootstrap setting and define an important quantity: the \textit{prevalence score}.
Suppose that $X$ is sub-sampled $n$ times at a rate $r = \abs{\hat{X}}/\abs{X}$, generating bootstraps $\left\{\hat{X}_i^{(r)}\right\}_{i=1}^n$.
\cite{Reani2021-mu} measure the bootstrap stability of a nontrivial generator $\gamma \in \PH_k(X)$ by its \textit{prevalence score}
\begin{equation}\label{eqn:prevscore}
\rho(\gamma | n, r) \coloneqq \frac{1}{n} \sum_{j=1}^n \alpha(\gamma, \hat{\gamma}_j),
\end{equation}
where $\hat{\gamma}_j$ is the match of $\gamma$ in the $j^\text{th}$ bootstrap $\hat{X}_j^{(r)}$.
This is just the average matching affinity (over all bootstraps) between $\gamma$ and its matches. 

Note that $\rho: \PH_k(X) \to [0,1]$ is a random variable on $\PH_k(X)$ parameterized by $n$ and the sub-sampling rate $r = \abs{\hat{X}_i}/\abs{X}$. 
For a specific realization of the random variable $\rho$, we take $n, r$ fixed and write $\rho_X(\gamma)$.
In the present study, we compute prevalence scores fixing $k=1$. 

\subsubsection{Implementation}

The specific challenges of our problem context required a mild extension of existing topological bootstrapping implementations.
Our implementation of the topological bootstrap builds on \cite{Garcia-Redondo2022-bv}'s work, which efficiently integrates cycle registration with Ripser and Ripser-image (\cite{Bauer2021-an,Bauer2022-sq}) and refines the cycle affinity measures proposed by \cite{Reani2021-mu}.
We extended their implementation by broadening the conditions under which topological bootstrapping may be applied to include the case of arbitrary, precomputed metrics.
We conducted cycle registration using $n=1000$ bootstraps per dataset at a sub-sampling ratio of $r=0.90$ (no replacement); we considered $k=1$-dimensional cycle registration. 
Our implementation is branched from \cite{Garcia-Redondo2022-bv} and available on Github (\cite{Easley_Co_homology_cycle-match_for_2024}).

\subsection{Defining the prevalence-weighted Wasserstein}\label{sec:def_prevwt_Wp}

The space $\sD$ of persistence diagrams is a metric space (\cite{skrabaWassersteinStabilityPersistence2021}) under the Wasserstein distance. 
We also wish to include information about the sub-sampling stability of the underlying persistence module when comparing persistence diagrams.

\begin{definition}[Bootstrapped persistence module]\label{def:bstp_pers_sp}
Suppose $X$ is a metric space with a collection of sub-samples $\{\hat{X}_i^{(r_X)}\}_{i=1}^{n_X}$ and persistent $k$-homology module $\PH_k(X)$ such that $\PH_k\left( \hat{X}_i^{(r_X)} \right)$ are cycle-registered bootstraps of $\PH_k(X)$. The cycle-registered bootstraps $\PH_k\left( \hat{X}_i^{(r_X)} \right)$ empirically determine the prevalence score $\rho_X$ on $\PH_k(X)$, giving the triple $(\PH_k(X), \PH_k\left( \hat{X}_i^{(r_X)} \right), \rho_X)$. Note that the prevalence score is a random variable $\PH_k(X) \to [0,1]$ and $\rho_X$ is a \textit{realization} of it fully determined by the collection of sub-samples $\{\hat{X}_i^{(r_X)}\}_{i=1}^{n_X}$. Dropping the full set of peristence modules of bootstrapped sub-spaces, we define the \textbf{\textit{bootstrapped persistence of $X$}} as 
\[
\cP\cH_k(X) \coloneqq (\PH_k(X), \rho_X).
\]
When is clear from context, we will suppress the subscript $k$ (denoting the homology dimension). Writing $\PD_X$ as the persistence diagram descending from $\PH_k(X)$, we also define the \textbf{\textit{bootstrapped diagram of $X$}} as
\[
\cP\cD_X \coloneqq (\PD_X, \rho_X).
\]
\end{definition}

\begin{definition}[Space $\sP$ of bootstrapped diagrams]\label{def:P}
We define the \textbf{\textit{bootstrapped persistence spaces}} of bootstrapped persistence modules as
\[
\sP_k^H = \left\{ \cP\cH_k(X) = (\PH_k(X), \rho_X) \mid \PH_k(X) \text{ is finitely generated and } \Pers_q(\PH_k(X)) < \infty \right\},
\]
where $\Pers_q$ denotes the "total persistence" operator (i.e., the $q^\text{th}$-power sum of lifetime values), and (similarly) of bootstrapped persistence diagrams as
\[
\sP = \left\{ \cP\cD(X) = (\PD_X, \rho_X) \mid \Pers_q(\PD_X)<\infty \text{ and } \abs{\PD_X \setminus \Delta} < \infty \right\},
\]
where $\Delta$ is the empty diagram.
\end{definition}

\begin{remark}
Note that Definition \ref{def:bstp_pers_sp}) given above distinguishes between persistence modules/diagrams generated by the same metric space when they are accompanied by different collections of sub-samples. In particular, if two different sub-samples of a metric space $X$ determine different prevalence scores $\rho_X$ and $\rho_X'$ on $\PH_k(X)$, then $(\PH_k(X), \rho_X) \neq (\PH_k(X), \rho_X')$ are distinct elements of $\sP_k^H$. By the same reasoning, $(\PD_X, \rho_X) \neq (\PD_X, \rho_X')$ also constitute distinct elements of $\sP$.
\end{remark}

\begin{remark*}
Suppose $\gamma \in \PH_k(X)$ is a (possibly trivial) generator of the persistence module $\PH_k(X)$.
When context is clear, we will refer interchangeably to the generator $\gamma$ and its basis-associated birth-death pair $\gamma \cong (b,d) \in \PD_X$ in the persistence diagram $\PD_X$.
\end{remark*}

Given some generator $\gamma \cong (b, b+\ell)$ with birth time $b$ and lifetime $\ell$, we write its projection to the diagonal $\Delta$ as $\Delta(\gamma)$, given by $\Delta(\gamma) = (b + \ell/2, b + \ell/2)$. If $\gamma$ is also equipped with a prevalence score $\rho(\gamma)$, we define
\begin{equation}\label{gamma_rho}
\gamma^\rho \coloneqq \rho(\gamma) \cdot \gamma + (1 - \rho(\gamma)) \cdot \Delta(\gamma) = \frac{\rho\ell}{2}(1,-1) + \Delta(\gamma)
\end{equation}
as the $\rho$-shrinkage of $\gamma$ to the diagonal. Now, given $\cP\cD = (\PD, \rho) \in \sP$, define the \textbf{\textit{diagram shrinkage map}} $S$ by
\begin{align}\label{dgm_shrink}
S : \sP &\to \sD \nonumber \\
(\PD, \rho) &\mapsto \PD^{\rho}, \\
\PD^{\rho} &\coloneqq \left\{ \gamma^\rho \mid \gamma \in \PD \right\}. \nonumber
\end{align}
Note that $\PD^\rho$ is simply a different persistent diagram.

Denote the set of augmented diagram bijections between $\PD_X$ and $\PD_Y$ as $\Gamma_{XY}$, and let $\pi_q^{\Delta}(\gamma)$ be the $q$-norm distance $\norm{\gamma - \Delta(\gamma)}_q$.
For convenience, write $\rho^\dag(\gamma) \coloneqq 1 - \rho(\gamma)$.
We then define the \textit{prevalence-weighted} Wasserstein distance $\hat{W}_{p,q}$ as
\begin{equation}\label{eqn:prevwt_Wp}
\begin{split}
\hat{W}_{p,q} \left( \cP\cH_k(X), \cP\cH_k(Y) \right)
&\coloneqq \inf_{\phi \in \Gamma_{XY}} \left( \sum_{\gamma \in \PD^{(k)}_X} K_{p,q}((\gamma, \rho_X), (\phi(\gamma),\rho_Y)) \right)^{\frac{1}{p}}, \\
K_{p,q}((\gamma, \rho_X), (\eta,\rho_Y))
&\coloneqq \norm{\gamma^{\rho_X} - \eta^{\rho_Y}}^p_q + \left\lvert \rho^\dag_X(\gamma)  \pi_q^{\Delta}(\gamma) - \rho^\dag_Y(\eta) \pi_q^{\Delta}(\eta) \right\rvert^p.
\end{split}
\end{equation}
Intuitively, we can think of $K_{p,q}((\gamma, \rho_X), (\phi(\gamma),\rho_Y))$ as defining an interpolated cost between the diagram Wasserstein $W_{p,q}$ and the cost of projection to the diagonal, where prevalence scores control interpolation weights.

The prevalence-weighted Wasserstein (Eqn. \ref{eqn:prevwt_Wp}) bears some similarity to the cycle-registered distance of \cite{Reani2021-mu}:
\[
d^Z_{IM_{p,q}}(\PH_k(X), \PH_k(Y )) \coloneqq \left( \sum_{\gamma \sim \eta} \norm{\pi(\gamma) - \pi(\eta)}_q^p + \sum_{\gamma \sim \emptyset} \pi_q^\Delta(\gamma)^p + \sum_{\eta \sim \emptyset} \pi_q^\Delta(\eta)^p \right)^{\frac{1}{p}},
\]
which sums over cycles $\gamma \in \PH_k(X)$ and $\eta \in \PH_k(Y)$ based on their match status (via the co-embedding $X \hookrightarrow Z \hookleftarrow Y$) in $\PH_k(Z)$. However, unlike $d^Z_{IM_{p,q}}$, $\hat{W}_{p,q}$ neither expects nor allows the incorporation of cycle registration information \textit{between} its input spaces. Instead, we emphasize that each input space $\cP\cH_k(X)$ and $\cP\cH_k(Y)$ incorporates, via the prevalence score, information about the cycle-registered stability of \textit{each} of the modules $\PH_k(X)$ and $\PH_k(Y)$.

Theorem \ref{thm:metric_prevwt_thm} (below) asserts that (under Definition \ref{eqn:prevwt_Wp}), $\hat{W}_{p,q}$ is a distance metric on the space $\sP$ (Definition \ref{def:P}). However, $\hat{W}_{p,q}$ likely does not admit a straightforward comparison to $W_{p,q}$ under most circumstances (see Proposition \ref{prop:hatW_vs_W_inp-stbl}).

\subsection{Metric properties of the prevalence-weighted Wasserstein}\label{sec:prevwtWp_proof}

In this subsection, we assume that all persistence diagrams have finitely many above-diagonal points and fix an arbitrary homology dimension $k \geq 1$. In addition, we will refer to $\gamma \in \PH_k$ and its birth-death point in $\PD$ interchangeably except when context renders it necessary to parse the distinction. Similarly, we use $\Delta$ to denote both the diagonal of an arbitrary diagram and the empty diagram (whose only component in the diagonal), except where the conflation would create ill-definition.

\begin{theorem}\label{thm:metric_prevwt_thm}
The prevalence-weighted Wasserstein is a metric on the space $\sP$ (Definition \ref{def:P}).
\end{theorem}

\begin{proof}
The symmetry and non-negativity of $\hat{W}_{p,q}$ follow immediately from the definition. It remains to show that it is non-degenerate and satisfies the triangle inequality.

\paragraph{Nondegeneracy.}

Suppose $\gamma \cong (b, b + \ell)$ and $\eta \cong (b', b' + \ell')$ have associated nonzero prevalence scores $\rho(\gamma) = r$ and $\rho(\eta) = s$. Note that $\Delta(\gamma) = (b+\ell/2, b+\ell/2)$ and $\Delta(\eta) = (b' + \ell'/2, b' + \ell'/2)$. We then have
\begin{align*}
K((\gamma,r), (\eta, s))
&= \norm{\gamma^\rho - \eta^\rho}_q^p + \abs{(1-r)\pi_q^\Delta(\gamma) - (1-s)\pi_q^\Delta(\eta)}_p \\
&= 2^{-p}\norm{(r\ell - s\ell')(1,-1) + 2(\Delta(\gamma) - \Delta(\eta))}_q^p + 2^{p/q-p}\abs{(1-r)\ell - (1-s)\ell'}^p \\
&= 0 \\
\iff 0
&= (r\ell - s\ell')(1,-1) + 2(\Delta(\gamma) - \Delta(\eta)) \\
\text{ and } 0
&= (1-r)\ell - (1-s)\ell'.
\end{align*}
Since the second equality holds if and only if $\ell - \ell' = r\ell - s\ell'$, we have
\begin{align*}
K_{p,q}((\gamma,r), (\eta, s))
&= 0 \\
\iff 0
&= \norm{(b-b') + (\ell-\ell'),(b' - b)}_q^p,
\end{align*}
which holds if and only $b = b'$ and $\ell = \ell'$. Because the kernel $K_{p,q}$ is nondegenerate, $\hat{W}_{p,q}$ is as well.

\paragraph{Triangle inequality.}

Suppose $\cP\cD_X$, $\cP\cD_Y$, and $\cP\cD_Z$ are in $\sP$ and $\phi_{ij}$ are the minimizing diagram bijections corresponding to $\hat{W}_{p,q}(\cP\cD_i, \cP\cD_j)$ for $i,j \in \{X,Y,Z\}$. Since $\phi_{ij}$ are diagram bijections, $\tilde{\phi}_{XZ}(\gamma) \coloneqq \phi_{XY}^*\phi_{YZ}$ is a diagram bijection as well. Define
\begin{equation}
S^{\tilde{\phi}_{XZ}}_{p,q} \left( \cP\cD_X, \cP\cD_Z \right)^p
\coloneqq \sum_{\gamma \in \PD_X} K_{p,q} \left( (\gamma, \rho_X), (\tilde{\phi}_{XZ}(\gamma), \rho_Z) \right),     
\end{equation}
whose terms expand as
\begin{align*}
K_{p,q} ( \gamma, \tilde{\phi}_{XZ}(\gamma) )
&= \norm{\gamma^{\rho_X} - \tilde{\phi}_{XZ}(\gamma)^{\rho_Z}}_q^p + \abs{\rho^\dag_X(\gamma)\pi_q^\Delta(\gamma) - \rho^\dag_Z(\tilde{\phi}_{XZ}(\gamma))\pi_q^\Delta(\tilde{\phi}_{XZ}(\gamma))}^p.
\end{align*}

Because $\tilde{\phi}_{XZ}$ is a diagram biejction, it follows immediately that $\hat{W}_{p,q} \left( \cP\cD_X, \cP\cD_Z \right)^p  \leq S^{\tilde{\phi}_{XZ}}_{p,q} \left( \cP\cD_X, \cP\cD_Z \right)^p$, and it suffices to show that
\[
S^{\tilde{\phi}_{XZ}}_{p,q} \left( \cP\cD_X, \cP\cD_Z \right)^p \leq \hat{W}_{p,q} \left( \cP\cD_X, \cP\cD_Y \right)^p + \hat{W}_{p,q} \left( \cP\cD_Y, \cP\cD_Z \right)^p.
\]
This follows from applying Minkowski's inequality separately to sums of each term of $K_{p,q}$. First, we have
\begin{align*}
&\left( \sum_{\gamma \in \PD_X} \abs{\rho_X^\dag(\gamma)\pi_q^\Delta(\gamma) - \rho_Y^\dag(\phi_{XY}(\gamma))\pi_q^\Delta(\phi_{XY}(\gamma))}^p \right)^{\frac{1}{p}} + \\
&\left( \sum_{\gamma \in \PD_X} \abs{\rho_Y^\dag(\phi_{XY}(\gamma))\pi_q^\Delta(\phi_{XY}(\gamma)) - \rho_Z^\dag(\tilde{\phi}_{XZ}(\gamma))\pi_q^\Delta(\tilde{\phi}_{XZ}(\gamma))}^p \right)^{\frac{1}{p}} \\
&\geq  \left( \sum_{\gamma \in \PD_X} \abs{\rho_X^\dag(\gamma)\pi_q^\Delta(\gamma) - \rho_Z^\dag(\tilde{\phi}_{XZ}(\gamma))\pi_q^\Delta(\tilde{\phi}_{XZ}(\gamma))}^p \right)^{\frac{1}{p}}.
\end{align*}
Another application of Minkowski's inequality, this time to the left-hand term of $K_{p,q}$, gives
\[
\norm{\gamma^{\rho_X} - \tilde{\phi}_{XZ}(\gamma)^{\rho_Z}}_q
\leq \norm{\gamma^{\rho_X} - \phi_{XY}(\gamma)^{\rho_Y}}_q + \norm{\phi_{XY}(\gamma)^{\rho_Y} - \tilde{\phi}_{XZ}(\gamma)^{\rho_Z}}_q.
\]
Taken together with the last inequality, this gives $K_{p,q}(\gamma, \tilde{\phi}_{XZ}(\gamma))^{1/p} \leq K_{p,q}(\gamma, \phi_{XY}(\gamma))^{1/p} + K_{p,q}(\phi_{XY}(\gamma), \tilde{\phi}_{XZ}(\gamma))^{1/p}$, which implies
\[
S^{\tilde{\phi}_{XZ}}_{p,q} \left( \cP\cD_X, \cP\cD_Z \right)^p \leq \hat{W}_{p,q} \left( \cP\cD_X, \cP\cD_Y \right)^p + \hat{W}_{p,q} \left( \cP\cD_Y, \cP\cD_Z \right)^p.
\]
Thus, it follows that $\hat{W}_{p,q}$ is a metric on $\sP$.
\end{proof}

\subsubsection{Simple properties of $\hat{W}_{p,q}$}

To prove a few of the basic properties of $\hat{W}_{p,q}$, we begin by defining some special subspaces of $\sP$.

\begin{definition}[The space $\sP_0$ of sample-degenerate diagrams]\label{def:sampdegen_dgm}
Suppose $\cP\cD = (\PD, \rho) \in \sP$ and $\PD$ is a diagram descending from a persistence module $\PH$ with finitely many nontrivial generators. We say that any $\gamma \in \PH$ with $\rho(\gamma) = 0$ is a \textbf{\textit{sample-degenerate generator}}. Similarly, $\cP\cH$ is a \textbf{\textit{sample-degenerate persistence module}} if $\rho \equiv 0$ and the corresponding $\PD$ is a \textbf{\textit{sample-degenerate diagram}}. We use $\sP_0$ to denote the subspace of sample-degenerate diagrams.
\end{definition}

\bigskip

\begin{proposition}
If $\cP\cD_X, \cP\cD_Y \in \sP_0$, then
\[
\hat{W}_{p,q}\left( \cP\cD_X, \cP\cD_Y \right) \leq 2^{1/q}(1 + 2^{-p})^{1/p} \cdot W_{p,1} (\PD_X, \PD_Y)
\]
\end{proposition}

\begin{proof}
For $\gamma = (b,b+ \ell)$ and $\eta = (b',b'+\ell')$, we compute 
\begin{align*}
K_{p,q}((\gamma, 0), (\eta, 0))
&= \norm{\Delta(\gamma) - \Delta(\eta)}_q^p + \abs{\pi_q^\Delta(\gamma) - \pi_q^\Delta(\gamma)}^p \\
&= 2^{p/q} \left(\abs{(b-b') + (\ell-\ell')/2}^p + 2^{-p}\abs{\ell-\ell'}^p\right).
\end{align*}
Since $\abs{(b-b') + (\ell-\ell')/2}^p \leq (\abs{(b-b')} + \abs{(\ell-\ell')/2})^p \leq \norm{\gamma - \eta}_1^p$ and $\abs{\ell-\ell'}^p \leq \norm{\gamma - \eta}_1^p$, we have
\[
K_{p,q}((\gamma, 0), (\eta, 0)) \leq 2^{p/q} \norm{\gamma - \eta}_1^p \left( 1 + 2^{-p} \right),
\]
which gives the upper bound.
\end{proof}

\bigskip

\begin{definition}[The space $\sP_1$ of repeatable diagrams]
In analogy to Definition $\ref{def:sampdegen_dgm}$, we say $\gamma \in \PH$ is a \textit{\textbf{repeatable generator}} if $\rho(\gamma) = 1$. Simiarly, we say $\cP\cD = (\PD, \rho)$ is a \textit{\textbf{repeatable diagram}} if $\rho(\gamma) = 1$ for all off-diagonal $\gamma \in \PD$.
\end{definition}
\bigskip

\begin{corollary}
If $\cP\cD_X, \cP\cD_Y \in \sP_1$, then
\[
\hat{W}_{p,q}\left( \cP\cD_X, \cP\cD_Y \right) = W_{p,q} (\PD_X, \PD_Y).
\]
In other words, $\left. \hat{W}_{p,q} \right|_{\sP_1} = \left. W_{p,q} \right|_{\sD}.$
\end{corollary}

\bigskip

\begin{remark*}
We identify the empty diagram $\Delta$ with the element $(\Delta, \rho_0: \Delta \to 0) \in \sP_0$, and refer to  $\Delta$ and $(\Delta, \rho_0)$ interchangeably. 
\end{remark*}

\begin{corollary}\label{thm:degen_equiv}
Suppose $\cP\cD = (\PD, \rho) \in \sP$.
\begin{enumerate}[label=(\alph*)]
    \item For all $1 < p < \infty$,
    \[
    \hat{W}_{p,q}(\cP\cD, \Delta) = \left( \sum_{\gamma \in \PD} \pi_q^\Delta(\gamma)^p \right)^{\frac{1}{p}}
    \]
    if and only if $\cP\cD \in \sP_0 \oplus \sP_1$.

    \item Suppose $\PD \in \sD$ and $\rho, \rho'$ are different prevalence functions on $\PD$. Then
\[
\hat{W}_{p,q} \left( \cP\cD, \cP\cD' \right) = \left( \sum_{\gamma \in \PD} 2 \abs{\rho(\gamma) - \rho'(\gamma)}^p \pi_q^\Delta(\gamma)^p \right)^{\frac{1}{p}}
\]
if and only if there exists no diagram bijection $\varphi: \PD \to \PD$ such that $S^{\varphi}_{p,q} < S_{p,q}^{\varphi_0}$, where $\varphi_0$ is the identity on $\PD$.
\end{enumerate}
\end{corollary}

\begin{proof}
Take $\cP\cD = (\PD, \rho) \in \sP$ with off-diagonal points $\gamma_i = (b_i, b_i + \ell_i)$ for $i=1, \ldots, n$.

\begin{enumerate}[label=(\alph*)]
\item Suppose $p > 1$. We then have
\begin{align*}
\hat{W}_{p,q} (\cP\cD, \Delta)^p
&= \sum_{i=1}^n \norm{\gamma^{\rho_X}_i - \Delta(\gamma_i)}_q^p + \abs{(1 - \rho_x(\gamma_i)) \norm{\gamma_i - \Delta(\gamma_i)}_q - \norm{\Delta(\gamma_i) - \Delta(\gamma_i)}_q}^p \\
&= \sum_{i=1}^n \rho_X(\gamma_i)^p \norm{\gamma_i - \Delta(\gamma_i)}^p + \abs{1 - \rho_X(\gamma_i)}^p\pi_q^\Delta(\gamma_i)^p \\
&= \sum_{i=1}^n \pi_q^\Delta(\gamma_i)^p \left( \rho_X(\gamma_i)^p + \abs{1 - \rho_X(\gamma_i)}^p \right),
\end{align*}
Since $p > 1$ and $\rho_X$ takes image in $[0,1]$, we have
\[
\hat{W}_{p,q} (\cP\cD, \Delta)^p = \sum_{i=1}^n \pi_q^\Delta(\gamma_i)^p
\]
if and only if $\rho_X(\gamma_i) \in \{0,1\}$ for all $i$. Thus, $\cP\cD \in \sP_0 \oplus \sP_1$.

\item By definition, the given condition is that under which the identity map $\varphi_0$ is the cost-optimal bijection taking $\cP\cD \to \cP\cD'$. A straightforward calculation gives
\begin{align*}
K_{p,q} \left( (\gamma, \rho), (\gamma, \rho') \right)
&= \norm{(\rho(\gamma) - \rho'(\gamma))\gamma + (\rho'(\gamma) - \rho(\gamma))\Delta(\gamma)}_q^p + \abs{(\rho'(\gamma)-\rho(\gamma))\pi_q^\Delta(\gamma)}^p \\
&= 2 \abs{\rho(\gamma) - \rho'(\gamma)}^p \pi_q^\Delta(\gamma)^p,
\end{align*}
and summing over $\gamma \in \PD$ gives the result.
\end{enumerate}
This completes the proof.
\end{proof}

\begin{remark}
Following from the proof of Corollary \ref{thm:degen_equiv}(a), note that
\[
\hat{W}_{1,q}(\cP\cD, \Delta) = W_{1,q}(\PD, \Delta) = \sum_{i=1}^n \pi_q^\Delta(\gamma_i)
\]
also holds for any $\cP\cD \in \sP$.
\end{remark}

\subsection{Limitations of the prevalence-weighted Wasserstein}

\begin{proposition}[Input Instability]\label{prop:hatW_vs_W_inp-stbl}
The prevalence-weighted Wasserstein distance $\hat{W}_{p,q}$ is not bounded above by a constant multiple of $W_{p,q}$. 
\end{proposition}

The proposition above is closely related to another difficulty that plagues $\hat{W}_{p,q}$ in practice: the relation
\[
\pi_q^\Delta(\gamma) \gg \norm{\gamma - \phi(\gamma)}_q
\]
tends to hold for most diagram bijections (i.e., is generic) in our problem setting. The domination of $\hat{W}_{p,q}$ by the diagonal-projection cost of unstable components makes it difficult to compare directly to the usual Wasserstein, limiting its practical application.

%% file: supplemental_investigation.tex
\section[Supplemental Investigation]{Supplemental Investigation: Including Other Measurements of Inter-Subject Variability and the Feature-Shuffled Null}\label{sec:supp_inv}

\subsection{Background}

Given a finite dataset $\hat{\cS} \subset \cS \subset \bbR^D$ sampled from an (abstract, possibly infinite and infinite-dimensional) space $\cS$ of possible samples of the datatype in $\hat{\cS}$ embedded in a real $D$-dimensional ambient space, we use the following data to define a "realization of subject space":
\begin{itemize}
    \item A dimension-reducing map $\phi_\alpha : \cS \to \bbR^{d_\alpha}$ and its image $\phi_i(\hat{\cS}) \coloneqq \hat{\cS}_\alpha$ on $\hat{\cS} \subset \cS$

    \item A dissimilarity (pre-)metric $\delta_\beta : \cS_\alpha \times \cS_\alpha \to \bbR^+$
\end{itemize}
In the case that $\hat{\cS}$ is a neuroimaging dataset, a "realization of subject space" corresponds to a joint choice of a brain representation and dissimilarity function.
We then consider $X_{\alpha\beta} \coloneqq (\hat{\cS}_\alpha, \delta_\beta)$ as a finite (pre-)metric space and consider its persistent homology under the Vietoris-Rips filtration.
As a reminder, we realize subject space in practice by computing pairwise distances between all subjects for each brain representation method, decomposition rank within a given representation, and considered feature type.
Each feature type under consideration is structured either as a vector (maps, amplitudes) or a symmetric positive semidefinite (SPSD) matrix (network matrix, connectivity network, etc).
If we wish to consider multiple types of inter-subject variability measurements for a given brain representation, we may define multiple choices $\delta_\beta$ of (pre-)metric for a given $\cS_\alpha$. 
In this supplemental investigation, we consider (a) the effect of varying choices of $\delta_\beta$ on the characterization of brain representation differences laid out in the primary manuscript and (b) the effect of comparing against the feature-shuffled null as opposed to the subject-shuffled null. 

\subsection{Methods}

For each brain representation method, decomposition rank within a given representation, and considered feature type, we compute pairwise distances between all subjects.
Each feature type under consideration is structured either as a vector (maps, amplitudes) or a symmetric positive semidefinite (SPSD) matrix (network correlation matrices).
The primary comparisons were performed using correlation correlation because it enabled comparisons across feature types, which was calculated for vectorized and network matrices as described in \S \ref{sec:dist_fns}.
In this investigation, we consider the impact of alternate measures for both vector and matrix datatypes.

\subsubsection{Inner product dissimilarity for vectorized datatypes}

Let $\langle \cdot , \cdot \rangle$ denote the usual inner product on $\bbR^d$.
We define
\begin{equation}
\delta_{v}(s_i, s_j)^2 = 1 - \langle s_i, s_j \rangle^2 \label{eqn:inner_dissim}
\end{equation}
after scaling the Gram matrix $\delta_{ij} = \langle s_i, s_j \rangle$ to have entries in $[0,1]$.

\subsubsection{Geodesic distance between correlation matrices}

In the SPSD matrix case, we consider the geodesic distance between matrices on the Riemannian symmetric positive-definite (SPD) cone.
The geodesic distance $\delta_{pd}$ on the symmetric positive definite cone (\cite{Moakher2005-rx,pennecRiemannianFrameworkTensor2006a}) is given by
\begin{equation}\label{Riem_dist_Pn}
\delta_{pd}(A_i, A_j) = \norm{\log\left( A_i^{-1}A_j \right)}_F
\end{equation}
and is efficiently implemented via the approximate joint diagonalizer (\cite{Congedo2015-ce}). This geodesic distance is sometimes used to compute the "geometric mean" of correlation matrices (\cite{dadiBenchmarkingFunctionalConnectomebased2019}). 

\begin{figure}[h]
\begin{center}
\textbf{Feature type by dissimilarity function} 
\end{center}
\bigskip
\hskip 0.25in
\tikzset{grow'=down,level distance=25mm}
\Tree 
[.$\delta_{v_1},\delta_{v_2}$ 
 Maps 
 Amplitudes 
]
\quad\vline
\tikzset{grow'=right,level distance=35mm}
\vskip -1 in
\hskip +2.25in
\Tree 
[.$\delta_{pd_1},\delta_{pd_2}$
 [.{Network Matrices}
 Spatial
 Full
 Partial
 ]
]
\vskip 0.21in
\caption[Dissimilarity functions and feature types]{Tree showing admissible feature types for each dissimilarity function discussed in subsection \ref{sec:dist_fns}.}
\label{fig_sppl:dist_tree}
\end{figure}

\subsubsection{The feature-shuffled null distribution}\label{sec:feat_shuff}
To complement the subject-shuffled null, we also constructed a null retaining community structure while removing projective structure: the "feature-shuffled" null. To generate "feature-type" null distributions, we held each subject's distribution of brain representation features fixed while independently shuffling feature order within each subject. This breaks the map of "global coordinates" assigned (as feature types) to the dimension-reduced dataset, but retains the individual variability in subjects' unique feature sets.
Formally, given a length-$N$ sequence $(\eta) \coloneqq \eta_1, \ldots, \eta_N$ of permutations (specifically, derangements) in $\Sym(d_k)$, we define a corresponding null dataset $X^{(\eta)}$
\begin{equation}\label{eqn:feat_perms}
    (\eta) : [X]_{ij} \mapsto [X]_{i \eta_i(j)}.
\end{equation}
Randomly selecting $R$ such sequences $(\eta)$ gives an approximate empirical "feature-null" distribution.
Because permutations shuffle feature order and fix per-subject feature distributions, no permutation sequence generating this null distribution needs to account for the family structure present in HCP data.
We investigate the implications of comparing against the feature-shuffled null in this supplement.

\subsection{Results}\label{sec:metric_supp_res}

\input{tables/significance_table}

\subsubsection{Topological signal and stability by feature and metric choice}\label{sec:brep_metric_stability}

We show significance results against the subject-shuffled null distribution in Table \ref{tbl:sig_solo-breps}.
Of the 72 brain representations screened by AUC significance, 45 showed significantly non-null inter-subject variability structure at $\alpha = 0.05$. 
The 27 brain representations that did not pass the threshold for topologically significant inter-subject variability structure included nearly all subject space realizations using the inner-product derived distance metric (Eq. \ref{eqn:inner_dissim}) and a few especially low- or high-dimensional partial correlation matrices (Table \ref{tbl:sig_solo-breps}). 

We also considered the geodesic distance on the positive-definite cone for correlation matrix-type features.
The profile of significantly non-null network matrices under the PSD cone geodesic metric departs distinctly from the significance profile of network matrices with respect to the correlation dissimilarity.
In particular, higher-dimensional network matrices (HCP-MMP 1.0, ICA300, and Schaefer200/600) that showed significance with respect to the correlation dissimilarity failed significance under the geodesic metric, while some low-dimensional network matrices (ICA15, ICA25, Yeo17) appeared significant under the geodesic metric and not the correlation dissimilarity. 
This result suggests that network matrix distances are more affected by the geometry of the positive definite at lower dimensions than higher ones.

\subsubsection{Evaluating dissimilarity with non-correlation metrics under the subject-shuffled null}

\begin{figure}
    \centering 
    \includegraphics[width=\linewidth]{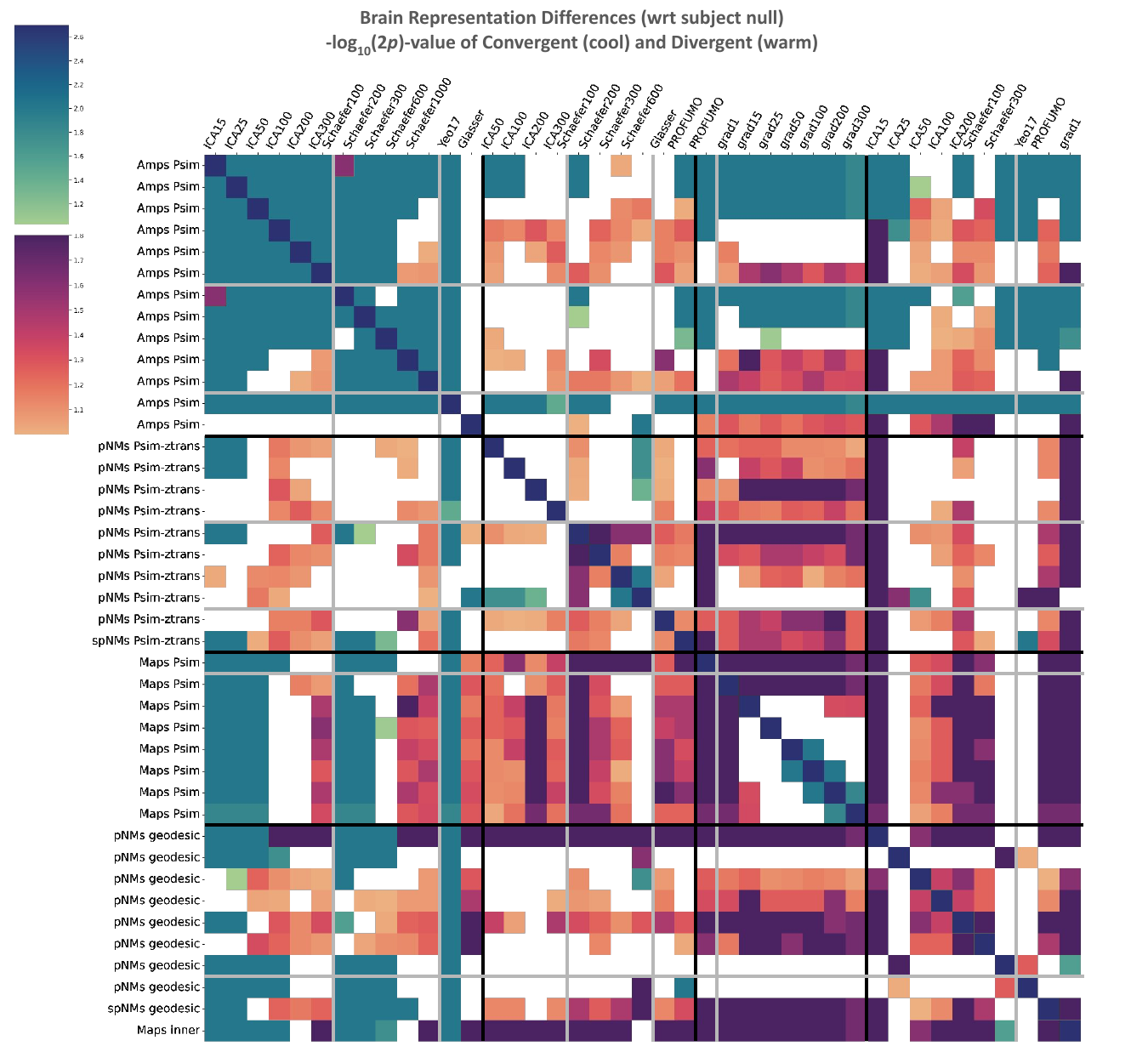}
    \caption[Significance results for all metric types against the subject null.]{Significance results for all metric types against the subject null. As in the main investigation, pairwise significance is evaluated at $\alpha=0.05$ as convergent, divergent, or incomparable. Overall results are similar to correlation-only when restricting to significantly non-null distributions; network matrices are more likely to be pairwise divergent under the geodesic metric and pairwise incomparable when computed from the correlation dissimilarity.}
    \label{fig_sppl:subject-pval_all-pairwise_grid}
\end{figure}

\begin{sidewaysfigure}
    \centering
    \includegraphics[width=\linewidth]{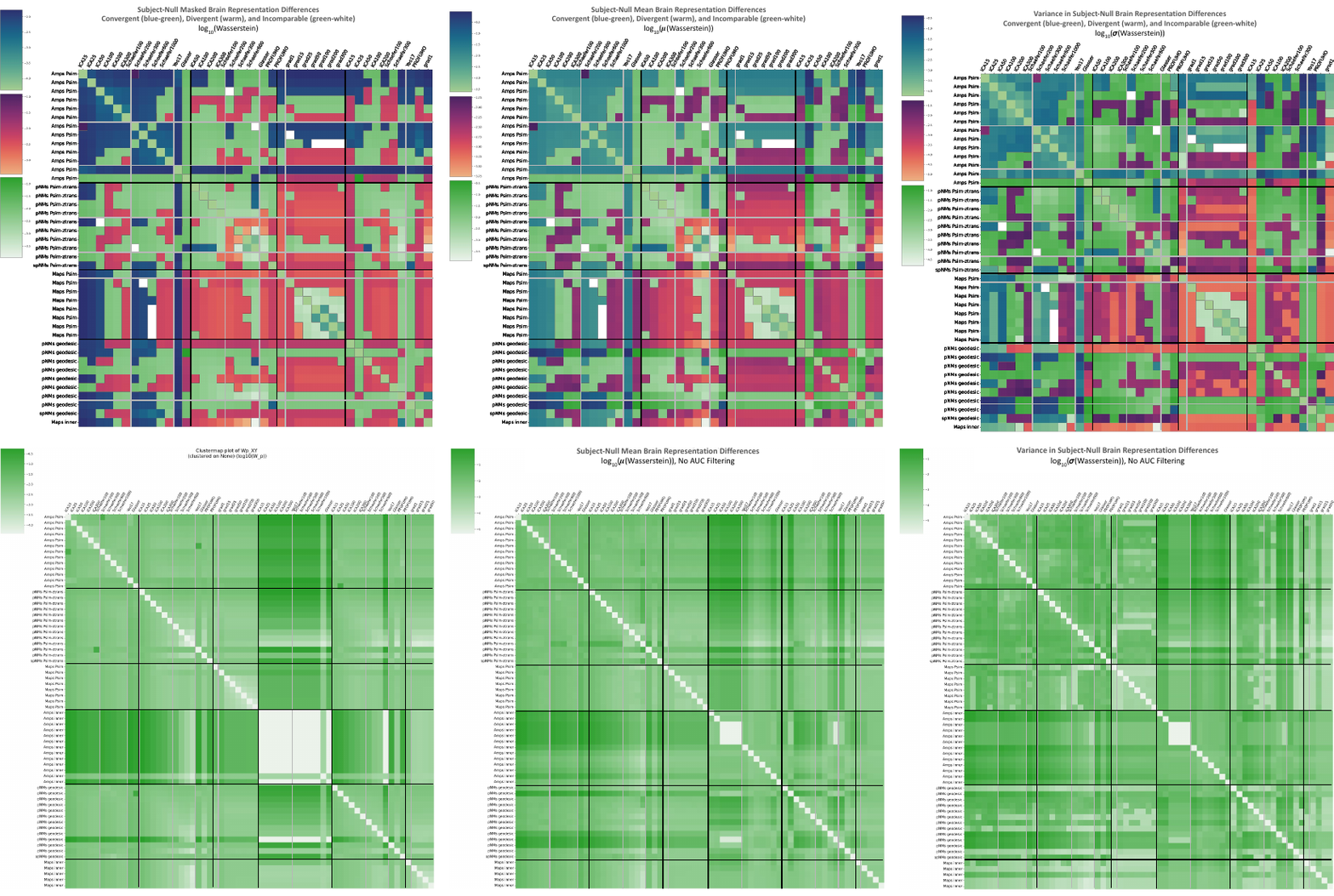}
    \caption[Wasserstein distance distributions, real and subject null.]{Wasserstein distance distributions from both real and \textbf{\textit{subject}} null data across all distance metrics. As when restricting to the correlation dissimilarity, the properties of the null differences between a pair of brain representations have a stronger determination of the significance of their relationship than to do the null distances. Surprisingly, null distances based on the inner-product cosine distance are \textit{less} likely to be trivial than their real-data counterparts.}
    \label{fig_sppl:subject-distances-pairwise_grids}
\end{sidewaysfigure}

\begin{figure}
    \centering
    \includegraphics[width=0.48\linewidth]{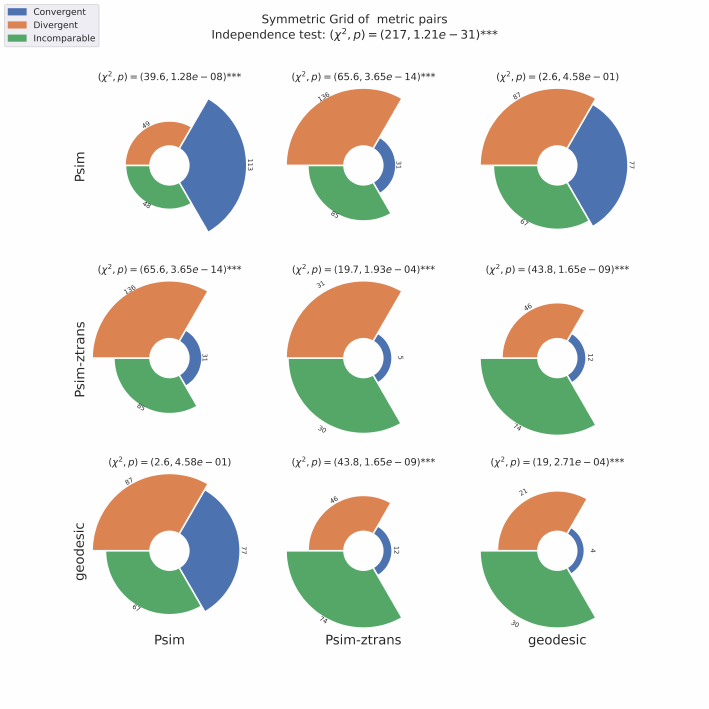}
    \includegraphics[width=0.48\linewidth]{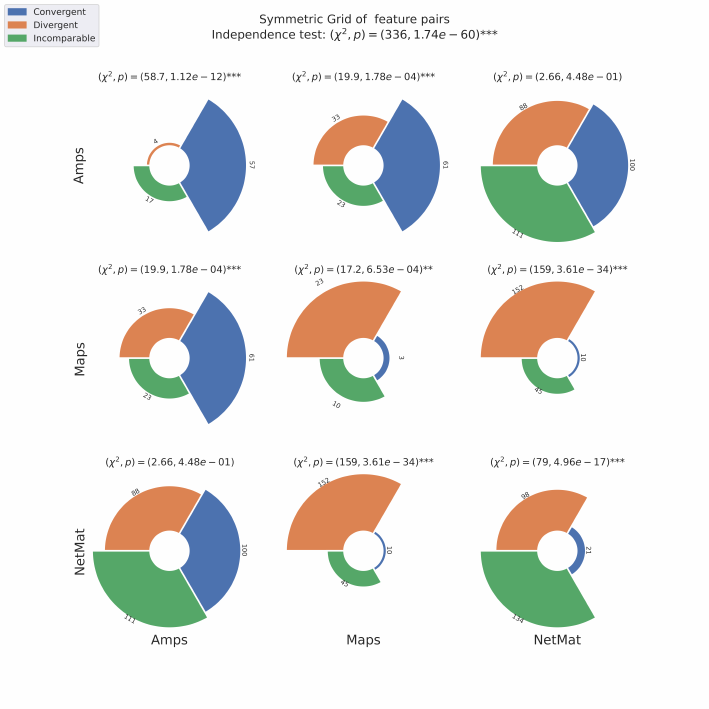}
    \caption[Significantly similar/different brain representations.]{Significance results for all metric types against the subject null. Aggregated by metric choice (left) and feature choice (right). Statistical sensitivity to analysis choices largely corroborates the insights gained when restricting to the correlation dissimilarity. However, the geodesic distance on connectivity matrices is more likely to generate divergent comparisons than the correlation dissimilarity, under which nearly all connectivity matrices are pairwise incomparable.}
    \label{fig_sppl:sym_feat_chi2_allres}
\end{figure}

Non-correlation dissimilarity functions largely corroborate insights from correlation case; see figure \ref{fig_sppl:subject-pval_all-pairwise_grid}. 
Maps-based and amplitudes-based feature comparison profiles remain largely unchanged (figure \ref{fig_sppl:sym_feat_chi2_allres}), and the statistical significance of brain representations differences remains much more driven by the behavior of the null distribution than the raw distances between representations (figure \ref{fig_sppl:subject-distances-pairwise_grids}).

\begin{figure}
    \centering
    \fbox{\includegraphics[width=0.48\linewidth]{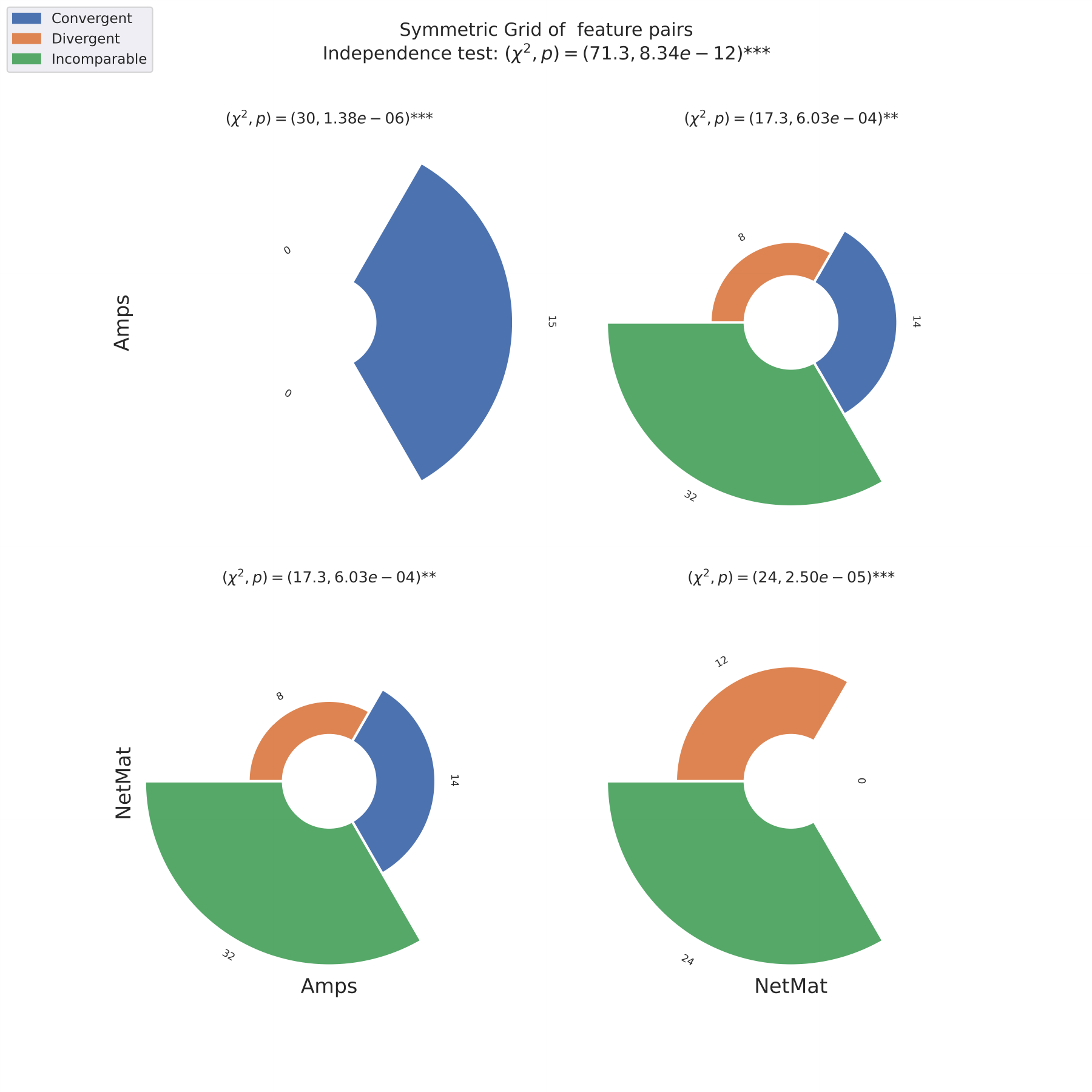}}
    \fbox{\includegraphics[width=0.48\linewidth]{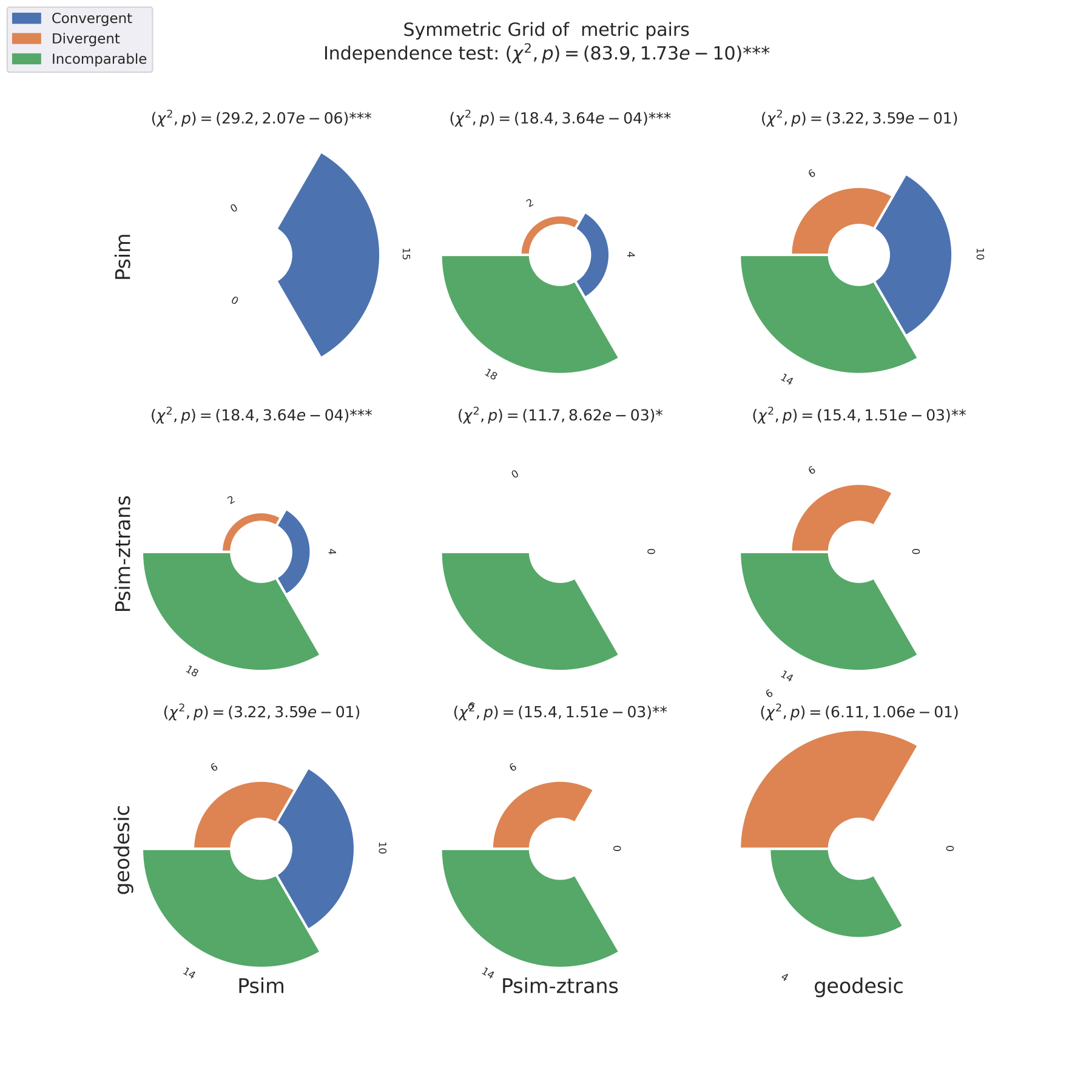}}
    \fbox{\includegraphics[width=0.48\linewidth]{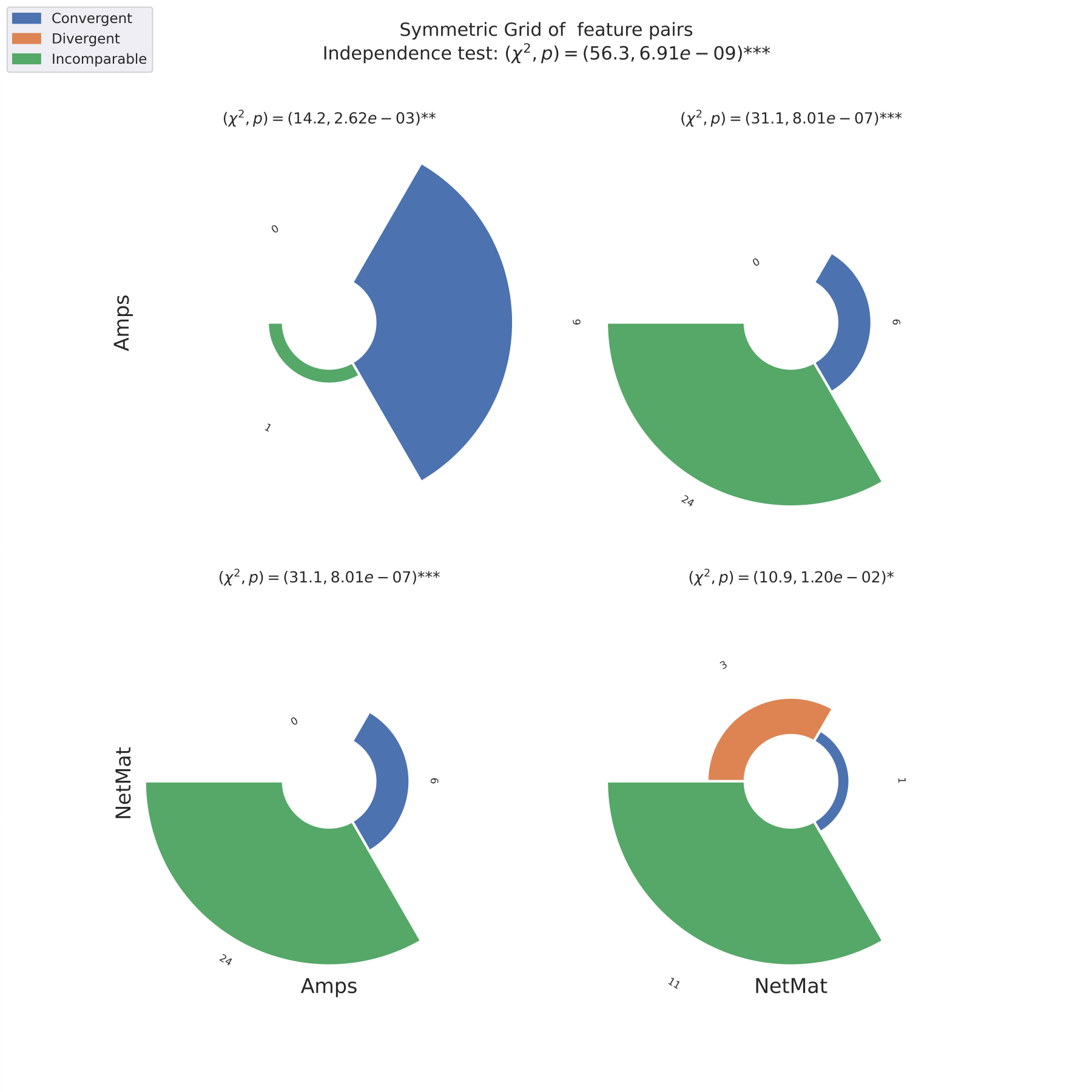}}
    \fbox{\includegraphics[width=0.48\linewidth]{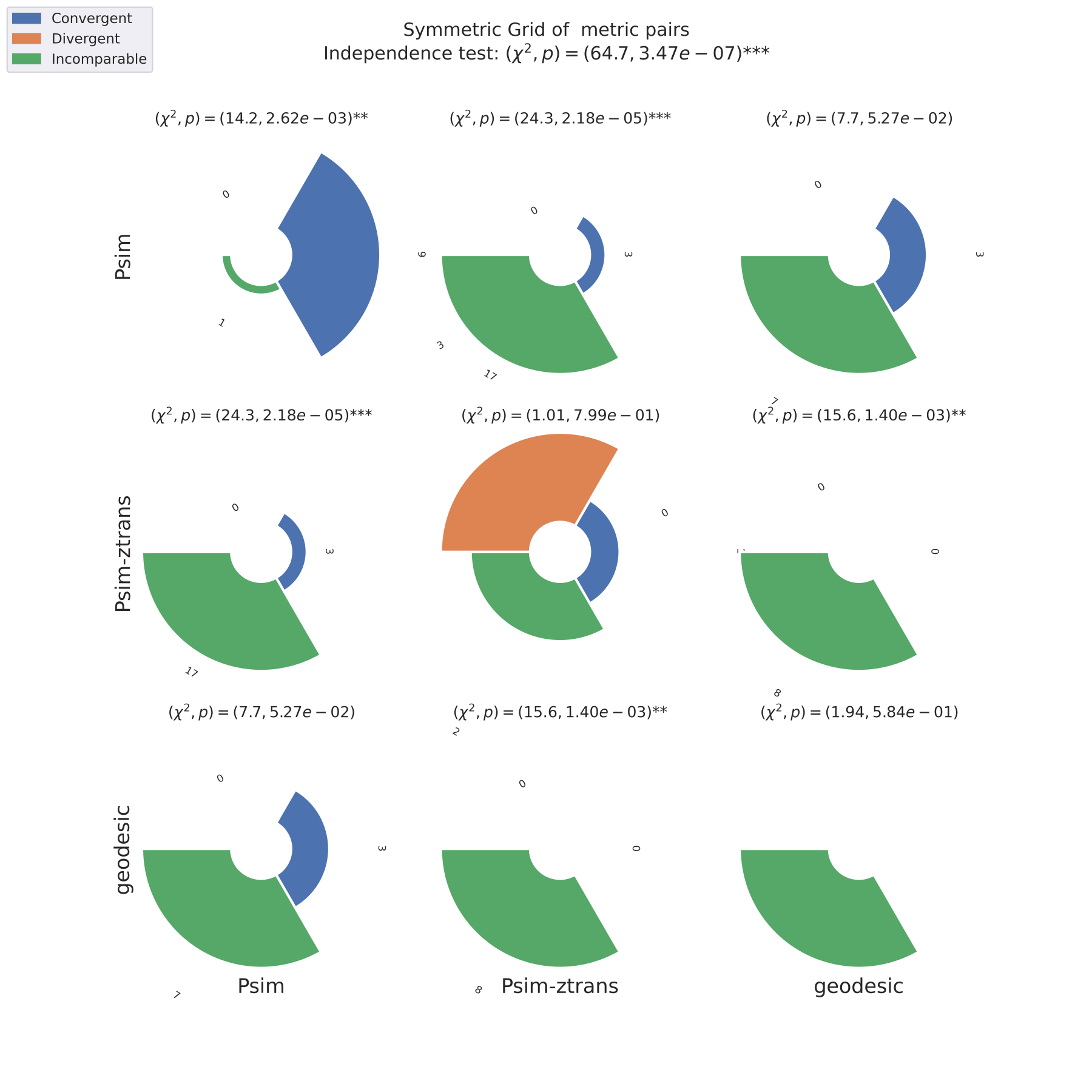}}
    \caption{Subtle differences arise between behavior of ICA- (top) and Schaefer-derived (bottom) network matrix differences when using the geodesic distance on the symmetric positive-definite cone. These differences, aggregated by metric (left) and feature (right), are not captured by the correlation dissimilarity.}
    \label{fig_sppl:NM_ICA-vs-Schaefer}
\end{figure}

However, the geodesic distance on connectivity matrices is more likely to generate divergent comparisons than the correlation dissimilarity, under which nearly all connectivity matrices are pairwise incomparable. 
These differences are driven by subtle differences in the responses of ICA- and Schaeffer-derived functional connectivity matrices to the geodesic distance on the symmetric positive-definite (SPD) cone, as shown in figure \ref{fig_sppl:NM_ICA-vs-Schaefer}.
The SPD geodesic captures differences in the organization of inter-subject variability when measured via ICA functional connectivity networks versus Schaefer functional connectivity networks, while the correlation dissimilarity registers them as incomparable.
This distinction in sensitivity between the geodesic metric and correlation dissimilarity is subtle but important, as it suggests a difference between the intra-comparability of functional connectivity measurements of hard and soft parcellations at different parcellation resolutions.

\subsubsection{Comparing against the feature-shuffled null}

\begin{figure}
    \centering 
    \includegraphics[width=\linewidth]{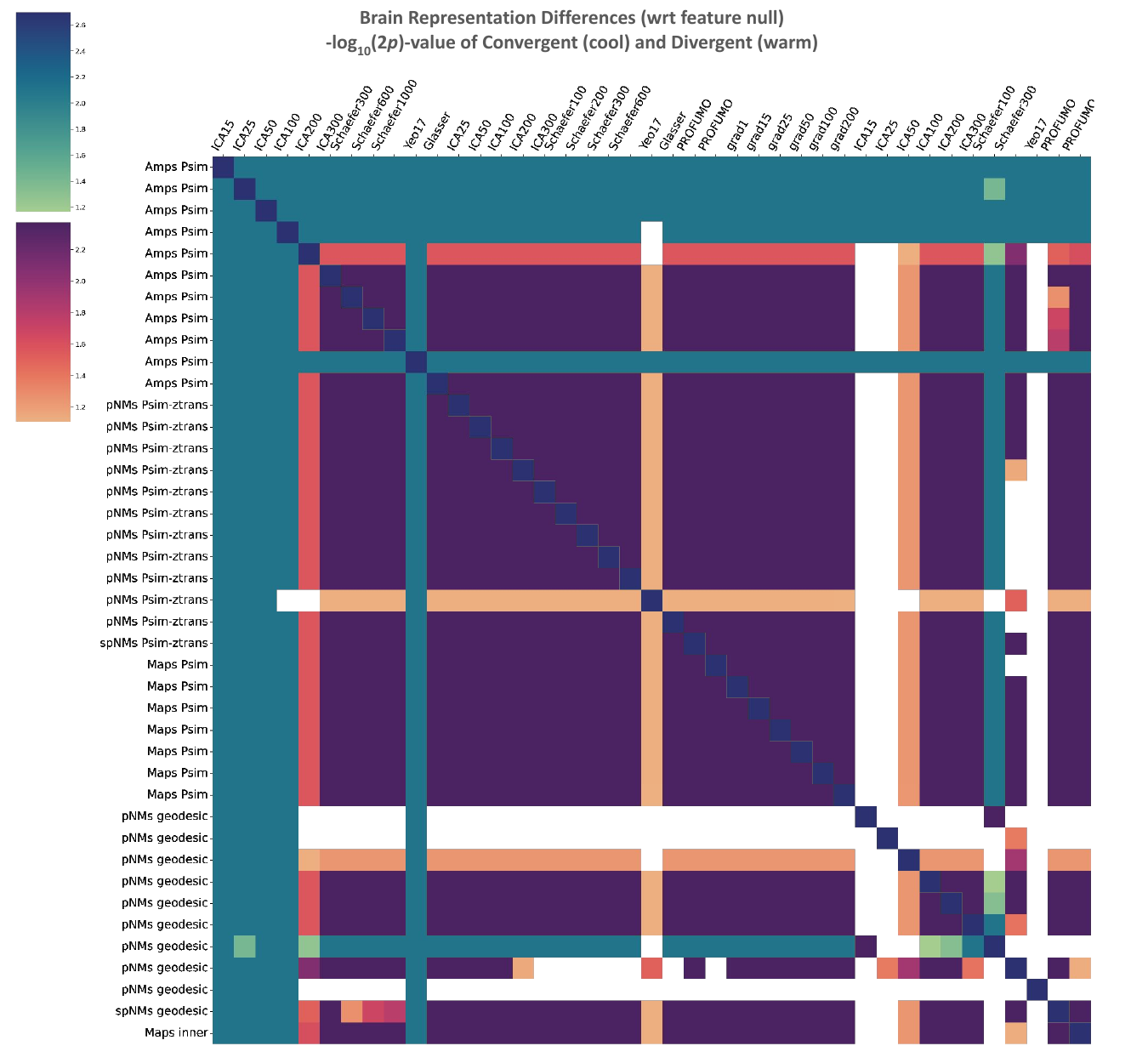}
    \caption[Significance results for all metric types against the feature null.]{Significance results for all metric types against the feature null. Note the relative lack of structure in comparison to the subject-null case. Distance significance exhibits far fewer distinctions among brain representation groupings, highlighting the lack of structure remaining in data after feature shuffling.}
    \label{fig_sppl:feature-pval_all-pairwise_grid}
\end{figure}

\begin{sidewaysfigure}
    \centering
    \includegraphics[width=\linewidth]{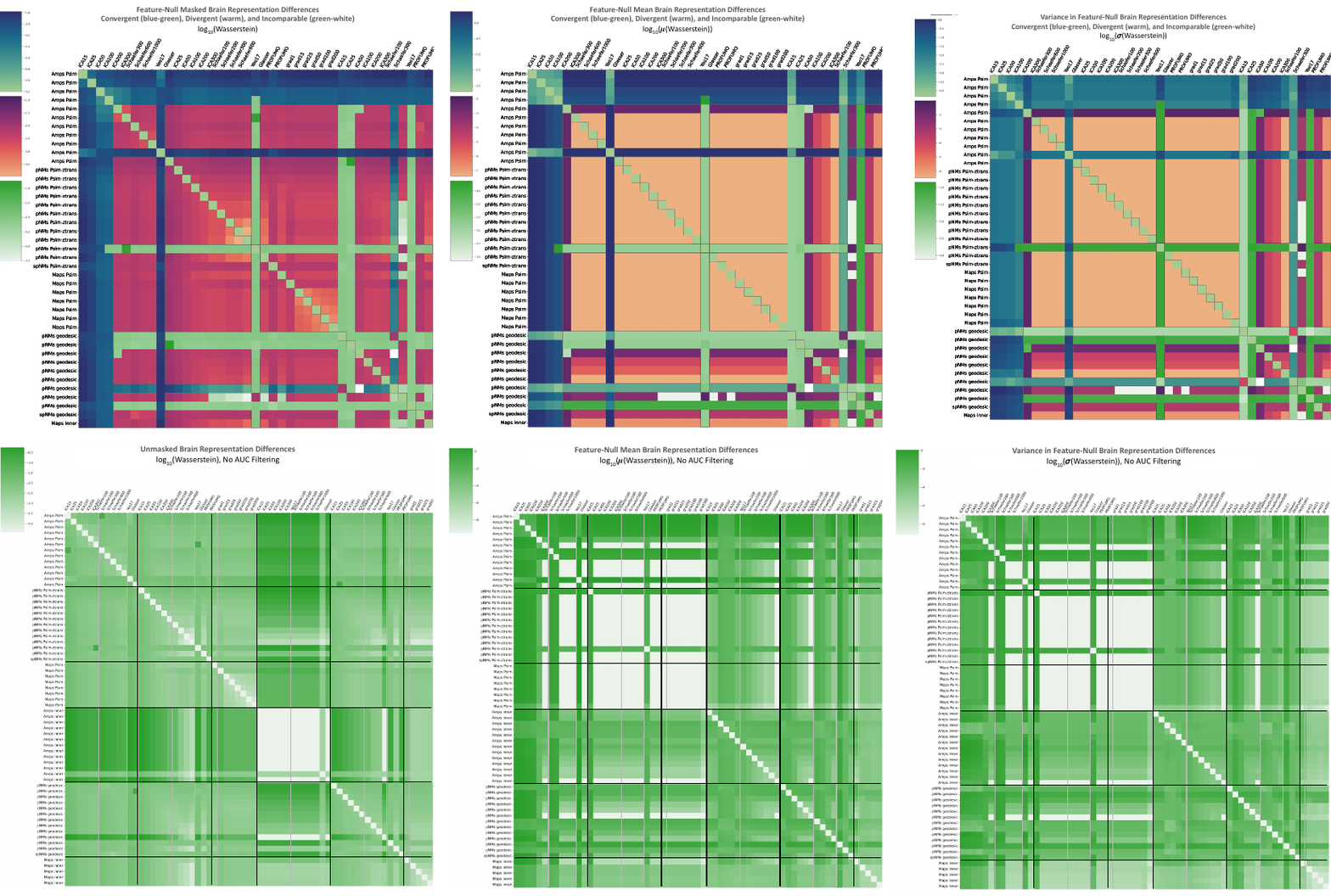}
    \caption[Wasserstein distance distributions, real and feature null.]{Wasserstein distance distributions from both real and \textbf{\textit{feature}} null data across all distance metrics. Most persistence diagrams produced by the feature null were trivial under the correlation dissimilarity. Surprisingly, the trivial diagrams produced by the inner-product cosine distance ("inner") from real data were matched with substantially \textbf{\textit{non}}-trivial diagrams in the null case.}
    \label{fig_sppl:feature-distances-pairwise_grids}
\end{sidewaysfigure}

As shown in figure \ref{fig_sppl:feature-pval_all-pairwise_grid}, the feature-shuffled null enforces are a vastly more strict convergence criteria than the subject-shuffled null.
This emerges, in part, as a the consequence of the large number of empty (or almost empty) diagrams generated by the feature-shuffling permutations, resulting a large number of zero-mean, zero-variance distance-pair distributions between brain representations (as shown in figure \ref{fig_sppl:feature-distances-pairwise_grids}).
The lack of structure (relative to the subject-shuffled null) in significance values demonstrates the relative stringency of the feature-shuffled null: much more data-specific information is destroyed compared to the subject-shuffled null.
However, it is still notable that the block of low-dimensional ICA amplitudes are significantly convergent with themselves and all other representations.
This suggests that the \textbf{\textit{unordered distribution}} of amplitude values carries nontrivial inter-subject variability information.
For example, these results suggest that subject "fingerprinting" should remain possible on order-permuted ICA amplitudes data (provided it is not also subject-permuted).

\subsection{Discussion}

This supplemental investigation underscores the primary findings of the main manuscript while elucidating the consequences of some of its key methodological decisions. 
The results above show that our choice of null model is crucial to the interpretability of the main results and suggest that further information can be gleaned from the exploration of between-subject distance/dissimilarity metrics beyond correlation.

The distance significance (figures \ref{fig_sppl:subject-pval_all-pairwise_grid} and \ref{fig_sppl:feature-pval_all-pairwise_grid}) and null distribution results (figures \ref{fig_sppl:subject-distances-pairwise_grids} and \ref{fig_sppl:feature-distances-pairwise_grids}) show a stark contrast of brain representation comparison summaries.
Care must be taken when interpreting the statistical significance of a distance with respect to a distribution of distances between null data.
As the significance threshold $\alpha \to 0$, Figure \ref{fig:AUC_significance_threshlines} shows that the number of significantly non-null brain representations is approximately constant regardless of choice of null distribution. 
However, Table \ref{tbl:sig_solo-breps} reveals that consensus (between choices of null model) on which brain representations contain non-null signal actually \textit{decreases} as $\alpha \to 0$. 
This carries the clear implication that brain representations cannot be meaningfully imposed with a strict ordering on information content. 
Rather, brain representations are differentially sensitive to varying information structures, including inter-subject variability.
Distance significance strongly depends on the choice of null distribution, and is largely uninformative when computed against the feature-shuffled null.
Comparing the subject- and feature-shuffled null results highlights the richness of data structure retained by the subject-shuffled null, emphasizing its utility as a subtle null model. 

Our results in this supplement also demonstrate the importance of the method used to quantify differences between subjects.
We compared the correlation dissimilarity (eq. \ref{eqn:corr_dissim}) to an inner-product derived quasi-cosine metric (eq. \ref{eqn:inner_dissim}) on vectorized data and the SPD geodesic (eq. \ref{Riem_dist_Pn}) on correlation matrices.
The differences in behavior of vectorized dissimilarity measures was most dramatic, with nearly all inner-product "distance"-derived persistence modules failing be significantly non-null (against either null).
The SPD geodesic showed less drastic a departure from the correlation dissimilarity, but still detected non-null structure in low-dimensional -- and obscured it in high-dimensional -- representations functional connectivity more often than the correlation dissimilarity.
Differences between the SPD geodesic and correlation dissimilarity also emerge when comparing representations, as the geodesic metric is more sensitive to variation in individual differences between functional connectivity derived from different parcellations.
A broader survey and deeper interrogation of measurements of inter-subject variability may be an important step in characterizing the recoverable information available in resting-state fMRI.

This supplement explored the implications of choosing different null distributions and inter-subject variability measurements. 
Both choices carry meaningful consequences for the analysis of post-processing pipeline flexibility we performed, and highlight the importance of \textit{a priori} experimental design with a principled alignment to the research question at hand.

\newpage 
\begin{spacing}{1.0}
\printbibliography[
heading=bibintoc,
title={Bibliography}
]

\end{spacing}

%% file: tables/significance_table.tex
\begin{table}
\centering
\begin{adjustbox}{height=0.45\textheight}
\begin{tabular}{|l|c|c|c|c|c|c|}
\hline
\textbf{\begin{tabular}[c]{@{}l@{}}Brain Representation\\ (reduction\_feature\_metric)\end{tabular}} & \textbf{\begin{tabular}[c]{@{}c@{}}nontrivial \\ (subject)\end{tabular}} & \textbf{\begin{tabular}[c]{@{}c@{}}nontrivial \\ (feature)\end{tabular}} & \textbf{\begin{tabular}[c]{@{}c@{}}$\alpha < 0.05$ \\ (subject)\end{tabular}} & \textbf{\begin{tabular}[c]{@{}c@{}}$\alpha < 0.05$ \\ (feature)\end{tabular}} & \textbf{\begin{tabular}[c]{@{}c@{}}$\alpha < \varepsilon$ \\ (subject)\end{tabular}} & \textbf{\begin{tabular}[c]{@{}c@{}}$\alpha < \varepsilon$ \\ (feature)\end{tabular}} \\ \hline
Glasser\_Amps\_Psim & \cmark & \cmark & \cmark & \cmark & \xmark & \cmark \\ \hline
Glasser\_Amps\_inner & \xmark & \xmark & \xmark & \xmark & \xmark & \xmark \\ \hline
Glasser\_NMs\_Psim-ztrans & \cmark & \cmark & \cmark & \cmark & \cmark & \cmark \\ \hline
Glasser\_NMs\_geodesic & \cmark & \cmark & \cmark & \cmark & \cmark & \cmark \\ \hline
Glasser\_pNMs\_Psim-ztrans & \cmark & \cmark & \cmark & \cmark & \cmark & \cmark \\ \hline
Glasser\_pNMs\_geodesic & \cmark & \cmark & \xmark & \xmark & \xmark & \xmark \\ \hline
ICA100\_Amps\_Psim & \cmark & \cmark & \cmark & \cmark & \cmark & \cmark \\ \hline
ICA100\_Amps\_inner & \xmark & \xmark & \xmark & \xmark & \xmark & \xmark \\ \hline
ICA100\_pNMs\_Psim-ztrans & \cmark & \cmark & \cmark & \cmark & \xmark & \cmark \\ \hline
ICA100\_pNMs\_geodesic & \cmark & \cmark & \cmark & \cmark & \cmark & \cmark \\ \hline
ICA15\_Amps\_Psim & \cmark & \cmark & \cmark & \cmark & \cmark & \cmark \\ \hline
ICA15\_Amps\_inner & \xmark & \xmark & \xmark & \xmark & \xmark & \xmark \\ \hline
ICA15\_pNMs\_Psim-ztrans & \cmark & \cmark & \xmark & \xmark & \xmark & \xmark \\ \hline
ICA15\_pNMs\_geodesic & \cmark & \cmark & \cmark & \cmark & \cmark & \cmark \\ \hline
ICA200\_Amps\_Psim & \cmark & \cmark & \cmark & \cmark & \cmark & \xmark \\ \hline
ICA200\_Amps\_inner & \xmark & \xmark & \xmark & \xmark & \xmark & \xmark \\ \hline
ICA200\_pNMs\_Psim-ztrans & \cmark & \cmark & \cmark & \cmark & \cmark & \cmark \\ \hline
ICA200\_pNMs\_geodesic & \cmark & \cmark & \cmark & \cmark & \xmark & \cmark \\ \hline
ICA25\_Amps\_Psim & \cmark & \cmark & \cmark & \cmark & \cmark & \cmark \\ \hline
ICA25\_Amps\_inner & \xmark & \xmark & \xmark & \xmark & \xmark & \xmark \\ \hline
ICA25\_pNMs\_Psim-ztrans & \cmark & \cmark & \xmark & \cmark & \xmark & \cmark \\ \hline
ICA25\_pNMs\_geodesic & \cmark & \cmark & \cmark & \cmark & \xmark & \xmark \\ \hline
ICA300\_Amps\_Psim & \cmark & \cmark & \cmark & \cmark & \xmark & \cmark \\ \hline
ICA300\_Amps\_inner & \xmark & \xmark & \xmark & \xmark & \xmark & \xmark \\ \hline
ICA300\_pNMs\_Psim-ztrans & \cmark & \cmark & \cmark & \cmark & \cmark & \cmark \\ \hline
ICA300\_pNMs\_geodesic & \cmark & \cmark & \xmark & \cmark & \xmark & \cmark \\ \hline
ICA50\_Amps\_Psim & \cmark & \cmark & \cmark & \cmark & \cmark & \cmark \\ \hline
ICA50\_Amps\_inner & \xmark & \xmark & \xmark & \xmark & \xmark & \xmark \\ \hline
ICA50\_pNMs\_Psim-ztrans & \cmark & \cmark & \cmark & \cmark & \xmark & \cmark \\ \hline
ICA50\_pNMs\_geodesic & \cmark & \cmark & \cmark & \cmark & \xmark & \cmark \\ \hline
PROFUMO\_Maps\_Psim & \cmark & \cmark & \cmark & \cmark & \cmark & \cmark \\ \hline
PROFUMO\_Maps\_inner & \cmark & \cmark & \xmark & \cmark & \xmark & \xmark \\ \hline
PROFUMO\_spNMs\_Psim-ztrans & \cmark & \cmark & \cmark & \cmark & \xmark & \cmark \\ \hline
PROFUMO\_spNMs\_geodesic & \cmark & \cmark & \cmark & \cmark & \cmark & \cmark \\ \hline
Schaefer1000\_Amps\_Psim & \cmark & \cmark & \cmark & \cmark & \xmark & \cmark \\ \hline
Schaefer1000\_Amps\_inner & \cmark & \cmark & \xmark & \xmark & \xmark & \xmark \\ \hline
Schaefer100\_Amps\_Psim & \cmark & \cmark & \cmark & \xmark & \cmark & \xmark \\ \hline
Schaefer100\_Amps\_inner & \xmark & \xmark & \xmark & \xmark & \xmark & \xmark \\ \hline
Schaefer100\_pNMs\_Psim-ztrans & \cmark & \cmark & \cmark & \cmark & \xmark & \cmark \\ \hline
Schaefer100\_pNMs\_geodesic & \cmark & \cmark & \cmark & \cmark & \cmark & \xmark \\ \hline
Schaefer200\_Amps\_Psim & \cmark & \cmark & \cmark & \xmark & \cmark & \xmark \\ \hline
Schaefer200\_Amps\_inner & \xmark & \xmark & \xmark & \xmark & \xmark & \xmark \\ \hline
Schaefer200\_pNMs\_Psim-ztrans & \cmark & \cmark & \cmark & \cmark & \xmark & \cmark \\ \hline
Schaefer200\_pNMs\_geodesic & \cmark & \cmark & \xmark & \xmark & \xmark & \xmark \\ \hline
Schaefer300\_Amps\_Psim & \cmark & \cmark & \cmark & \cmark & \cmark & \cmark \\ \hline
Schaefer300\_Amps\_inner & \xmark & \xmark & \xmark & \xmark & \xmark & \xmark \\ \hline
Schaefer300\_pNMs\_Psim-ztrans & \cmark & \cmark & \cmark & \cmark & \cmark & \cmark \\ \hline
Schaefer300\_pNMs\_geodesic & \cmark & \cmark & \cmark & \cmark & \xmark & \xmark \\ \hline
Schaefer600\_Amps\_Psim & \cmark & \cmark & \cmark & \cmark & \xmark & \cmark \\ \hline
Schaefer600\_Amps\_inner & \xmark & \xmark & \xmark & \xmark & \xmark & \xmark \\ \hline
Schaefer600\_pNMs\_Psim-ztrans & \cmark & \cmark & \cmark & \cmark & \cmark & \cmark \\ \hline
Schaefer600\_pNMs\_geodesic & \xmark & \xmark & \xmark & \xmark & \xmark & \xmark \\ \hline
Yeo17\_Amps\_Psim & \cmark & \cmark & \cmark & \cmark & \cmark & \cmark \\ \hline
Yeo17\_Amps\_inner & \xmark & \xmark & \xmark & \xmark & \xmark & \xmark \\ \hline
Yeo17\_NMs\_Psim-ztrans & \cmark & \cmark & \cmark & \cmark & \xmark & \xmark \\ \hline
Yeo17\_NMs\_geodesic & \cmark & \cmark & \cmark & \cmark & \cmark & \cmark \\ \hline
Yeo17\_pNMs\_Psim-ztrans & \cmark & \cmark & \xmark & \cmark & \xmark & \xmark \\ \hline
Yeo17\_pNMs\_geodesic & \cmark & \cmark & \cmark & \cmark & \xmark & \xmark \\ \hline
DiffGradient100\_Maps\_Psim & \cmark & \cmark & \cmark & \cmark & \xmark & \cmark \\ \hline
DiffGradient100\_Maps\_inner & \cmark & \cmark & \xmark & \xmark & \xmark & \xmark \\ \hline
DiffGradient15\_Maps\_Psim & \cmark & \cmark & \cmark & \cmark & \xmark & \cmark \\ \hline
DiffGradient15\_Maps\_inner & \cmark & \cmark & \xmark & \xmark & \xmark & \xmark \\ \hline
DiffGradient1\_Maps\_Psim & \cmark & \cmark & \cmark & \cmark & \cmark & \cmark \\ \hline
DiffGradient1\_Maps\_inner & \cmark & \cmark & \cmark & \xmark & \xmark & \xmark \\ \hline
DiffGradient200\_Maps\_Psim & \cmark & \cmark & \cmark & \cmark & \xmark & \xmark \\ \hline
DiffGradient25\_Maps\_Psim & \cmark & \cmark & \cmark & \cmark & \xmark & \cmark \\ \hline
DiffGradient25\_Maps\_inner & \cmark & \cmark & \xmark & \xmark & \xmark & \xmark \\ \hline
DiffGradient300\_Maps\_Psim & \cmark & \xmark & \cmark & \xmark & \cmark & \xmark \\ \hline
DiffGradient50\_Maps\_Psim & \cmark & \cmark & \cmark & \cmark & \xmark & \cmark \\ \hline
DiffGradient50\_Maps\_inner & \cmark & \cmark & \xmark & \xmark & \xmark & \xmark \\ \hline
\end{tabular}
\end{adjustbox}
\caption[Table of significant brain representations.]{All brain representations and topological non-null significance. A green check-mark \cmark indicates that the brain representation is significant at the given threshold; a red 'X' \xmark indicates that it is not.}
\label{tbl:sig_solo-breps}
\end{table}